\def\sfh{\hbox{$\ \!\!^{\rm h}$}}
\def\sfm{\hbox{$\ \!\!^{\rm m}$}}

\newcommand{\corot}[1][]{CoRoT{#1}}
\newcommand{\galaxia}[1][]{\textit{Galaxia}{#1}}

\documentclass{aa}  

\usepackage{txfonts}
 \usepackage[utf8]{inputenc}
\usepackage{graphicx}
\usepackage{supertabular}
\begin{document}

\title{Stellar classification of CoRoT targets}
\author{C. Damiani\inst{1,2}, J.-C. Meunier\inst{1}, C. Moutou\inst{1}, M. Deleuil\inst{1}, N. Ysard\inst{2}, F. Baudin\inst{2}, and H. Deeg\inst{3,4}}
\authorrunning{C. Damiani, J.-C. Meunier, C. Moutou et al.}
\institute{Aix-Marseille Universit{\'e}, CNRS, Laboratoire d\rq{}Astrophysique de Marseille, UMR 7326, 13388, Marseille, France\\
\email{cilia.damiani@ias.u-psud.fr}
\and
Universit{\'e} Paris-Sud, CNRS, Institut d'Astrophysique Spatiale, UMR8617, 91405 Orsay Cedex, France
\and
Instituto de Astrof{\'i}sica de Canarias (IAC), Calle V{\'i}a L{\'a}ctea s/n, 38200 La Laguna, Tenerife, Spain
\and
Universidad de La Laguna, Dept. de Astrof\'\i sica, E-38206 La Laguna, Tenerife, Spain}

\date{Accepted xxx. Received xxx; in original form xxx}
\abstract
{The \corot{} mission was the first dedicated to the search for exoplanets from space. The \corot{} exoplanet channel observed about 163 600 targets to detect transiting planetary companions. In addition to the search for exoplanets, the extremely precise photometric time series provided by \corot{} for this vast number of stars is an invaluable resource for stellar studies. Because \corot{} targets are faint ($11\leq r \leq16$) and close to the galactic plane, only a small subsample has been observed spectroscopically. Consequently, the stellar classification of \corot{} targets required the design of a classification method suited for the needs and time frame of the mission.}
{We describe the latest classification scheme used to derive the spectral type of \corot{} targets, which is based on broadband multi-colour photometry. We assess the accuracy of this spectral classification for the first time.}
{We validated the method on simulated data. This allows the quantification of the effect of different sources of uncertainty on the spectral type. Using galaxy population synthesis models, we produced a synthetic catalogue that has the same properties as the \corot{} targets. In this way, we are able to predict typical errors depending on the estimated luminosity class and spectral type. We also compared our results with independent estimates of the spectral type. Cross-checking those results allows us to identify the systematics of the method and to characterise the stellar populations observed by \corot{}.}
{We find that the classification method performs better for stars that were observed during the mission-dedicated photometric ground-based campaigns.The luminosity class is wrong for less than 7\% of the targets. Generally, the effective temperature of stars classified as early type (O, B, and A) is overestimated. Conversely, the temperature of stars classified as later type tends to be underestimated. This is mainly due to the adverse effect of interstellar reddening. We find that the median error on the effective temperature is less than 5\% for dwarf stars classified with a spectral later than F0, but it is worse for earlier type stars, with up to 20\% error  for A and late-B dwarfs, and up to 70\% for early-B and O-type dwarfs. Similar results are found for giants, with a median error that is lower than 7\% for  G- and later type  giants, but greater than 25\% for earlier types. Overall, we find an average median absolute temperature difference $|\Delta T_{\rm eff}| = 533\pm6$~K for the whole sample of stars classified as dwarfs and $|\Delta T_{\rm eff}| = 280\pm3$~K for the whole sample of giant stars. The corresponding standard deviation is of about $925\pm 5$~K for dwarfs and $304\pm4$~K for giants. Typically for late-type stars, this means that the classification is accurate to about half a class.}
{}
\maketitle
\section{Introduction}
During its 2729 days in orbit, the faint stars channel of the \corot{} \citep[COnvection ROtation \& planetary Transits; ][]{Baglin2006} mission observed about 163 600 targets over 26 runs, 18 of which were long runs with durations of more than 150 days,  while 8 were short runs with typical duration of 30 days. This resulted in the detection of 33 planets to date. The extremely precise photometric time series provided by the faint stars channel can also serve a large number of additional scientific objectives. To ensure the selection of the targets and the exploitation of the light curve, a minimum knowledge of the targets is required, such as their colours or spectral type and luminosity class.

\begin{figure*}
\begin{center}
\includegraphics[width=0.32\textwidth]{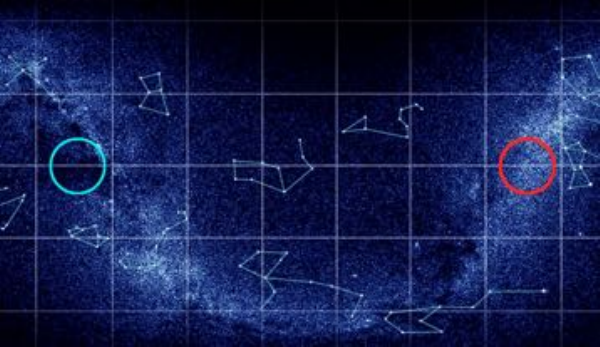}\quad
\includegraphics[width=0.32\textwidth]{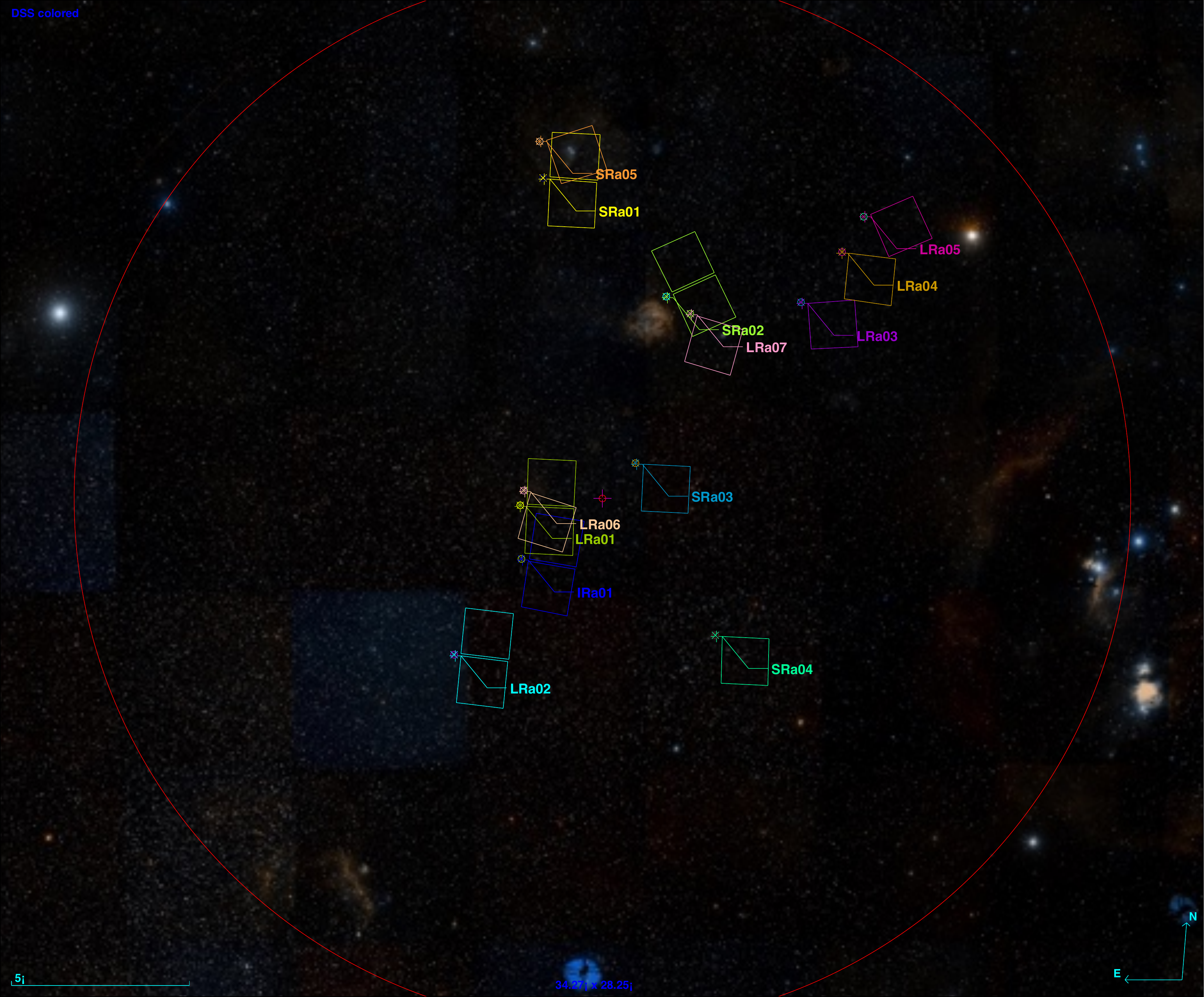}\quad
\includegraphics[width=0.32\textwidth]{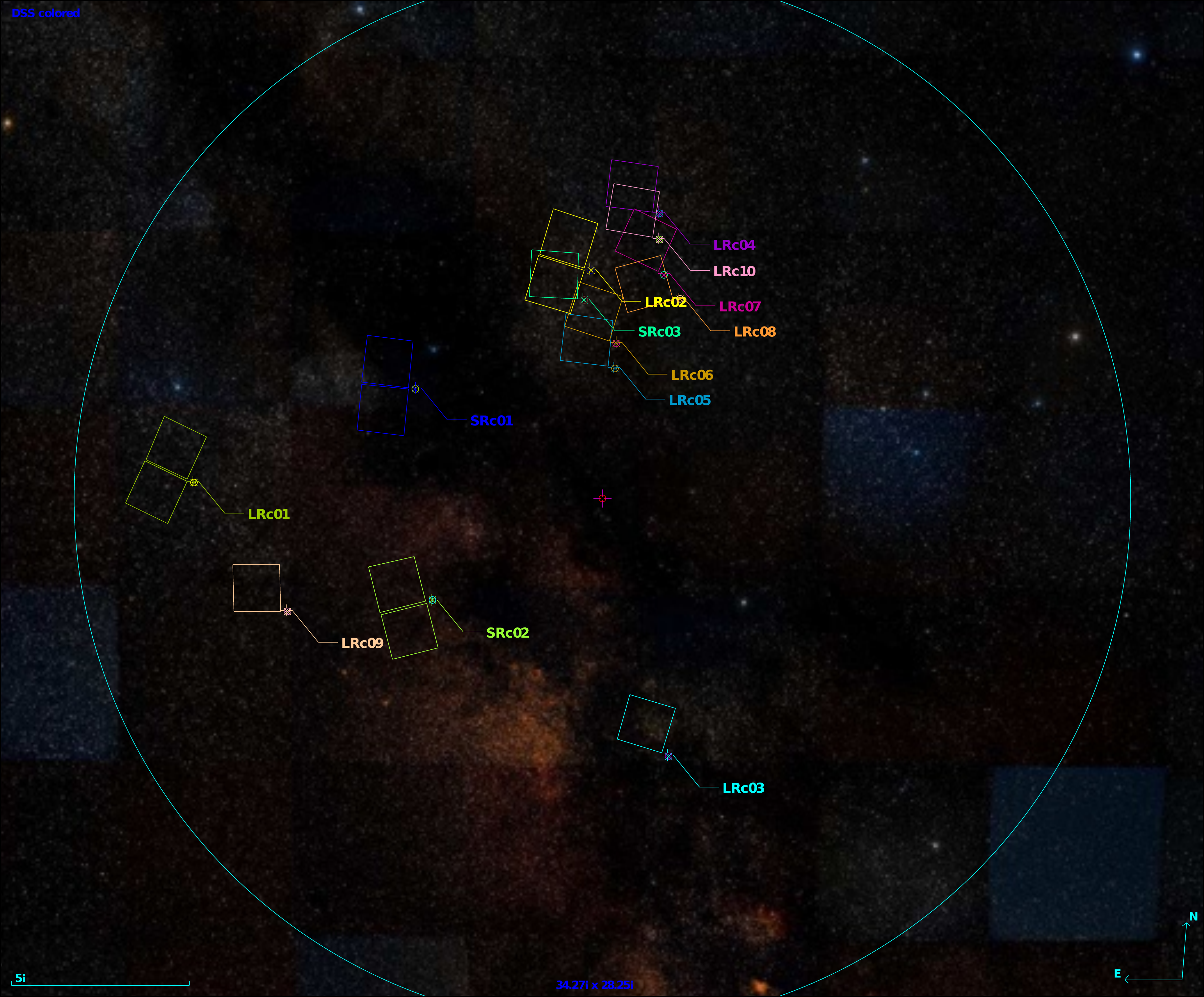}
\caption[So-called CoRoT eyes.]{So-called CoRoT eyes. Left: the two observable zones of CoRoT are highlighted by two circles in this map of the sky; the red and blue circles are in the so-called centre and anticentre directions, respectively. Credits: CNES. Middle: STSci Digitized Sky Survey negative-coloured view of the anticentre direction. The footprints of the CCDs of the different runs of observation are given in different colours. Right: same image but for the centre direction.}
\label{fig:corot_eyes}
\end{center}
\end{figure*}

This is one of the reasons why the Exodat \citep{Deleuil2009} database was built. Its prime objective is to provide positions, colours, and stellar classification for the stars in the observable zones of \corot{}, the so-called \corot{} eyes. Those eyes are two circular regions of about $15^\circ$ radius each, centred on the Galactic plane (i.e. in the direction perpendicular to the polar orbit of CoRoT), in opposite directions at $6\sfh50\sfm$ and $18\sfh50\sfm$ in right ascension (RA) (see Fig. \ref{fig:corot_eyes}). While not close to the centre of the Galaxy, these two directions of pointing are referred to as the anticentre and centre directions.  
Exodat is an information system and a database that gathers an extended body of data of various origins. It includes a search engine for exploring and displaying data, which is available to the community through a user friendly web interface\footnote{This service was described as Exo-Dat by  \citet{Deleuil2009}; it can be found at \url{http://cesam.lam.fr/exodat/}}\citep{2014ASPC..485..203A}. It contains data for more than 51 million stars to provide information for both the potential targets of the exoplanet programme (with $11 \leq r \leq 16$), and also the fainter background sources (down to about the $19^{\rm th}$ V magnitude) to estimate the level of light contamination of the targets. 

The stellar classification of the potential targets required the design of a classification method suited for the needs and time frame of the mission. Considering the extent of the database, the use of spectroscopy was deemed impossible. Consequently, a first order stellar classification (FOSC) was conducted with the aid of broadband multi-colour photometry in a wide spectral range. Of course, any detailed analysis of targets of particular interest would require complementary observations for a better characterisation. Indeed this has been applied on a target-to-target basis, especially for planet hosting stars, but also on more consequential samples (see Sec.~\ref{sec:spectro}). Overall, no more than 15~\% of the targets observed over the  lifetime of the mission were the object of complementary observations.  

Notwithstanding, stellar parameters are essential to the scientific exploitation of the \corot{} light curves. For the vast majority of targets, to this day, the only source of information concerning the spectral type of the star is Exodat. The content of Exodat is also used to feed the complementary information delivered together with the light curves\footnote{available at \url{http://idoc-corot.ias.u-psud.fr}}  that are now publicly available for all CoRoT targets.

 While an overview of the FOSC method is given in \citet{Deleuil2009}, to date there is no description in the literature of the latest classification scheme, and more importantly, nor is there an estimation of the reliability of the final results. In this paper, we give a thorough description of the method and its inputs in Sec.~\ref{sec:method}. We present an estimation of the precision and accuracy of the FOSC using simulated data in Sec.~\ref{sec:accuracy}. Then in Sec.~\ref{sec:gal}, we use a synthetic galactic population model to obtain statistically relevant estimates of the typical error made on the spectral type for each spectral class. In Sec.~\ref{sec:spectro} we compare the results of the classification method to the classification obtained by spectroscopic campaigns. We discuss the properties of the stellar populations observed by \corot{} in Sec.~\ref{sec:stelpop}. We give our conclusions in Sect.~\ref{sec:conclusion}.

\section{Classification method}\label{sec:method}
 The FOSC method is adapted from the automatic classification technique developed by \citet{Hatziminaoglou2002} for celestial objects classification and quasar identification. This method consists in fitting the apparent broadband photometry magnitudes of the target to the spectral energy distribution (SED) of template stars of different spectral types.
 \subsection{Spectral templates}
The templates are taken from the Pickles stellar spectral flux library UVKLIB \citep{Pickles1998}. This library, put together using observed spectra, has a wide spectral coverage that ranges from the near-UV up to 2.5 $\mu$m with a spectral resolution of about 500. It is organised following the Morgan-Keenan system, encompassing spectral types from O to M and luminosity classes from supergiants to dwarfs with good completeness and uniformity \citet{Pickles1998}. For the G-K dwarf and giant branches, the library also has a few templates for non-solar chemical abundances. To avoid introducing too many degeneracies, only the solar abundance templates are used in the FOSC.
\begin{figure}
\begin{center}
\includegraphics[width=1.0\columnwidth]{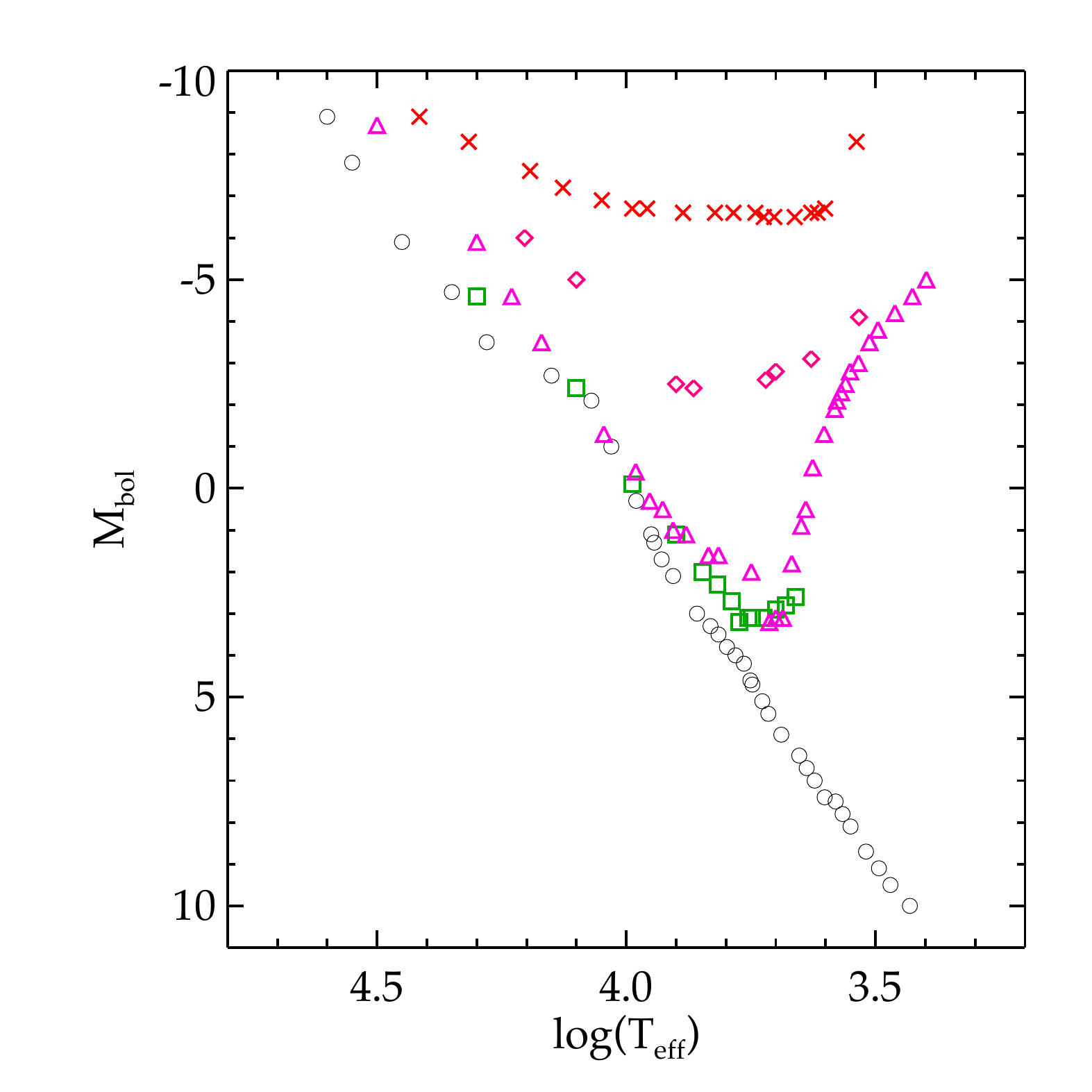}
\caption{Bolometric magnitude as a function of the logarithm of the
effective temperature for the templates of the Pickles library for luminosity class V (black
circles), IV (green squares), III (pink triangles), II (magenta diamonds), and I (red
crosses).}
\label{fig:pickles}
\end{center}
\end{figure}

Templates of extragalactic sources, white dwarfs, and brown dwarfs are not used in the FOSC because it is expected that the number of such objects in the considered range of magnitudes and galactic regions is small enough to be safely neglected. Templates of binary stars are not used either even though the proportion of binary stars in the sample is expected to be significant. Again, this limitation is conceded to avoid too many degeneracies because broadband photometry does not allow us to clearly distinguish the binary nature of the target. If the spectral type of the two components are similar, the resulting spectral classification would give the same result when using the template of a single star since it is insensitive to the absolute magnitude of the target. On the other hand, if the spectral type of the two components are very different, the colours of the binary would largely be undistinguishable from the colours of the brightest component because of the extremely different luminosities of the two components in the considered wavelength range. In those cases, the spectral type of a binary determined with the FOSC method thus corresponds to that of the primary star. In intermediate cases, we expect the spectral type determination to be degraded when the target is a binary. The binary nature of the target should be known before attempting the classification, however, to ensure the reliability of the result. This was not the case prior to the mission, but the results of the planetary transit search in \corot{} light curves has identified a number of eclipsing binaries. In this case, a second order classification that is suited for binary targets would produce better results. This is out of the scope of this paper and will be tackled in a dedicated study of \corot{} binaries. 

Overall, the classification is carried out over 106 different spectral types distributed in 5 luminosity classes. There are 36 templates for dwarfs stars, which roughly corresponds to a sampling of about one template every two subclasses in the main sequence. The templates of the Pickles library are given with an effective temperature scale, but the corresponding surface gravity is not available and consequently it is not an output of the FOSC. For reference, the correspondence between spectral type and effective temperature is given in Appendix \ref{appA}. The bolometric magnitude and effective temperature of the templates used by the FOSC are plotted on Fig.~\ref{fig:pickles}.
\subsection{Fitting the observed SED to the templates}    
  The templates of the UVKLIB library are used to produce a library of reddened reference SEDs to take the effect of interstellar absorption into account. They were created for a range of colour excess from $E_{B-V}=0$ to $E_{B-V}=4$ in steps of 0.05. The extinction law is taken from \citet{Fitz1999} using the average value of the ratio of total to selective extinction $R_V = \frac{A_V}{E_{B-V}}=3.1$. The actual value of $R_V$ does not matter here given that in the wavelength range considered in this study ($350\ {\rm nm}\leq\lambda \leq 2.2\,\mu$m), the extinction curve is insensitive to $R_V$ variations. For a given value of the colour excess $E_{B-V}$, the reddened flux distribution is then multiplied with the appropriate relative spectral response curves of the different bands. The set of filters and their characteristics are described in Sect.~\ref{sec:phot}. Integrating the transmitted flux over the wavelength  range of the filter provides the reddened template fluxes in the different bands. We decided to use magnitudes in the Vega system and observed magnitudes are converted to this system if required. The spectral energy distribution of $\alpha$~Lyrae (Vega) was treated in the same way to obtain the reference flux in every band.

For each star, the apparent magnitudes are converted to fluxes that are normalised to the flux in the band where the photometric error is the smallest. The photometric error $\epsilon_i$ is also converted to an error in flux $\sigma_i$, taking into account the error on the zero point $\epsilon_{0i}$ for the corresponding filter $i$, using the relationship
\begin{equation}\label{eq:errortheo}
\sigma_i= \frac{2.3}{2.5}\sqrt{\epsilon_i^2 + \epsilon_{0i}^2}
.\end{equation}
The numerical factor in the right-hand side of Eq.\ref{eq:errortheo} comes from the Taylor expansion of the decimal logarithm used in the computations of magnitudes. The observed flux is then confronted with the computed fluxes for all the reddened templates by computing
 \begin{equation}\label{Eq:chi2}
 \chi^2=\sum_{i=1}^n \frac{\left(F^i_{\rm obs} - F^i_{\rm mod}(E_{B-V})\right)^2}{\sigma_i^2}
 ,\end{equation}
 where $F^i_{\rm obs} $ and $F^i_{\rm mod}$ are the observed and template fluxes in the $i$ band, respectively, $\sigma_i$ is the estimated error of the observed flux in this band, and $n$ is the number of bands that are used.

\begin{figure}
\begin{center}
\includegraphics[width=0.45\columnwidth]{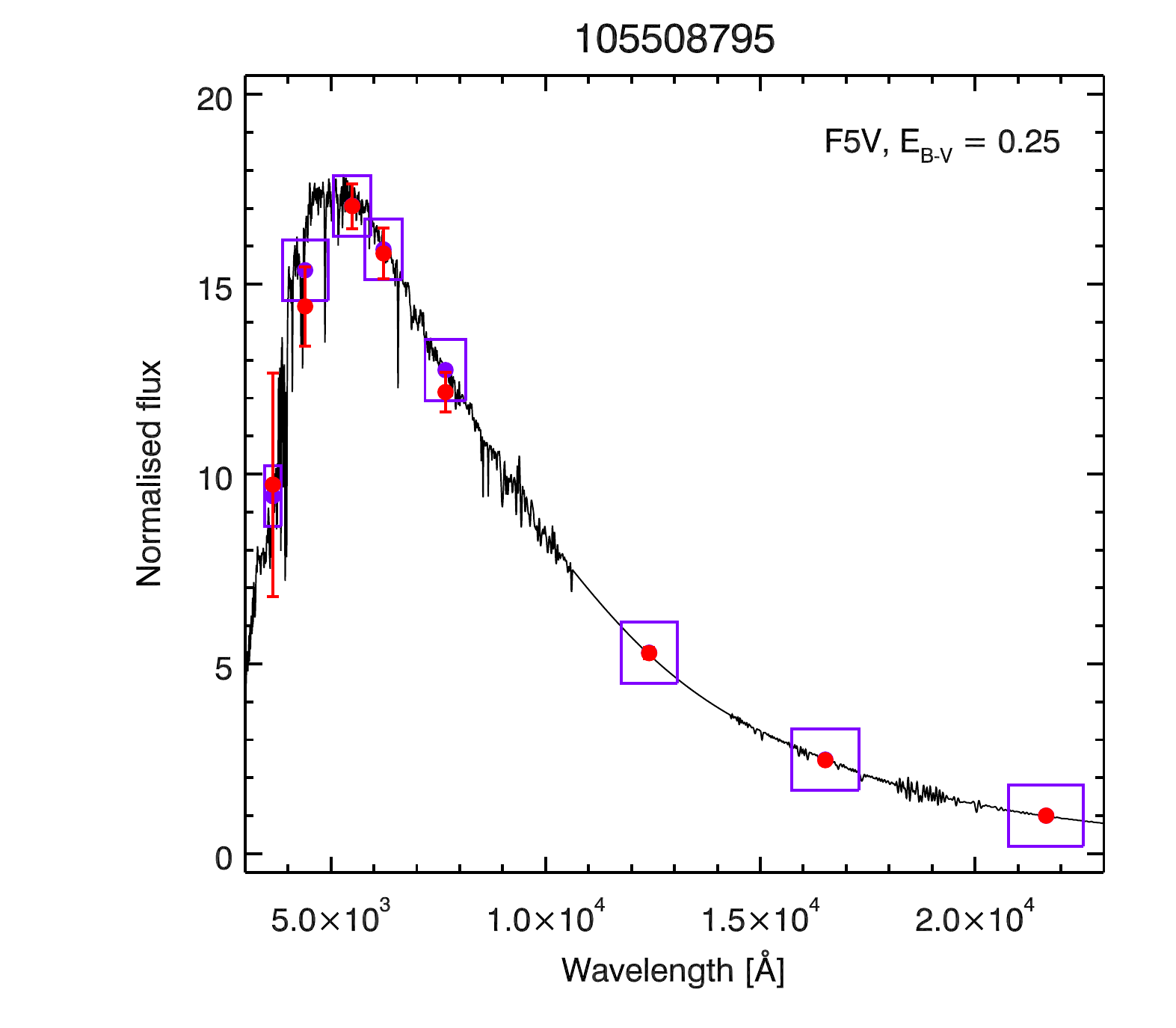}\quad
\includegraphics[width=0.45\columnwidth]{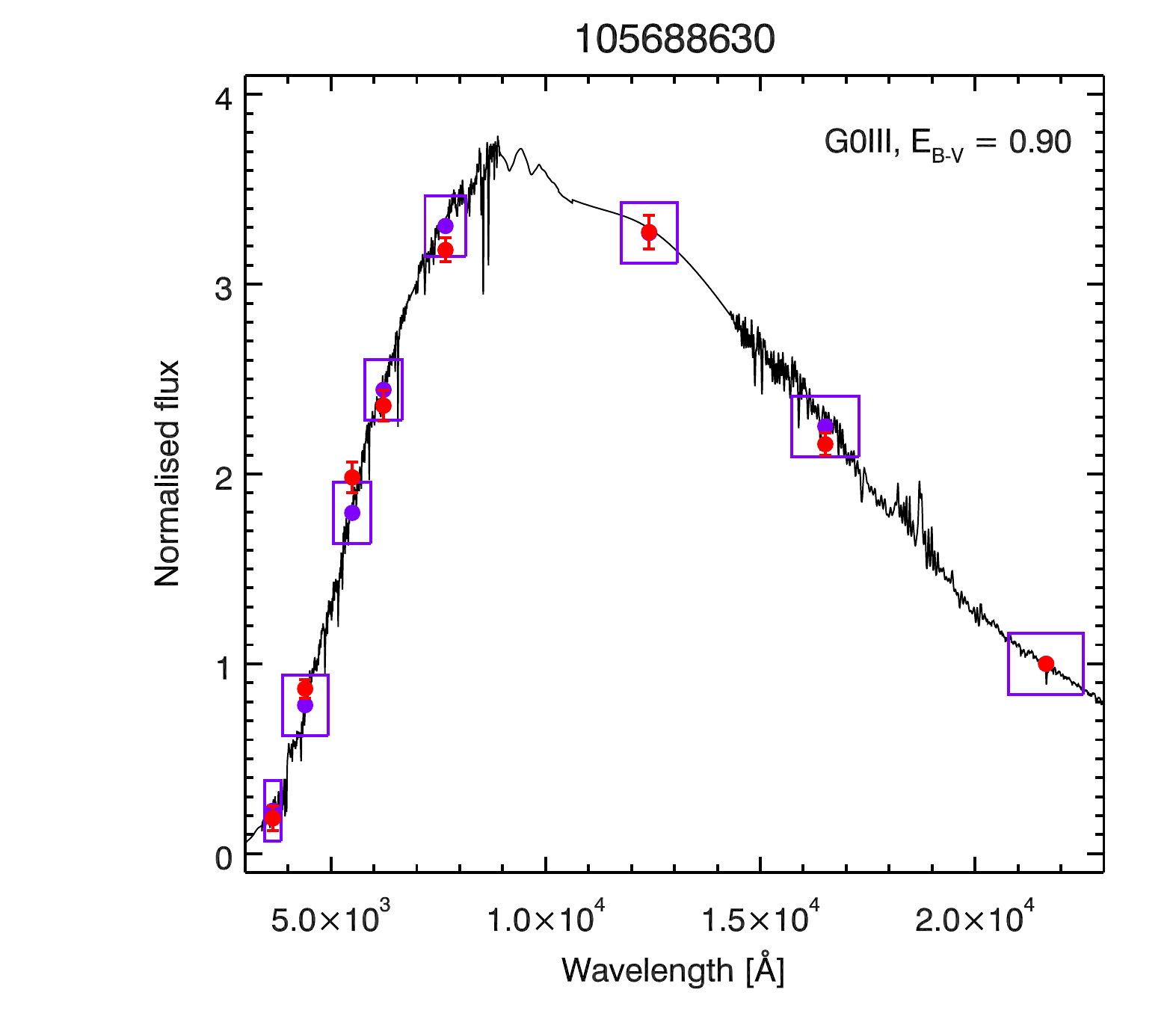}\\
\includegraphics[width=0.45\columnwidth]{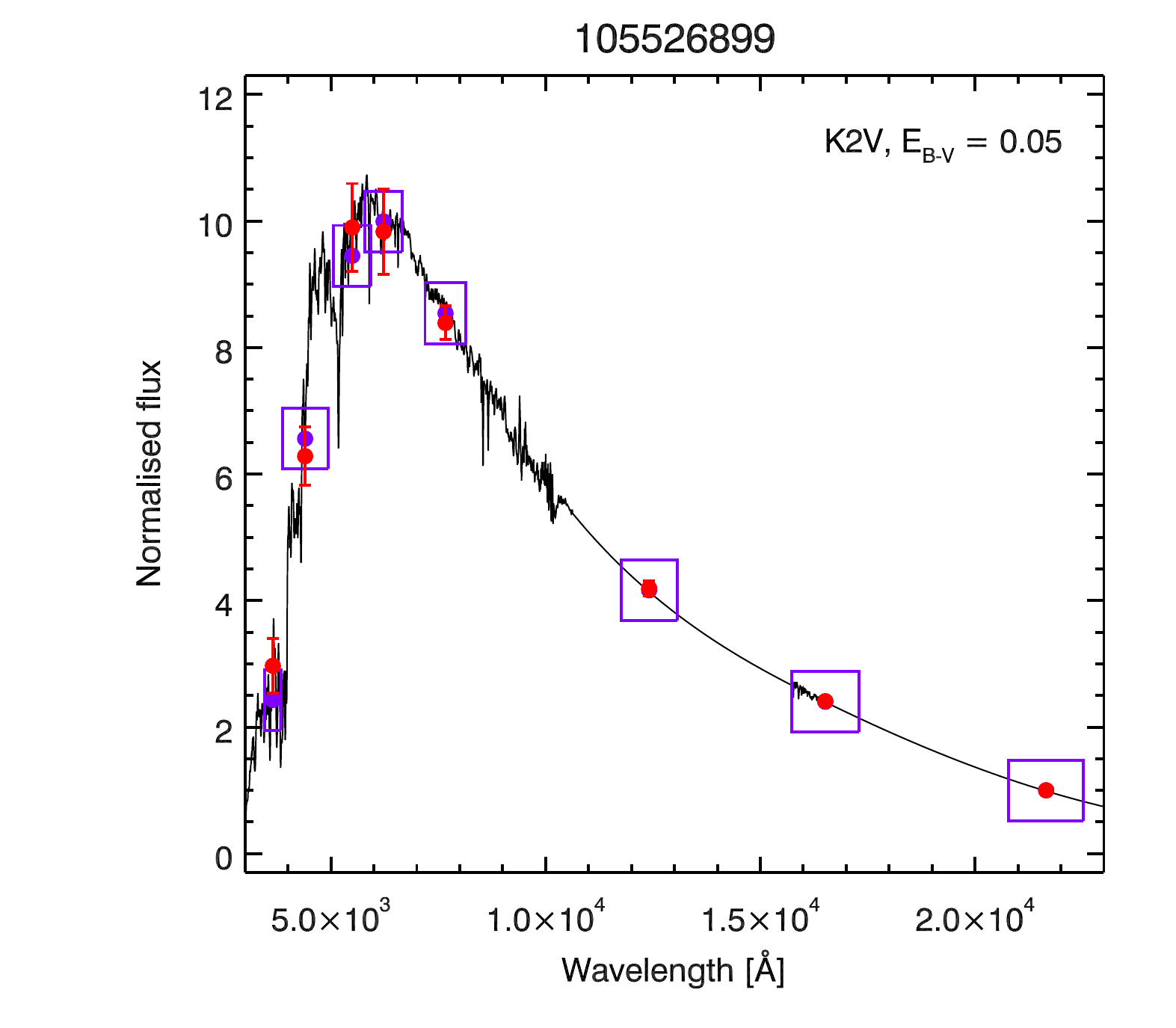}\quad
\includegraphics[width=0.45\columnwidth]{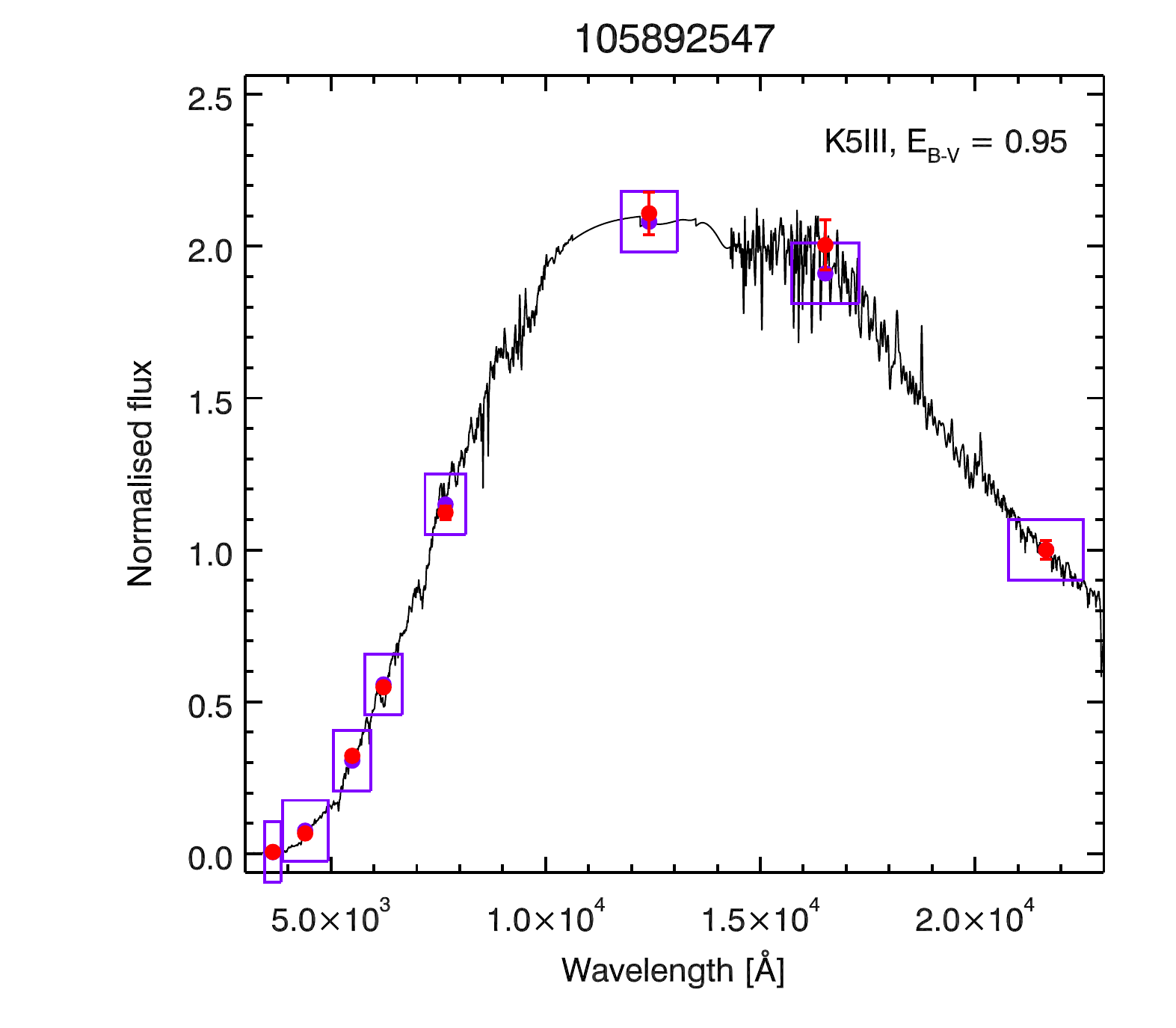}
\caption{Spectral energy distribution of four \corot{} targets. The observed magnitudes are converted to fluxes and normalised to the band with the smallest photometric error (red dots with error bars). The FOSC returns the best matching pair of spectral type and $E_{B-V}$ in the library of templates. Their values are given in the top right corner of every plot, and the corresponding spectrum is over-plotted in solid black line. The match is obtained by computing the flux of the reddened templates convolved by the filters' response, $F^i_{\rm mod}(E_{B-V})$, represented as purple dots. The purple box width gives the bandwidth of the corresponding filter.}
\label{fig:exSED}
\end{center}
\end{figure}

Finding the minimum $\chi^2$ then leads to the attribution of a spectral type and a colour excess that best reproduce the observed magnitudes. The reduced $\chi^2$ is used as quality index QI. We use $n$ magnitudes to determine two parameters: the spectral type and $E_{B-V}$. In addition, Eq.~\ref{Eq:chi2} is normalised to the passband with the smallest photometric error. This subtracts one degree of freedom from the $\chi^2$ analysis. Hence, QI is computed as
\begin{equation}\label{eq:QI}
QI=\chi^2/(n - 3).
\end{equation}
Examples of the result of the FOSC are given in Fig.~\ref{fig:exSED}. This figure shows the reddened SED of the best-fitting spectral type and $E_{B-V}$ together with the observed magnitudes of four targets.

As already pointed out by \citet{Hatziminaoglou2002}, a classification scheme based on the $\chi^2$ technique may lead to degeneracies that are class dependent. Multiple minima may occur in the parameter space and such degeneracies can only be solved by including additional information in the classification procedure. In the following we first describe the inputs used by the FOSC, and then how we use additional information to avoid the main degeneracies.

\subsection{Broadband multi-colour photometry}  \label{sec:phot}
    Very little was known about the faint stars in the \corot{} eyes prior to the launch. Thus a large observational campaign was performed to prepare the pointings and target selection of the satellite. Photometric observations started in June 2002, with observing campaigns in winters and in summers in order to cover the centre and anticentre directions. These observations were described in \citet{Deleuil2009}. They formed the first catalogue of the database that we name  \lq\lq{}Obscat\rq\rq{}\footnote{\citet{Deleuil2009} used the term "Exo-Cat" for this set of data. Obscat is available online at \url{http://cesam.lam.fr/exodat/}}. This catalogue provides photometry in Harris $B$ and $V$, Sloan-Gunn $r'$ and $i'$ filters. An RGO $U$ filter was also used, but because long exposures are required, Obscat only provides U-band photometry over small areas. The magnitudes are calibrated in the Vega system. To ensure a better classification of late-type stars, the spectral coverage was extended, and the infrared magnitudes of the point source catalogue issued by the 2 Micron All Sky Survey were also used \citep[2MASS; ][]{Kleinmann1994}. This formed the second catalogue of the database that was crossmatched with Obscat. 
    
    To fulfil the objectives of other scientific programmes, some runs were chosen in regions that were not covered by Obscat. Photometry in these regions needed to be obtained from publicly available catalogues that could complement Obscat, and therefore more catalogues were added to the database. Eventually, the PPMXL catalogue was chosen as the reference catalogue for Exodat when it became available. The PPMXL catalogue \citep{Roeser2010} combines the USNOB1.0 \citep{Monet2003} and 2MASS catalogues and provides accurate position and mean motions in the ICRS. The PPMXL catalogue aims to be complete from the brightest stars down to about $V\approx20$, and covers both \corot{} eyes. Essentially, USNO-B1.0 is  a three-colour (photographic \lq{}\lq{}blue", \lq{}\lq{}red", and \lq{}\lq{}infrared"), two epoch catalogue. It is one of the largest surveys in the visible that is publicly available, and the only survey that is deep enough to fulfil the two main purposes of the database: spectral typing and assessing of the level of contaminating light from background sources. To ensure the homogeneity of the whole database, Obscat was carefully crossmatched with PPMXL.
    
     Thus, the FOSC is carried out using visible magnitudes from Obscat  when available and from PPMXL when not available from Obscat, and infrared magnitudes from 2MASS are used. Since PPMXL gives blue and red magnitudes for two epochs, we use in priority the more precise photometry of the second epoch, and only resort to using the first epoch magnitude when no other blue or red magnitude is available. Overall, the statistics of the magnitudes used as input for the FOSC are listed in table~\ref{tab:magst}.
\begin{table}
        \centering
        \caption{Magnitudes used for the FOSC. The \lq{}\lq{}blue", \lq{}\lq{}red" and \lq{}\lq{}infrared" magnitudes are only used when the equivalent magnitude is missing in Obscat.}
        \begin{tabular}{lrr}
        \hline
        \hline
        \rule{0pt}{2ex}
        Filter     &  Number of stars & \% of all targets \\
        \hline
        \hline
        \rule{0pt}{2ex} 
        RGO $U$& 5 495 & 3\%\\
        \hline
        \rule{0pt}{2ex}
        Harris $B$&100 145&62\%\\
        \lq{}\lq{}Blue" second epoch&52 502&36\%\\
        \lq{}\lq{}Blue" first epoch&2 635&2\%\\
        \hline
        \rule{0pt}{2ex}
        Harris $V$&100 362&62\%\\
        \hline
        \rule{0pt}{2ex}
        Sloan-Gunn $r'$&100 692&62\%\\
        \lq{}\lq{}Red" second epoch&37 309&23\%\\
        \lq{}\lq{}Red" first epoch&23 050&14\%\\
        \hline
        \rule{0pt}{2ex}
        Sloan-Gunn $i'$&100 536&62\%\\
        \lq{}\lq{}Infrared"&58 347&36\%\\
        \hline
        \rule{0pt}{2ex}
        2MASS $J$&158 863& 98\%\\
        \hline
        \rule{0pt}{2ex}
        2MASS $H$& 159 066&98\% \\
        \hline
        \rule{0pt}{2ex}
        2MASS $K_s$& 158 415&98\%\\     
        \hline
        \hline
        \end{tabular}
        \label{tab:magst}
\end{table}

The responses of the filters used to produce the computed flux $F^i_{\rm mod}(E_{B-V})$ were thus chosen accordingly. For the $U$, $B$, $V$, $r'$,  and $i'$, we used the overall response of the filters convolved with the CCD efficiency of the Wide Field Camera made available by the Cambridge Astronomical Survey Unit for the Wide Field Survey\footnote{Details about the Wide Field Survey are available in the web page \url{http://www.ast.cam.ac.uk/~mike/wfcsur/}}.
We also used the zero point errors given in these web pages. For the infrared magnitudes $J$, $H$, and $K_s$, we used zero point error and relative spectral response curves of \citet{Cohen2003}. These include the effects of the optics and camera, but also a model of atmospheric transmission representative of typical survey conditions. This model includes the effects of aerosols and water vapour content appropriate to the desert conditions of the 2MASS sites, as detailed in \citet{Cohen2003}.

The USNO-B1.0 photometry is compiled from the digitisation of various photographic sky survey plates. To produce the computed flux $F^i_{\rm mod}(E_{B-V})$, we thus use the resultant spectral response of the original plates, measured to account for the photographic emulsion, filters, and instrument and atmospheric transmission as found in \citet{Lund1973} and \citet{Reid1991} for the two epochs. 

The photometric calibration of the USNO-B1.0 catalogue is of marginal quality. As detailed in \citet{Monet2003}, the root of the problem is the lack of suitable faint photometric standards. The calibration strategy varies from plate to plate, and combining all the 632 827 calibration stars spread over 3281 plates has a standard deviation of 0.25 mag. According to those authors, many plates show substantially smaller errors, but solutions for plates without photometric calibrators may be worse, and these represent 56\% of all plates. It is thus difficult to estimate the error on the zero points. Using the value of 0.25 mag cited earlier makes the classification largely insensitive to the corresponding data. In order to give some weight to USNO-B1.0 magnitudes, we set the zero point error arbitrarily to a value of 0.1 mag in all filters. Moreover, the original catalogue does not provide an error for each individual measurement, but is estimated to have an accuracy of about 0.3 mag. Acknowledging that the magnitudes from USNO-B1.0 should be used with care, we used a relative error of 3\% for all filters. On the other hand, the typical relative error in Obscat is about 1\%. Consequently, the quality of the fit when using USNO-B1.0 photometry is expected to be degraded. As can be seen in Tab.~\ref{tab:magst}, this concerns about a third of all the \corot{} targets.

\subsection{Dwarfs and giants degeneracy}\label{dwarfVSgiants}
\begin{figure}
\begin{center}
\includegraphics[width=0.45\columnwidth]{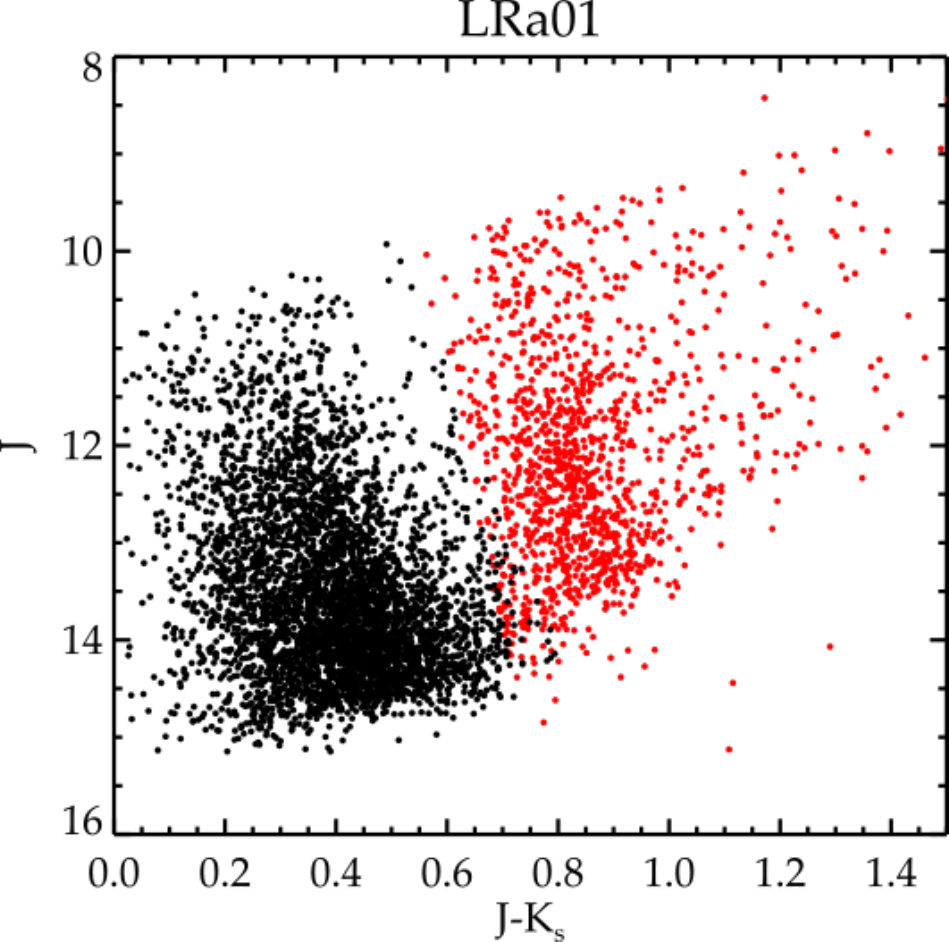}\quad
\includegraphics[width=0.45\columnwidth]{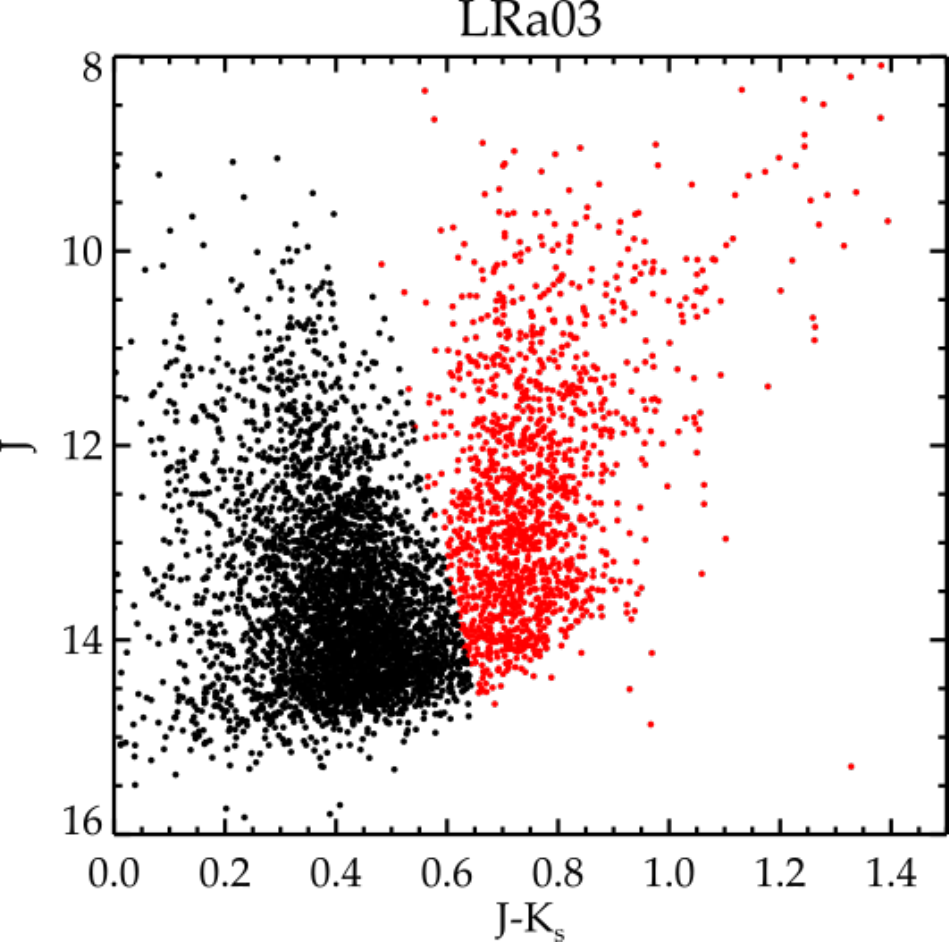}\\
\includegraphics[width=0.45\columnwidth]{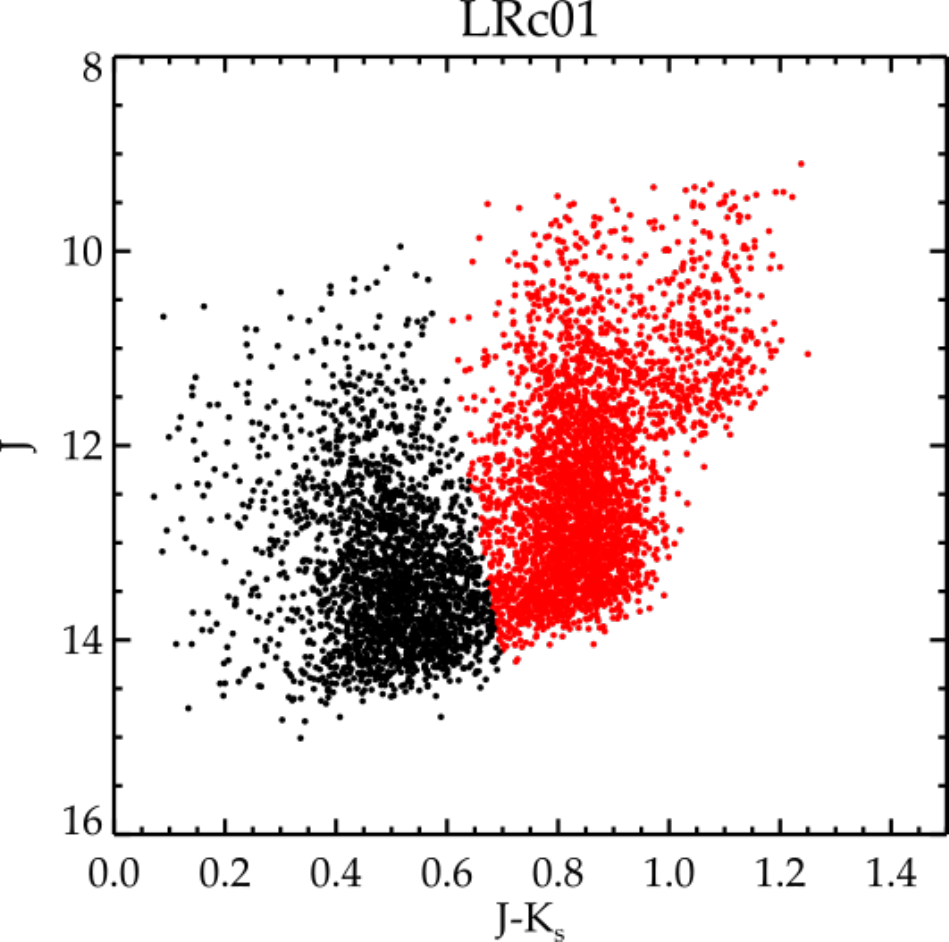}\quad
\includegraphics[width=0.45\columnwidth]{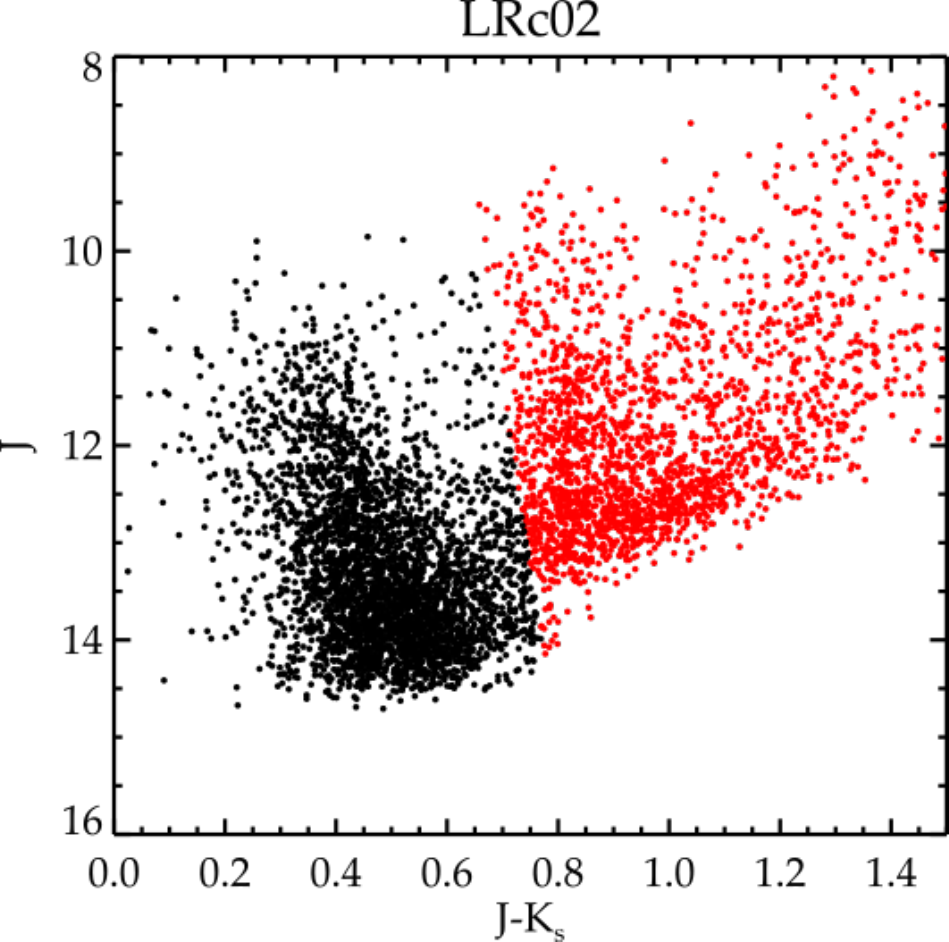}
\caption{$J$ vs. $J-K_{s}$ colour-magnitude diagrams for targets observed during different runs, as indicated on the figure. Black dots correspond to targets classified as "dwarfs" (luminosity class IV or V) and red dots are classified as "giants"  (luminosity class I, II, or III) .}
\label{fig:croc}
\end{center}
\end{figure}

The effect of luminosity on stellar energy distributions is often seen in some metallic lines that are usually shallow and/or narrow. It is thus very difficult to distinguish dwarf and giant stars when adjusting SEDs to broadband photometry alone. Near-infrared magnitudes can however be used to separate the two populations \citep{Ruphy1997} using colour-magnitude diagrams (CMD). \citet{Ruphy1997} used both colour-coulour and colour-magnitude diagrams to distinguish dwarfs and giants. The method is particularly reliable close to the Galactic plane where the typical extinction is enhancing the gap between the bluer dwarfs and redder giants. Specifically, the bluer part of the diagram should not contain any giants, while the redder may well contain cool dwarfs, especially at faint magnitudes (see also Sec.~\ref{sec:accuracy}). To avoid strong degeneracies with the FOSC method, we first use a preselection tool based on a CMD before the identification of the best-fitting stellar template. Depending on their position in the CMD, stars are hitherto identified as dwarfs or giants and their subsequent classification is carried out separately. Thus one pipeline only uses \lq{}\lq{}dwarfs" templates, which includes luminosity classes IV and V, while the other only uses \lq{}\lq{}giants" templates, which includes luminosity classes I, II, and III. Several combinations of magnitudes and colour can be chosen to identify the two populations. Besides, the location of dwarfs and giants in a CMD is a function of the interstellar extinction and thus, galactic coordinates.   We use the magnitudes from 2MASS in a $J$ versus $J-K_s$ colour-magnitude diagram to make the preselection tool
efficient for the maximum number of targets. For each observed field, the CMD plane is divided into two regions defined by a linear function that is adapted to the mean extinction of the field. Fig.~\ref{fig:croc} gives examples of the distinction of dwarfs and giants in the CMD for four typical runs. If the 2MASS magnitudes are not available, the classification is carried out using both dwarf and giant templates. In some rare instances, a particular target may be observed in two different runs (or more) and would result in different pre-classifications. In this case, the target is considered a dwarf by default and the corresponding spectral type is adopted for all the runs. This can be seen in Fig.~\ref{fig:croc} for the case of LRa01 where some black dots overlap the red dots region. This is intended to be conservative for exoplanet search, since it was deemed preferable to include false positives (when the star is classified as dwarf but is actually a giant) than to reject false negatives (when the dwarf star is wrongly classified as giant). The misclassifications resulting from using this preselection tool are described in Sec.~\ref{sec:gal} and Sec.~\ref{sec:spectro}.

\subsection{Interstellar reddening}\label{sec:reddening}
The main problem one faces when dealing with stellar classification is the role of interstellar extinction, which modifies the observed colours of stars. This problem is particularly important in our study, since the \corot{} fields are located within a few degrees from the Galactic plane. A distant, strongly reddened early-type star could be identified as a closer, cooler (intrinsically redder) star. One way to handle this would be to observe the \corot{} fields in a set of narrow- or medium-band filters, which can produce reddening-free colour indices. Considering the faintness of the targets, the observing time required to obtain good quality data on photometric nights would be preposterous. Another way would be to use an independent measurement of the extinction along the line of sight and set it as fixed parameter in the fitting procedure. However, this would require that we know the distance to the target, which is not the case here. Consequently, we let the reddening be a free parameter in the classification scheme, but limit it within a reasonable range. The following subsections describe the procedure adopted to set the upper bound of the explored range of $E_{B-V}$.
\subsubsection{From dust emission to interstellar extinction}
The reddening range may be constrained by far-infrared measurements of dust emission, which is then converted to extinction. For example, \citet[][hereafter SFD98]{Schlegel1998} produced an all-sky reddening map, based on satellite observations of far-infrared emission from dust, and calibrated this map using colour excesses of extragalactic sources at high galactic latitude. Their spatial resolution is low because they used the 42 arcmin beam of the Diffuse Infrared Background Experiment (DIRBE) on board the \textit{COBE} satellite to derive their dust temperature map.

Dust emission is one of the major foregrounds hampering the study of the cosmic microwave background (CMB). Thus one of the products associated with the 2013 release of data from the \textit{Planck} mission\footnote{See http://www.esa.int/Planck} was a new parametrisation of dust emission that covers the whole sky, based on data from the High Frequency Instrument (HFI)  \citep{Planck11}. The HFI frequency range does not extend to the peak of thermal dust emission close to 2000~GHz. The combination of \textit{Planck} and data from the \textit{Infrared Astronomical Satellite} offers a new view on interstellar dust by allowing the sampling of the dust spectrum from the Wien to the Rayleigh-Jeans sides, at 5 arcmin resolution over the whole sky. Following the same direct approach as SFD98 and using dust optical depth, the \textit{Planck} team produced maps of $E_{B-V}$ based on the \textit{Planck} dust emissions maps. To relate dust emission and absorption, they use a correlation calibrated on reddening measurements of galaxies from the Sloan Digital Sky Survey (SDSS). Many studies based on SDSS data have shown that the shape of the extinction curve in the diffuse interstellar medium (ISM) is compatible with that for stars from \citet{Fitz1999} with $R_V=3.1$ \citep{Planck11}. The \textit{Planck} reddening maps thus offer a better resolution than SFD98 maps, but suffer from the same limitations. The calibrators, although more numerous in the case of the \textit{Planck} maps, are always located at high galactic latitudes and probe the diffuse ISM. This means that the calibration may be unsuited for denser regions of the ISM, where the opacity of the grains is expected to be enhanced. Besides, by construction, those measurements can only provide the total integrated dust extinction on a given line of sight, but the dependence on the radial distance is lost.

\subsubsection{Estimating the maximum $E_{B-V}$ of \corot{} pointings}

\citet{Planck11} provides two maps that may be used to infer the $E_{B-V}$ maps. One gives a measure of the dust radiance or integrated intensity $\mathcal{R}$, and the other is the optical depth $\tau_{353}$ estimated from the measured emission at 353~GHz (850~$\mu$m). While the first map is less affected by the contamination at small angular scale from the cosmic infrared background anisotropy, the second map  might be a better tracer of $E_{B-V}$ in the denser regions of the ISM. So we decided to use the $\tau_{353}$ maps to constrain the range of $E_{B-V}$ in our stellar classification. In Appendix \ref{app:ebv}, Figure~\ref{fig:red} shows the $\tau_{353}$ maps we used and the locations of the different runs.

\begin{figure}
\begin{center}
\includegraphics[width=0.45\columnwidth]{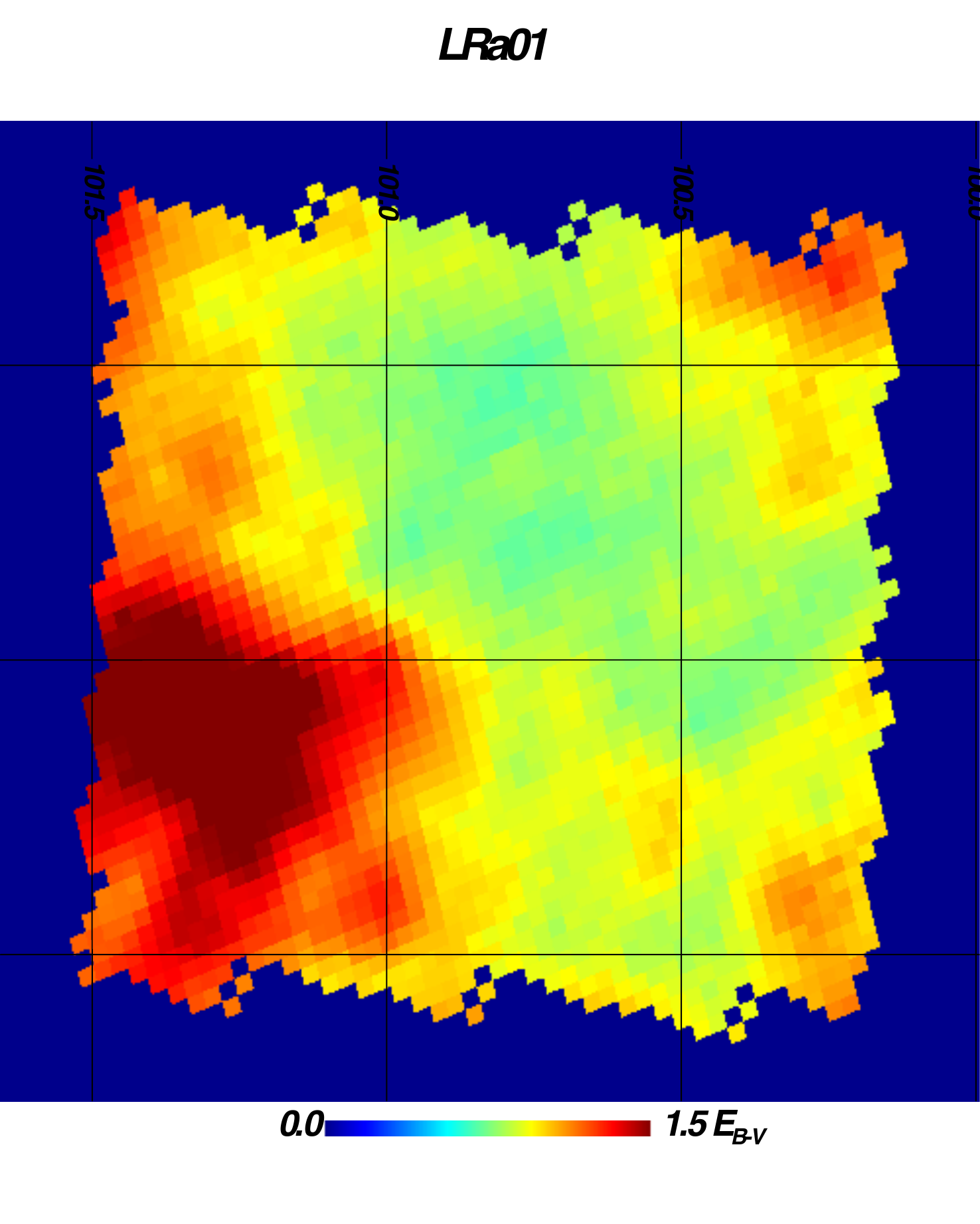}\quad
\includegraphics[width=0.45\columnwidth]{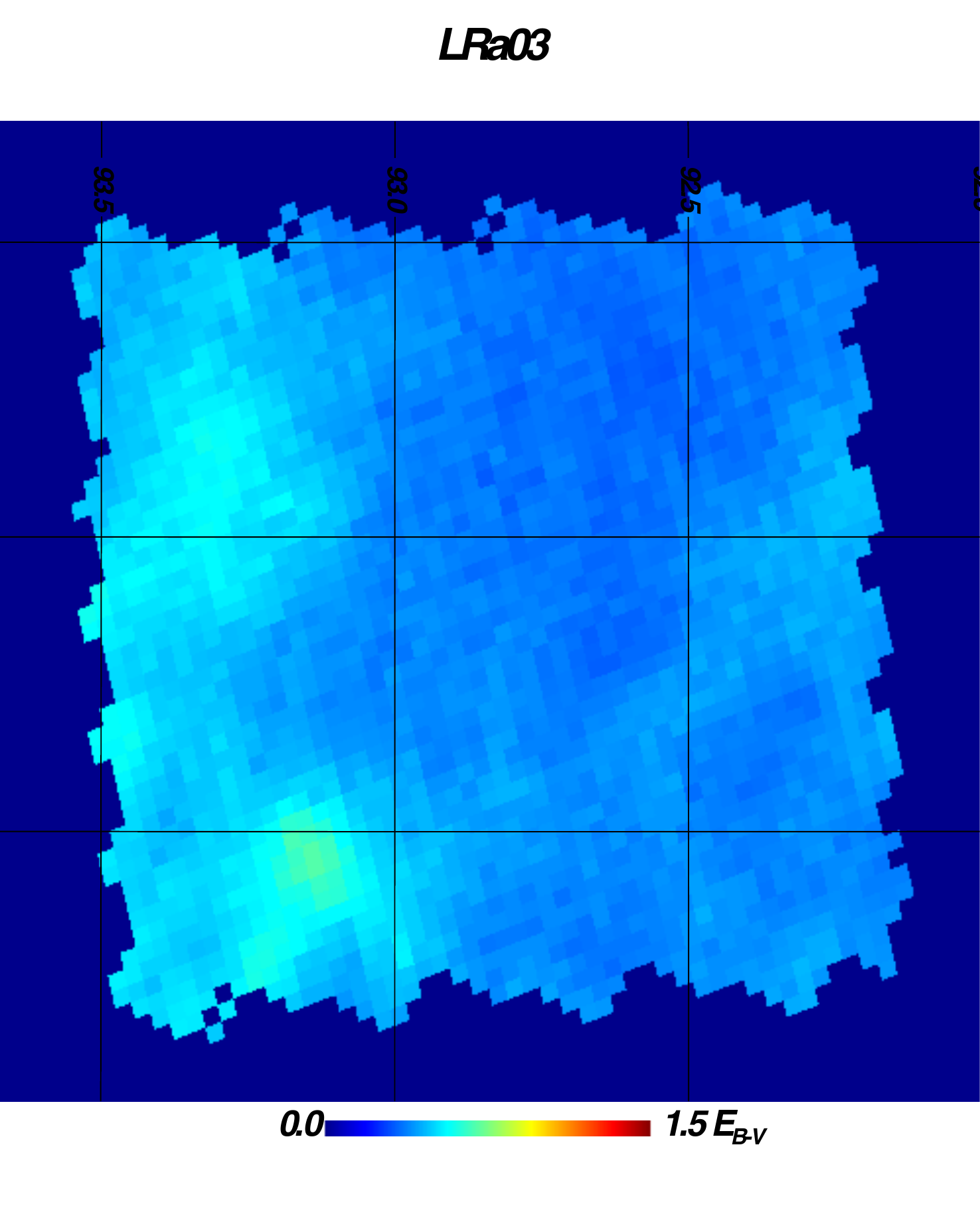}\\
\includegraphics[width=0.45\columnwidth]{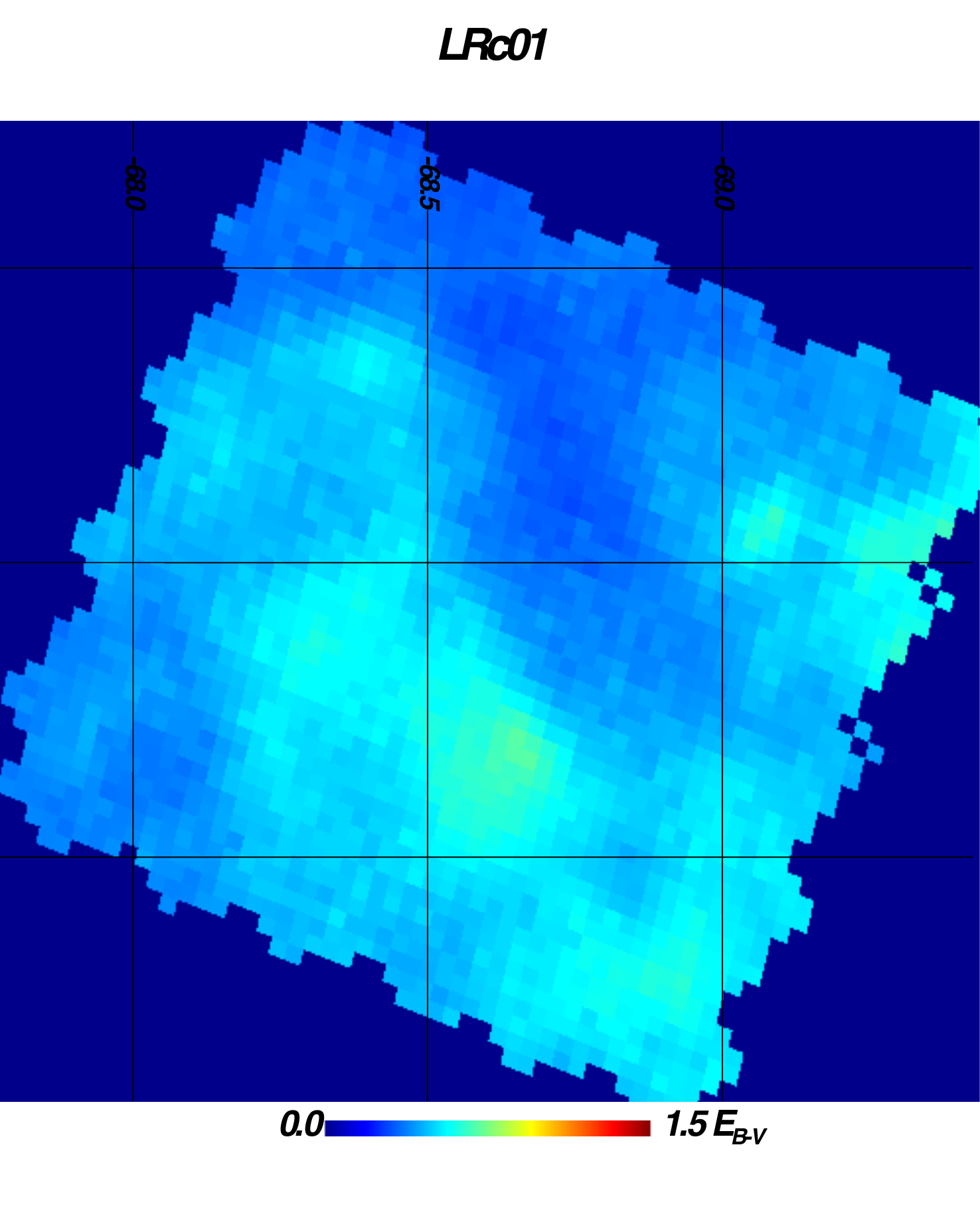}\quad
\includegraphics[width=0.45\columnwidth]{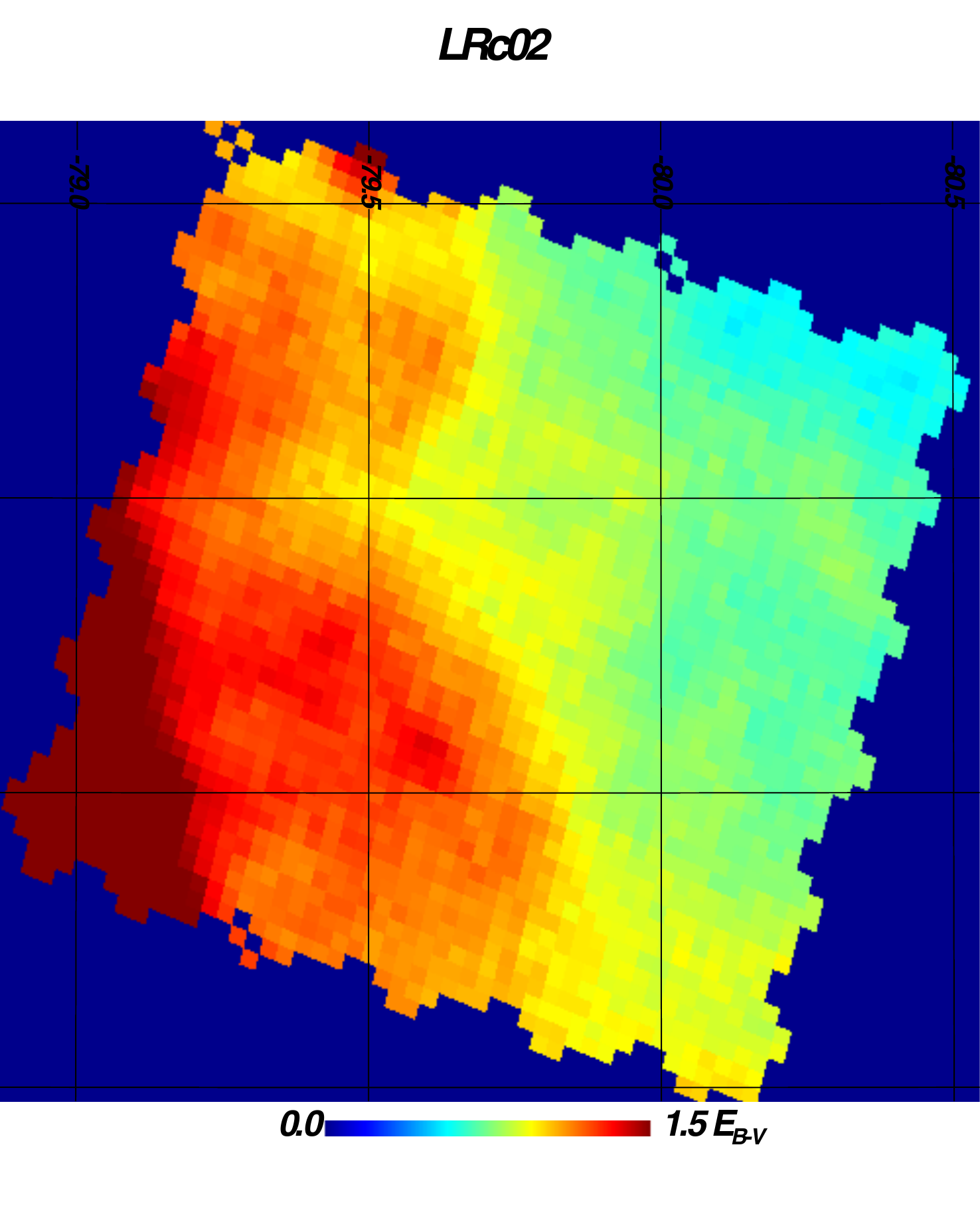}
\caption{Gnomonic projections of $E_{B-V}$ maps obtained by extracting the region observed by \corot{} from the $\tau_{353}$ maps of \textit{Planck}. The four panels correspond to the CCD footprint for four runs: LRa01, LRa03, LRc01, and LRc02 as indicated on the figure. The colour scale is the same for all runs; the maximum value was arbitrarily set to 1.5 to allow for an easy comparison of the different panels.}
\label{fig:redmaps}
\end{center}
\end{figure}

For every run, the region observed by \corot{} was extracted from the $\tau_{353}$ maps and converted to $E_{B-V}$ using the scaling factor $E_{B-V}/\tau_{353} = 1.49 \times 10^4$ \citep{Planck11}. Some examples are given on Fig.~\ref{fig:redmaps}. \corot{} field of view is well resolved at the 5 arcmin spatial resolution of \textit{Planck} data. In general, the variations of $E_{B-V}$ over the area of a field of view are small, but the maximum values of $E_{B-V}$ vary greatly depending on the direction of observation. Consequently, to save computing time, the FOSC is run using the same value of the maximum $E_{B-V}$ for all the stars observed in the same field of view. We must stress that adopting the scaling factor estimated using the correlations with quasars in directions of diffuse ISM could systematically overestimate $E_{B-V}$ in denser regions. Thus, from the spatial distribution of values of $E_{B-V}$, we computed the histogram of $E_{B-V}$  for a given field of view and took the value of the 80\% percentile to set the upper bound of the range of $E_{B-V}$ values allowed in the fitting procedure. The 80\% percentile permits a faithful representation of the maximum value of $E_{B-V}$ over the field of view, while excluding the densest regions where reddening may be overestimated. For the runs shown in Fig.~\ref{fig:redmaps}, this results in values of $max(E_{B-V}) =0.6$ for LRa03 and LRc01 of $max(E_{B-V}) =1.4$ for LRa01 and LRc02. For two runs in the centre direction located near dense clouds, this value was too high to constrain the fitting procedure significantly, and we set the upper bound to the more reasonable value of $E_{B-V} \leq 4$. Table~\ref{tab:red} in the Appendix gives the values of $max(E_{B-V})$ that were used for each run. 

\section{Validation through simulations}\label{sec:accuracy}
Given that our method is based on $\chi^2$ minimisation, it is expected that the greater source of uncertainty will come from the intrinsic degeneracy between spectral type and reddening. However, even if we set the reddening to its actual value, the intrinsic limitations of the method (such as the sampling in effective temperature and in luminosity classes, overlooking metallicity or missing magnitudes) are bound to introduce some systematic errors. We can expect that those systematic errors are correlated with stellar properties. This can be quantified by running the FOSC on a controlled set of synthetic magnitudes that are computed using stellar models distributed over the HR diagram. This provides a general estimation of the performances of the FOSC method, which are discussed in this section. The details of the different tests are given in Appendix~\ref{App:accuracy}. But \corot{} observations are magnitude limited, and they are surveying diverse regions of the Milky Way. Thus the populations actually observed with \corot{} have a specific range of stellar properties that must be taken into account to evaluate the statistical uncertainty of the classification in the \corot{} targets sample. This is tackled in Sec.~\ref{sec:gal}.

\subsection{Precision and accuracy of the FOSC}\label{subsec:accuracy}
The FOSC was validated on a synthetic set of stellar magnitudes. To explore the systematic effect of the method and its overall precision and accuracy, we used the Padova evolutionary tracks \citep[][and references therein]{Marigo2008} to simulate magnitudes of stars of known properties. The Padova isochrones are available for several photometric systems. We used the filters of the Jonhson-Cousin $U, B, V, R, I$ and Jonhson-Glass $J, H, K$ systems because they are very similar to the photometry that is available for \corot{} targets. We produced samples with different stellar properties to test different sources of uncertainty.

We checked the error introduced by the sampling of the Pickles library of templates. To this end, we produced a sample of synthetic absolute magnitudes that are not affected by reddening. We ran these through the FOSC, fixing the $E_{B-V}$ parameter to its actual value of 0 during the fitting procedure. This also enabled us to assess the limit of our validation method because there is an inherent mismatch between the synthetic absolute magnitudes (computed through stellar models and stellar atmosphere models) and those of the library (obtained through the compilation of real observed spectra or magnitudes). A detailed description of the results is given in Appendix~\ref{sec:temperror}. 

To test the effect of classifying dwarfs and giants separately, we also ran the FOSC on the synthetic dwarf sample but purposely using the giant templates and vice versa. Although this obviously returns the wrong luminosity class, the spectral class determination is overall as good as when the correct luminosity class is assumed. This is expected since the effect of luminosity on stellar energy distribution is very difficult to distinguish when using broadband photometry. This is why the only way to assign the luminosity class is by selecting dwarfs and giants based on colours before running the FOSC, as detailed in Sec.~\ref{dwarfVSgiants}. 

Another limitation of the method is the use of solar-metallicity templates. To test the effect of metallicity on our classification, we generated a synthetic sample of stars with low metallicity ($Z=0.008$, $Y=0.263$) and another sample of metal-rich stars ($Z=0.05$, $Y=0.34$).  For dwarf and subgiant stars, non-solar metallicity has no effect on the effective temperature determination, except for early metal-rich M stars, which tend to be classified as cooler stars. This effect is of the order of two subclasses in spectral type in the worst cases. For giant stars, there is also no significant effect induced by non-solar metallicity, except for late metal-rich M stars, which tend to be classified as hotter stars. This may be due to the lack of bright giants and supergiants templates of type later than M3 in the library, but this results in a misclassification of up to three subclasses in spectral type for those stars.

Overall, when reddening is negligible, the FOSC is capable of determining the spectral type down to the subclass level for late-type stars. In the case of metal-rich stars of spectral class M, the uncertainty is greater and it reaches several subclasses. The precision is also of a few subclasses for O and B stars, but it is expected because the library of templates is coarsely sampled for these spectral types (see Fig.~\ref{fig:pickles}). 

\subsection{Degeneracy between spectral type and reddening, missing magnitudes}\label{sec:degstred}
The \corot{} mission is probing regions of the galaxy that are close to the galactic plane and may be dense in dust. A strong reddening dims the bluer part of the SED of stars and increases the flux in the redder part, to a point where the shape of the SED of hotter stars becomes similar to that of cooler objects.
The validation process revealed that the limitations of Pickles library may be a major source of uncertainty in the classification of reddened stars (see App.~\ref{app:degstred}). Indeed, only eight of the stellar templates are based on observed reference spectra in the wavelength range between 1.06 - 1.43 $\mu$m (right around the $J$ band). For all the other templates, the spectrum in this interval is a smooth curve obtained by interpolation. Moreover, for about half of the templates, the relevant UVKLIB spectrum has no observed infrared extension at all and that part of the library spectrum consists solely of the smooth energy distribution matched to standard type colours \citep{Pickles1998}. Obviously, the lack of observed reference spectrum hinders the recognition of the rich complexity of spectral features essential to the MK classification of stars. When the reddening is weak, the information contained in the visible part of the spectrum is rich enough to make this effect negligible. But when $E_{B-V}\gtrsim 3$, the optical part of the spectrum is dimmed to a point where the main features allowing us to classify the star are in the infrared part of the spectrum, whatever the spectral type of the target. Any difference between the infrared extension of the Pickles library and infrared colours of the target contribute significantly to the result of the fitting procedure.

The lack of observed magnitudes in the bluer part of the spectrum also has an adverse effect on the classification. As can be seen in Table~\ref{tab:magst}, very few stars observed with \corot{} have a magnitude measured in the $U$ band. This is because observations were hindered by the faintness of \corot{} targets in this band, which requires particularly long exposures times. Also, the $V$ magnitude is only available for stars from the Obscat catalogue, which amounts to 62\% of all the targets. Nevertheless, the method remains robust to missing $U$ and $V$ magnitudes for non-reddened stars with $\log T_{\rm eff} \lesssim 4.0$. On the other hand, O and B stars are systematically misclassified and are impossible to differentiate when the $U$ and $V$ magnitude are missing. The impact of missing $U$ and $V$ magnitudes is also important for reddened G-type stars and results in an error in the temperature ranging from -10\%, to up to -30\% when the reddening is strong. 

If the magnitudes are affected by severe reddening and/or if some bands are missing, the accuracy of the FOSC can be severely limited, especially for early-type dwarf stars.  The stars observed by \corot{} are limited in apparent magnitudes by the instrumental capacities and are not randomly distributed over the HR diagram. Moreover, the magnitudes used for the FOSC come from several catalogues, with heterogeneous properties. Thus to estimate the average error made on the spectral type of \corot{} targets, it is necessary to include those effects in the synthetic sample. 

\section{Comparison with synthetic galactic populations}\label{sec:gal}
\begin{figure}[b]
\begin{center}
\includegraphics[width=1.0\columnwidth]{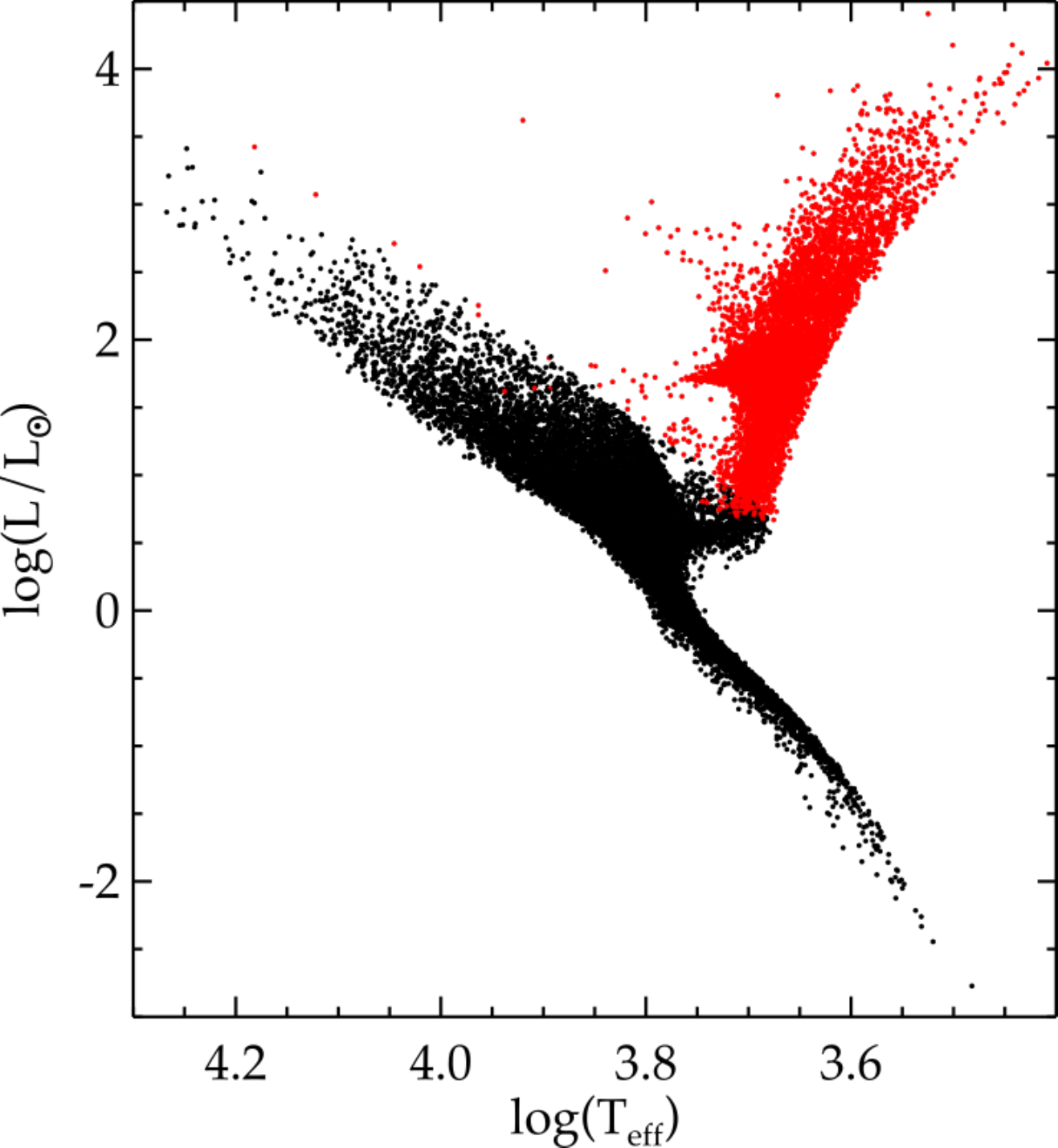}
\caption{Luminosity as a function of effective temperature of the synthetic target sample. Black dots correspond to dwarf stars (i.e. with $\log g \geq 3.5$ in c.g.s units).  Red dots indicate giant stars (i.e with $\log g < 3.5$).}
\label{fig:galHR}
\end{center}
\end{figure}

To assess statistical error estimates valid for the \corot{} targets, the stellar classification method was run on a simulated catalogue with properties close to the sample of stars observed by \corot{}. The spatial distribution of \corot{} targets spans over a wide area and different runs have very different galactic longitudes, latitudes, and interstellar extinction. We used the \galaxia{} code to generate synthetic galactic populations in each of the directions observed during the 26 runs of \corot{}. \galaxia{} is a publicly available code based on the Besançon model of the Galaxy \citep{Robin2003} that is particularly suited to generate synthetic surveys over wide areas. We made no modifications to the source code and use it as it is described in \citet{Sharma2011}. Section~\ref{sed:synthtarg} provides a summary of the main properties of the simulated stellar population.

\subsection{Synthetic \corot{} targets catalogue}\label{sed:synthtarg}
 In \galaxia, the stellar content of the Galaxy is modelled as a set of distinct components, namely a thin disk, a thick disk, and a bulge. For a given component, the stellar formation rate and the mass distribution (IMF) are assumed to be a function of age alone. Also, the present-day spatial distribution of stars is a function of age alone. The velocity and metallicity distribution are only functions of age and position. The parametrisation of the distribution functions for each galactic components are detailed in \citet{Sharma2011}. The morphology of the thin and thick disks are assumed to have a wrap and flare following the prescription of \citet{Robin2003}. 
 
 The Padova isochrones \citep{Marigo2008} are used to assign stellar parameters such as luminosity, effective temperature, magnitude, and colour to a star of a given age, metallicity, and mass. The Padova isochrones are limited to stellar mass greater than 0.15~M$_\odot$. For lower masses extending up to the hydrogen mass burning limit (0.07~M$_\odot < m < 0.15$~M$_\odot$; i.e. late M stars), the isochrones from \citet{Chabrier2000} are used. There is no binary star in the population synthesised with Galaxia. Since we do not use the binary nature in our classification, we deemed that it is not a priority to build a synthetic population of binaries.  

The synthetic model of galaxy produces a population of stars that has the same statistical properties as stars of the Milky Way in terms of distance to the Sun, age, IMF, and heavy elements distribution, as a function of the pointing direction. The only major unknown is interstellar absorption. This may induce systematic differences in stellar properties because the observed sample is magnitude limited. 

\galaxia{} includes a 3D model of dust, which provides $E_{B-V}$ for each star by integrating the dust density along the line of sight. The dust is modelled as a double exponential disk with a wrap and a flare that are assumed to be the same as those of the stellar thin disk. The two scale heights of the dust disk are obtained by fitting the extinction at infinity of the resulting dust column density to the SFD98 reddening map. This model is expected to produce very uncertain results near the galactic plane \citep{Sharma2011}. Indeed, assuming an average $R_V=3.1$, this model produces magnitudes that are not always consistent with the magnitudes that are actually observed. Using the \textit{Planck} reddening maps instead of SFD98 would not produce better constraints to the 3D model because the model is mainly constrained by large-scale structures, over which \textit{Planck} and SFD98 reddening maps are equivalent.

Consequently, we opted to adjust the synthetic 3D model of $E_{B-V}$ using the observed magnitudes of stars in the \corot{} fields. Each star in the galactic population synthesised with  \galaxia{} is referenced with a radial distance and a galactic longitude and latitude. For each run, we used the coordinates of the CCD footprints of \corot{} to select the synthetic stars that would be positioned in the corresponding region of the sky. Then we adjusted the synthetic magnitudes to those available in our photometric catalogues. This is carried out by introducing a correction factor that scales the synthetic $E_{B-V}$ produced by \galaxia. This factor is obtained by minimising the difference between the synthetic distribution of $J-K$ and the synthetic distribution of the stars in the field. Next, the errors on synthetic magnitudes are generated to reproduce the systematics and random behaviour of the photometric errors. The magnitudes and errors of the synthetic catalogue is then confronted to the real magnitudes and errors by performing a Kolmogorov-Smirnov test. In all instances, the null hypothesis, i.e. the two samples are drawn from the same population, cannot be rejected at the 0.5\% level. The number of potential targets, i.e. stars with $11 \leq r \leq 16$, is in general greater than the maximum number of targets actually observed because of the limitations imposed by \corot{} telemetry. Owing to the versatility of the mission, the specifications for target selection are not necessarily the same for all runs because they may be driven by different scientific objectives. In general, this results in an overestimation of the number of giants in the synthetic sample, but we have not taken this effect into account. This is a conservative approach, since giants within the specified apparent magnitude range are generally more reddened. Since we have seen that reddening is a major source of error in the FOSC, the classification of the observed sample should have a better precision than that given by the synthetic sample.

As summarised in Sec.~\ref{subsec:accuracy}, we extensively tested the method on a set of synthetic stellar magnitudes for which we were able to control the different sources of error and assess their relative importance. This showed that the main factors affecting the performance of the classification are the value of $max(E_{B-V})$ and the number of magnitudes used in the fitting procedure. The former  affects targets in a uniform manner amongst the different runs. The latter ranges from 0.4 to 4 for the 26 observing runs of \corot{} (see Tab.~\ref{app:ebv}). Only 3 runs have $max(E_{B-V})\geq 1.8$, and the others are either in the range $0.4\leq max(E_{B-V})\leq 0.7$ or $1.1\leq max(E_{B-V})\leq 1.7$, with the two most frequent values being $max(E_{B-V})= 0.6$ (6 runs) or $max(E_{B-V})= 1.4$ (5 runs). To present our results, we limit our illustrations to four runs, of which two are in the centre direction and the other two are in the anticentre direction. The runs LRc01 and LRa03 have similar low maximum values of interstellar extinction $max(E_{B-V}) =0.6$, LRc02 and LRa01 are representative of a higher extinction with $max(E_{B-V}) =1.4$. Those runs are thus representative of the typical properties of \corot{} pointings as far as the performance of the classification is concerned. The results we present here are valid for other runs of similar direction and $max(E_{B-V})$. The HR diagram of the synthetic sample of \corot{} targets generated with \galaxia{} for those four runs is given in Fig.~\ref{fig:galHR}.
\begin{figure*}
\begin{center}
\includegraphics[width=0.8\textwidth]{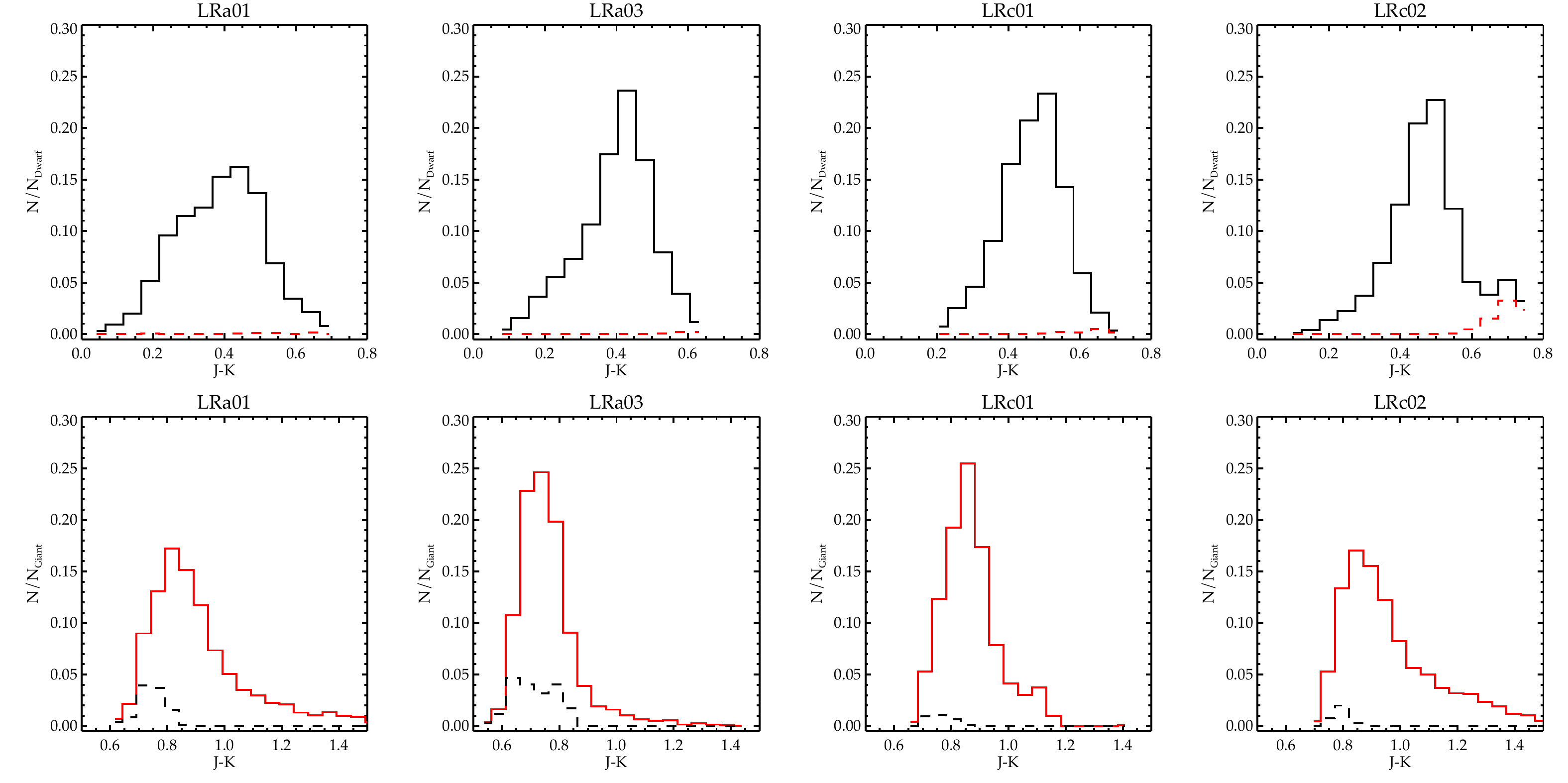}
\caption{Histogram of J-K for stars classified as dwarfs (top, black solid line) and giants (bottom, red solid line) for different runs. The distribution of giants (stars with $\log g < 3.5$) misclassified as dwarfs is indicated in the top plots with a red dashed line. The distribution of dwarfs (stars with $\log g \geq 3.5$) misclassified as giants is indicated in the bottom plots with a black dashed line.}
\label{fig:DandG}
\end{center}
\end{figure*}

\subsection{Luminosity class}\label{sec:LCgal}
The luminosity classification of the MK system gives an estimate of the surface gravity of the star ($\log g$) for a given metallicity, but other parameters such as microturbulence can mimic luminosity features and lead to erroneous luminosity classifications. As such, there is a significant scatter in the absolute magnitude calibration of the MK classes \citep{Gray2001}. This limited accuracy of the calibration is a hindrance when one wishes to use stellar models to check the classification. Here, we limit ourself to assess the quality of the basic distinction between dwarfs and giants. We identify synthetic stars as dwarfs or giants depending on their $\log g$, simply taking $\log g \geq 3.5$ for dwarfs and $\log g <3.5$ for giants. 

As previously described, the distinction between dwarfs and giants is not in itself a result of the $\chi^2$ minimisation of the FOSC, but it is based on the position of the target in a colour-magnitude diagram. The significance of this classification can be assessed using the synthetic catalogue, but the result is very sensitive to space absorption. Considering our lack of knowledge of the dust properties, we have chosen to compute the significance of the classification assuming that the value of the reddening of the synthetic catalogue is correct. This overlooks the uncertainty on the actual value of space absorption, but it provides a reasonable quantification of the expected statistical error on the luminosity class because we use the value of $E_{B-V}$ that matches best the observed population. Thus we used the synthetic catalogue to produce a ($J-K$, $J$) CMD and we applied the separation criterion used for the targets. We can then compare the classification as dwarf of giants to the $\log g$ of the synthetic stars. The results are shown in Fig.~\ref{fig:DandG}, which gives the percentage of misclassified stars as a function of $J-K$ for dwarfs and giants.

\begin{figure}[h]
\begin{center}
\includegraphics[width=1.0\columnwidth]{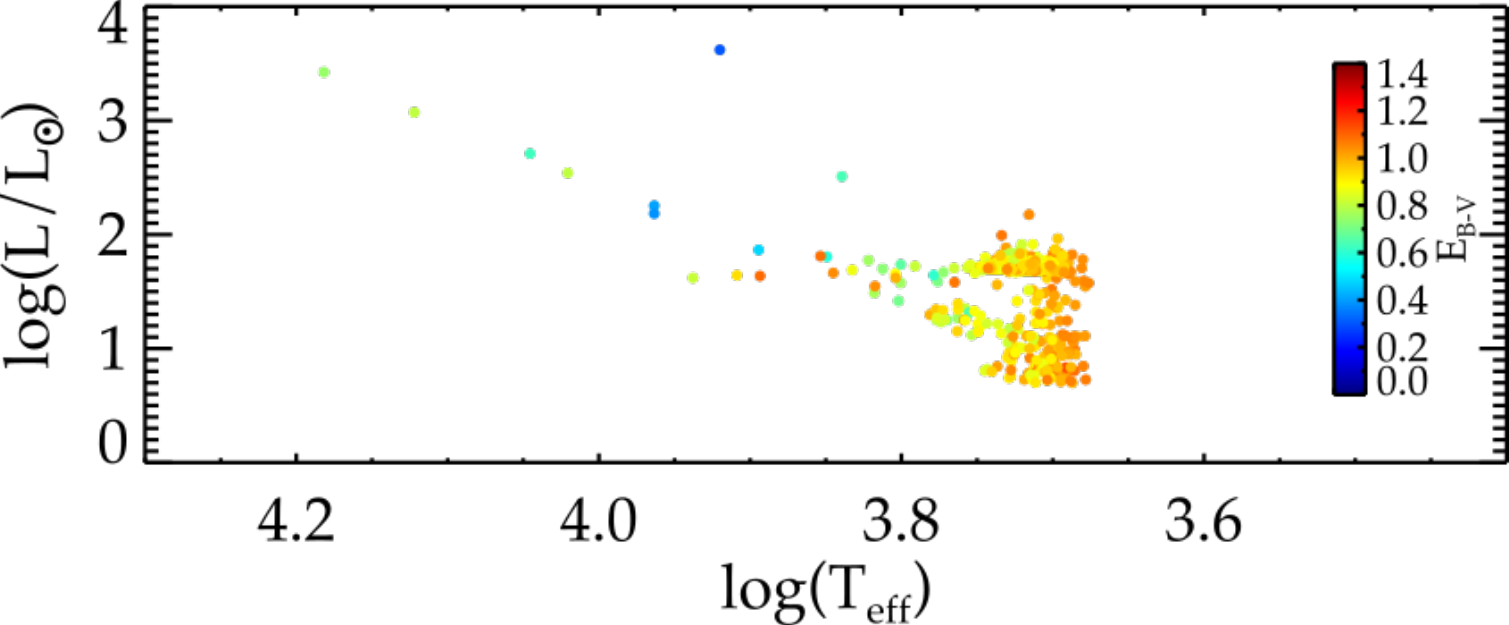}\\
\includegraphics[width=1.0\columnwidth]{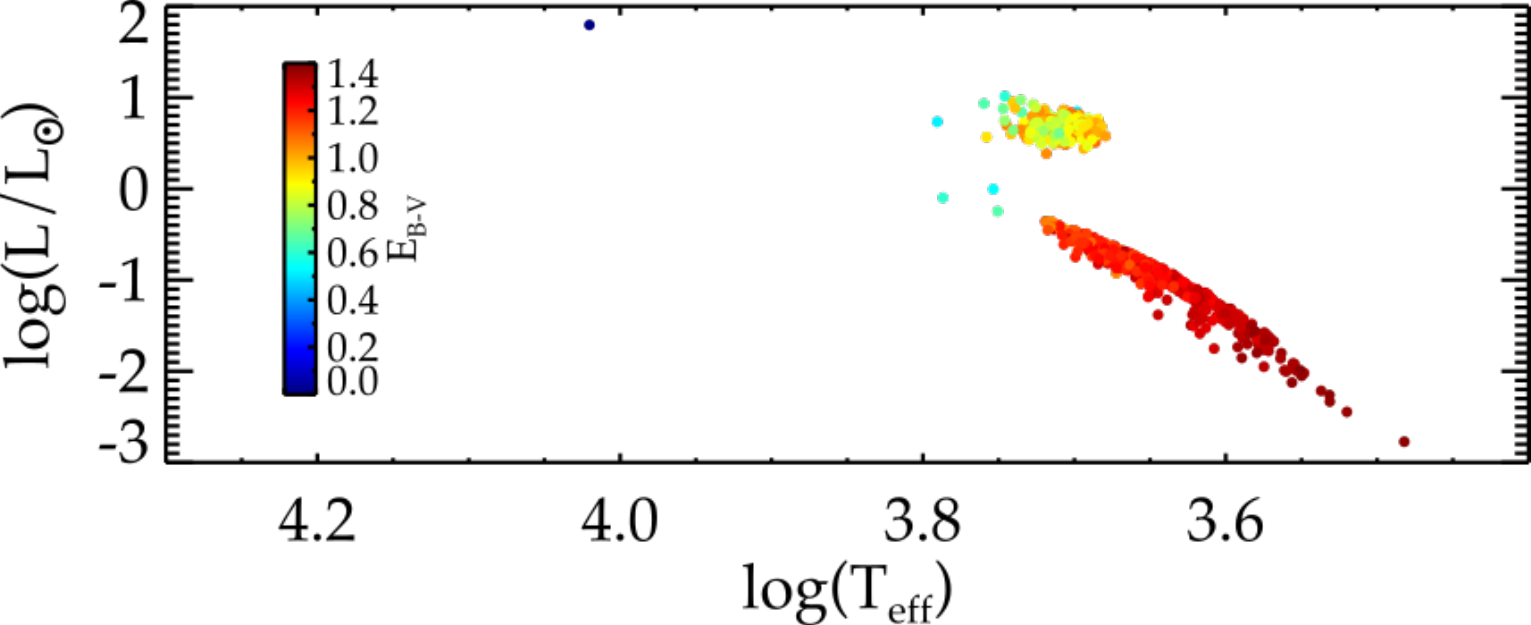}\\
\caption{Luminosity as a function of effective temperature for the sample of misclassified stars. The colours of the point indicate the $E_{B-V}$ computed with the dust model.  Top: giant stars (i.e with $\log g < 3.5$) that are classified as dwarfs. Bottom: dwarf stars (i.e. with $\log g \geq 3.5$ in c.g.s units) that were classified as giants.}
\label{fig:galHRwLC}
\end{center}
\end{figure}

For $J-K<0.5$ and $J-K>0.85$, there are a negligible number of stars that are misclassified because there is no overlap of the two populations in this part of the diagram. In this case, the luminosity classification is correct to a very high level of significance. The greater uncertainty is evidently found for stars that have a $J-K$ close to the criterion used for the selection, some of which are almost certainly misclassified. Interestingly, we see that there are more "lost dwarfs", i.e. dwarf stars that are wrongly classified as giants for the anticentre fields than for the centre fields. However, the contamination of the sample of stars classified as dwarfs with stars that are actually giants is more important in the centre fields. Overall the number of misclassified stars represents a very small percentage of the total sample. The runs displayed here are representative of all \corot{} observations with respect to the variables that affect the most the performance of the classification. Integrating over the values of $J-K$, we find that on the whole there are from  approximately 0.5\% to 8\% (depending on the run) of giant contamination in the sample of stars classified as dwarfs. Respectively, between 3 and 15\% of stars classified as giants are actually dwarfs. Overall, the percentage of wrongly classified stars in the entire sample of \corot{} targets is less than 7\%. As can be seen in Fig.~\ref{fig:galHRwLC}, separating dwarfs and giants using their infrared colours results in misclassifying relatively well-defined stellar populations. Not surprisingly, lost dwarfs are typically strongly reddened mid-G to M dwarfs, as well as mildly reddened late-G to early-K subgiants. The bulk of the population of wrongly classified giants are in the same temperature range but slightly more reddened. 

Finally, when the $J$ or $K$ magnitude, or both, are missing, the preselection between dwarfs and giants cannot be performed. In this case, that is for about 2\% of the targets, the FOSC is run using templates for dwarfs and giants both, and the luminosity class is thus decided by the $\chi^2$ minimisation. This results in an uncertainty on the luminosity class that is entirely different than that described here; this uncertainty is examined in section \ref{sec:horsclass}.

\subsection{Spectral type and reddening for the Obscat+2MASS catalogue}
\begin{figure*}[h]
\begin{center}
\includegraphics[width=0.99\textwidth]{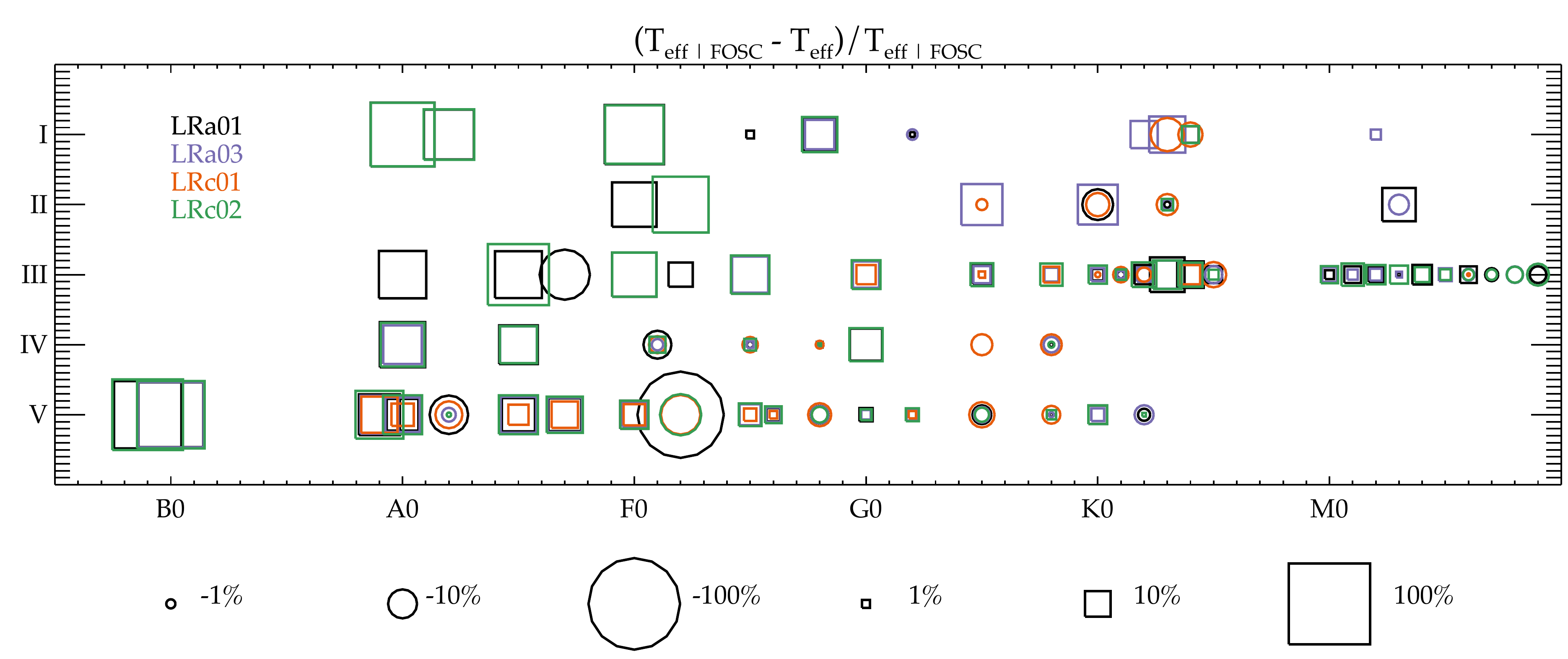}\\
\caption{Median of the relative error in effective temperature $\frac{T_{\rm eff\ |\ FOSC}\ -\ T_{\rm eff}}{T_{\rm eff\ |\ FOSC}}$ for bins of spectral type found by the FOSC, in a spectral class vs. luminosity class diagram. The colours of the symbols correspond to different runs. As indicated at the bottom, the size and shape of the symbols give the median value of $\frac{T_{\rm eff\ |\ FOSC}\ -\ T_{\rm eff}}{T_{\rm eff\ |\ FOSC}}$ computed over the sample of stars that are classified in the spectral type (i.e. a combination of a spectral class and a luminosity class) of the corresponding abscissa and ordinate. The synthetic magnitudes reproduce the characteristics of the Obscat catalogue.}
\label{fig:percentsobscat}
\end{center}
\end{figure*}

Using the synthetic catalogue of magnitudes, we reproduced the classification procedure for each field. First the dwarfs and giants are separated using the  ($J-K_s$, $J$) CMD, and the FOSC is run for dwarfs and giants using the corresponding templates, and setting the range of exploration of $E_{B-V}$ according to the run. As seen in the previous section, the synthetic catalogues contain a small proportion of stars that are wrongly classified. However, as observed in Sec.~\ref{sec:temperror}, the effective temperature determination is almost independent of the luminosity class, so this does not produce an important contribution to the determination of the error in $T_{\rm eff}$. This effect is nonetheless included in our estimation of the statistical error in the effective temperature determination so that it can be adopted for the actual classification. 

The results are shown in Fig.~\ref{fig:percentsobscat}. This figure gives the median values of the relative error for the different classes determined with the FOSC. In this figure, we used a synthetic catalogue that has the characteristics of the Obscat catalogue. We omitted the $U$-band magnitude because it is available for only $3\%$ of observed targets.
The largest differences are seen for the early types and especially for early-type giants. There is a general trend of better relative error for latter spectral types and higher luminosity class. This is clearly an effect of the missing $U$ magnitude (see App.~\ref{app:missmags}) with a clear trend to match cooler, weakly reddened stars with a reddened early-type template and thus overestimating both the reddening and the temperature. The median value of the error can be up to  70\% for O- and early B-type dwarfs, but reduces to about 20\% for late B- and A-type dwarfs and decreases to less than $\pm5\%$ for later type dwarfs.

 There does not appear to be a systematic effect on the accuracy of the temperature determination due to the run directions, but the runs with the greatest maximum extinction (LRa01 and LRc02, in black and green, respectively) tend to have lower precision. There is a conspicuous lack of classified M dwarfs for all runs. Because of the intrinsic faintness of those stars, it is expected that only a few would be observed (see Fig.~\ref{fig:galHR}), but none of these are found by the FOSC. This can be explained by the fact that reddened late-K and M dwarfs tend to be classified as giants. Worthy of noting, there is also the peculiar behaviour of stars classified as F2V. The median error for LRa01 appears be notably larger than for the other runs. The runs LRa01 and LRc02 have the same $max(E_{B-V})$ and, in principle, we would expect to find comparable performances for both runs; this is the case for the other spectral types (see the black and green symbols in Fig.~\ref{fig:percentsobscat}). This is not the case and is the result of two joint causes. Firstly there is a small sample effect. There are very few stars classified as F2V in general (see also Fig.~\ref{fig:hsc}). Here, there are 19 stars classified as F2V in LRa01 but only 2 in LRc02. It is not surprising that LRc02 has a median error value that is not representative of the general behaviour. Secondly, the high relative error found for LRa01 is a direct consequence of the high reddening affecting this field and the inadequacy of the infrared part for the spectrum of early-type stars in the Pickles library. Indeed, as can be seen on Fig.~\ref{fig:correltempdred014}, reddened stars that are classified as early-F type are generally actually earlier type stars with an incorrect estimation of $E_{B-V}$. This is consistent with the large negative value of the median error found for LRa01. Indeed interstellar extinction absorbs the bluer part of the spectrum of an early-type target, so that it behaves like that  of a less-reddened F-type star. Thus this part of the spectrum does not allow us to distinguish between the two solutions. As a consequence, the infrared part play a major role in deciding the spectral type. Since the effect is systematic and is found for synthetic stars and for observed stars (see Fig.~\ref{fig:hsc} ), we can conclude that the problem comes from the library and is clearly a consequence of an inadequate extension of the spectrum of early-type stars in the infrared domain, where there is no observed spectra of reference.
 
\begin{figure}
\begin{center}
\includegraphics[width=1.0\columnwidth]{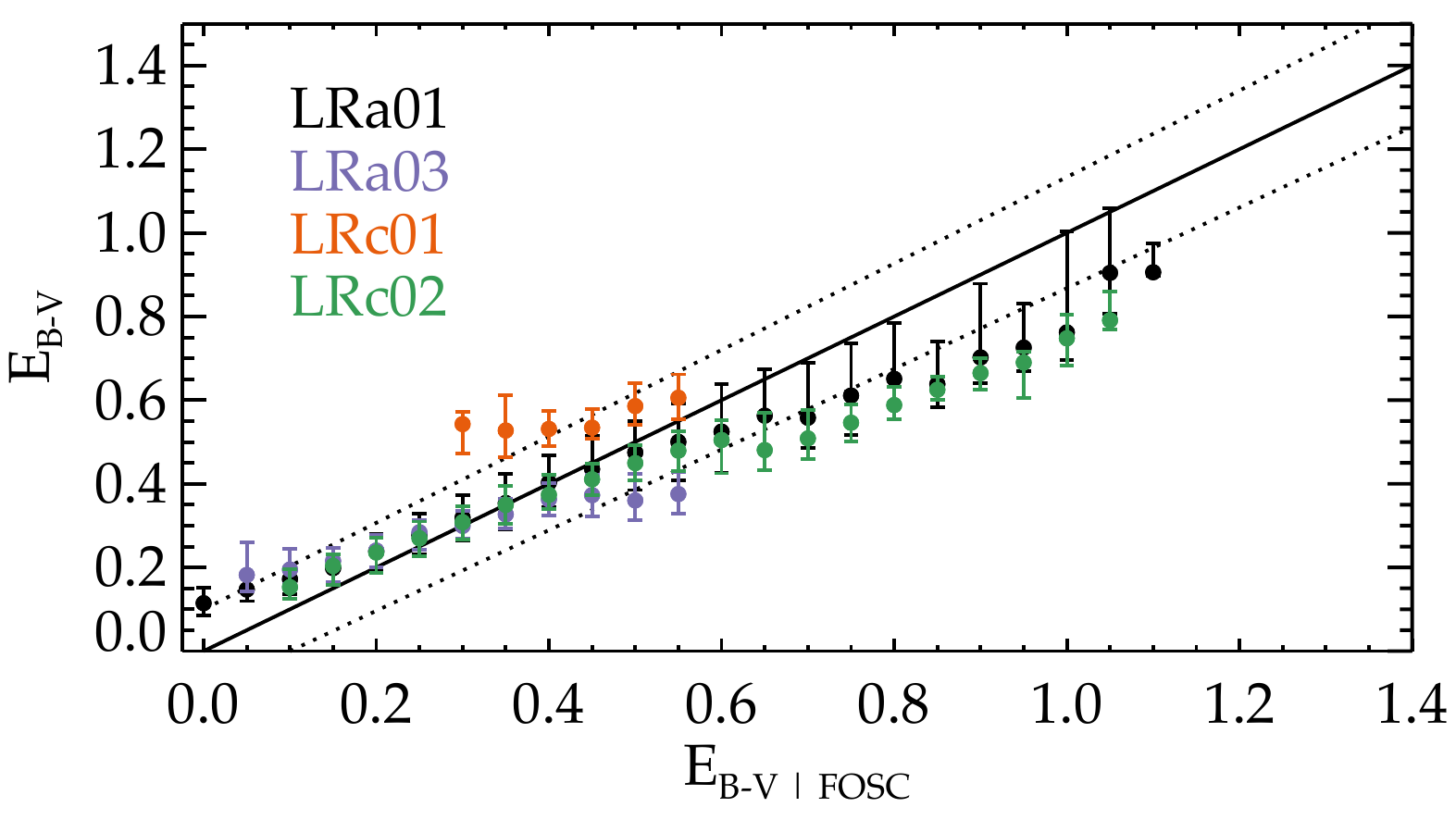}\\
\includegraphics[width=1.0\columnwidth]{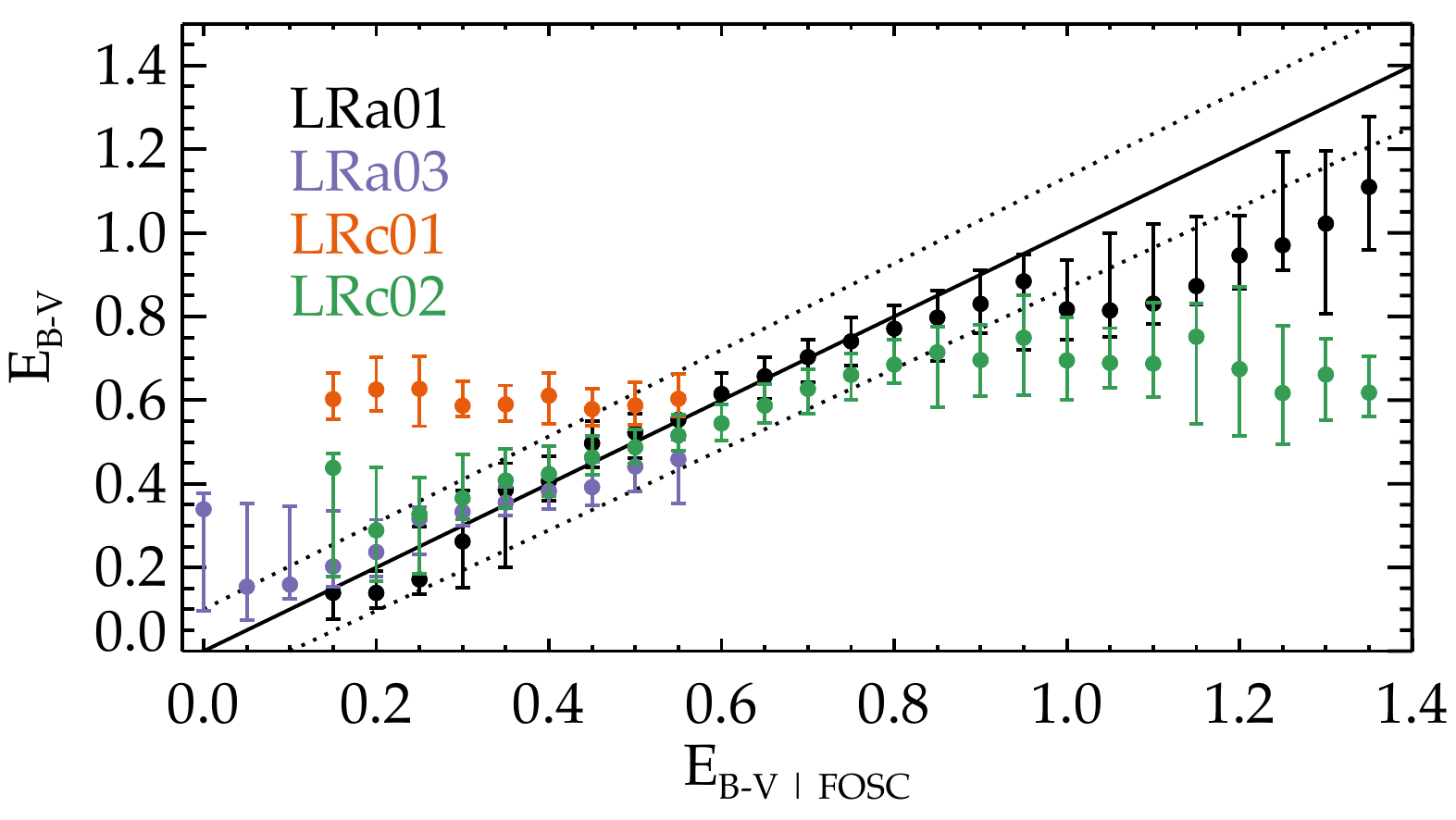}
\caption{Median values of $E_{B-V}$ used to produce the synthetic catalogues vs.that found with the FOSC for dwarfs (top) and giants (bottom). The error bars give the value of the first and third quartile of the distribution for each bin. The solid black line materialised perfect correlation and the dotted lines show the $\pm 10\%$ interval around it. The colours of the symbols correspond to the different runs, as indicated. The synthetic magnitudes reproduce the characteristics of the Obscat catalogue.}
\label{fig:ebvobscat}
\end{center}
\end{figure}

Lastly, the classification also returns an estimation of $E_{B-V}$ for each target as a by-product. Fig.~\ref{fig:ebvobscat} shows the median values of the synthetic $E_{B-V}$ by bins of values returned by  the FOSC for dwarfs and giants. The Pearson correlation coefficient between the values of synthetic $E_{B-V}$ and those found by the FOSC is between 0.3 and 0.8, depending on the run for the sample of stars classified as dwarfs and between 0.04 and 0.7 for stars classified as giants. There is a trend to overestimate the reddening as previously noted, and the precision and accuracy on $E_{B-V}$ clearly decreases with increasing $E_{B-V}$. 

We have shown that the FOSC is sensitive to the value of reddening with degraded performances when the reddening is high. Because giant stars are intrinsically more luminous than dwarf stars, the magnitude limit of \corot{} targets naturally selects giants that are at greater radial distance from the Sun and tend to be observed through higher values of interstellar absorption. This is why the precision of spectral type and $E_{B-V}$ is generally not as good for giants as for dwarfs.

\subsection{Spectral type and reddening for the PPMXL catalogue}
\begin{figure*}
\begin{center}
\includegraphics[width=0.99\textwidth]{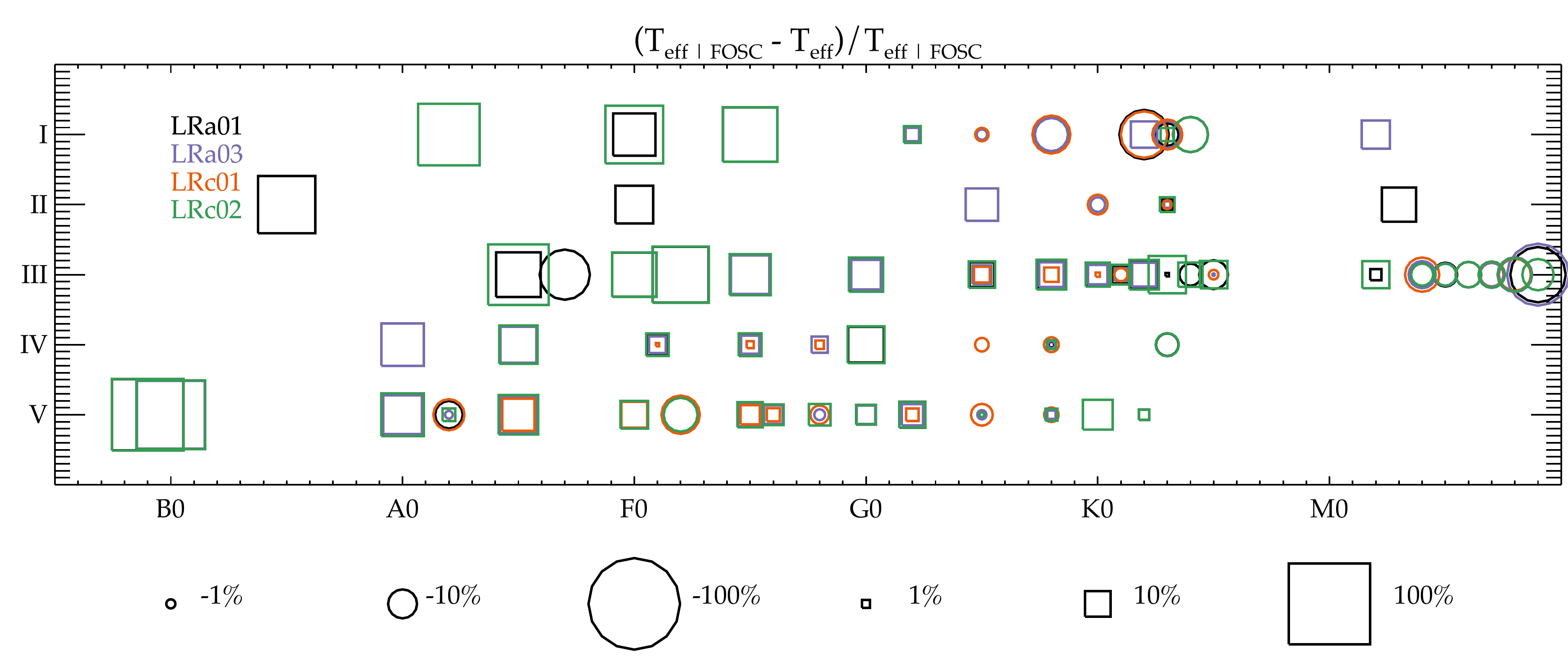}
\caption{Same as Fig.~\ref{fig:percentsobscat} but here the synthetic magnitudes reproduce the characteristics of the PPMXL catalogue.}
\label{fig:percentsusno}
\end{center}
\end{figure*}

Some targets do not have available Obscat magnitudes and, as explained in Sec.~\ref{sec:phot}, they were taken from the PPMXL catalogue with their B, R and I band photometry coming from the USNO-B1.0 catalogue. To illustrate the expected performance of the FOSC, in this case, we used the set of synthetic stars obtained with \galaxia{} (see Sec.~\ref{sed:synthtarg}); but this time we computed the synthetic magnitudes in the filters of PPMXL. The results are shown in Fig.~\ref{fig:percentsusno}. 
Overall, the same global behaviour is observed as for Obscat magnitudes, but the error on the temperature is in general greater by a factor of about two for late-type stars. The greatest effect is seen for late-type giants, especially at low temperature, where the precision of the classification is degraded by a factor four.

Finally, the input catalogue does not have a significant effect on the estimation of the reddening. Computing the Pearson coefficient of linear correlation between the known and found $E_{B-V}$ gives values that are comparable to those found using the Obscat magnitudes.

\subsection{Stellar classification without infrared magnitudes}\label{sec:horsclass}
For the marginal cases of stars that do not have 2MASS magnitudes ($\sim 2\%$), we tested the FOSC on the synthetic catalogue described above, but without using either the J, H, or K bands. We did not perform the preselection between dwarfs and giants, and we let the minimisation determine the luminosity class. As could be expected, the error on the luminosity class is consequently much more important. The proportion of dwarf stars (with $\log g \geq 3.5$) that is correctly classified is consistently of about  64\%  for the different runs. For giants stars, the proportion of correctly classified stars is much more variable with values of 85\% for LRa01, 52\% for LRa03, 65\% for LRc01, and 75\% for LRc02. This means that in the final population classified as dwarfs, from 10 to 50\% are actually giants, depending on the run. Conversely, there are from 30 to 65\% of stars classified as giants that are actually dwarfs. Moreover, the spectral classification is also affected. There are no stars classified as O or B dwarfs, and the median error on O and B giants increases to 80\%. For other spectral classes, there is no significant difference between the precision of the classification for dwarfs and giants. For A-type stars, the median error is about 35\%, and it ranges between 10 and 30 \% for F- and G-types. Later type stars are better classified with a typical median error  between 5 and 15\%. We advise that caution should be taken when dealing with those few stars for which infrared magnitudes are unavailable. We stress again that this concerns only a very small proportion of the whole sample of \corot{} targets.

\subsection{Statistical error on the temperature of \corot{} targets}
The majority of targets have Obscat magnitudes, which produce a smaller uncertainty on the spectral type compared to what is obtained when using the lower quality USNO-B1.0 magnitudes. We have created a population that is representative of the whole sample by merging the catalogues obtained for the different runs, and performing the FOSC taking magnitudes from Obscat or UNSO-B1.0, in proportion to what they are in the population of targets (see Sec~\ref{sec:method}). Synthetic stars are randomly attributed to one or the other catalogue and the process is repeated 100 times. Each repetition is a realisation of a randomly generated synthetic \corot{} targets catalogue. This is carried out to smear out any systematics arising from the correlation between the available magnitudes and the spectral type determination. For every realisation of the synthetic catalogue, the FOSC is run using the synthetic magnitudes. Then, the median of the relative error on the effective temperature is computed by bins of spectral types. As a measure of the spread of the distribution of the relative error, we compute the difference between the values of the 15th percentile and the 85th, which we note as $2\sigma$. Finally, we take the average of the median relative error and spread found over the 100 trials for each bin of spectral types. The results are presented in Table~\ref{tab:mederro}.
\begin{table}[h]
        \centering
        \caption{Average median values of $\Delta T_{\rm eff}/T_{\rm eff | FOSC}$ and the corresponding $2\sigma$ spread for different spectral types}\label{tab:mederro}
        \begin{tabular}{lll|ll}
        \hline
        \hline
         & \multicolumn{2}{c|}{Dwarfs} &  \multicolumn{2}{|c}{Giants} \\
        \hline
        Spectral type  &$ \Delta T_{\rm eff}/T_{\rm eff}$   & $2\sigma$     &   $\Delta T_{\rm eff}/T_{\rm eff }$   & $2\sigma$\\
        \hline
        \hline
O5 to B5& 70\% & 10\%& - & -  \\
B5 to A9 & 20\% &25\%& 44\% & 30\%\\ 
F0 to F9& 3\% &13\%& 24\% & 7\%\\  
G0 to G9& 1\% & 10\%& 7\% & 9\%\\
K0 to M9& -0.05\% & 10\%&1\%  & 12\%\\ 
        \hline
        \hline
        \end{tabular}
\end{table}

The median error on the temperature is in agreement with what was found for the Obscat magnitudes, which is expected since about $60\%$ of the targets have Obscat magnitudes. The error on the determination of the effective temperature varies greatly across the temperature range. Yet, stars in the temperature range in which the FOSC performs poorly are not frequently encountered in the \corot{} targets sample. Overall, we find an average median absolute temperature difference $|\Delta T_{\rm eff}| = 533\pm6$~K for the whole sample of stars classified as dwarfs and $|\Delta T_{\rm eff}| = 280\pm3$~K for the whole sample of giant stars. The corresponding standard deviation is of about $925\pm 5$~K for dwarfs and $304\pm4$~K for giants. This is the performance expected from classification methods using broadband photometry. For example, \citet{Farmer2013} give a median difference of $\Delta  T_{\rm eff} \sim500$~K with a corresponding standard deviation of $\sim920$~K for the dwarfs of the Kepler input catalogue, but they perform better for giants with a reported median  $\Delta T_{\rm eff}$ of 50~K and standard deviation of $\sim200$~K. This is probably due to their use of different filters and especially to their custom D51 filter, which is sensitive to surface gravity \citep{Brown2011}. Note that \citet{Farmer2013} only give the median of the temperature difference and not the median of the absolute temperature difference. For our simulations of the \corot{} target catalogue, we find a median temperature difference $\Delta T_{\rm eff} = 365\pm5$~K for dwarfs and $\Delta T_{\rm eff} = 220\pm4$~K for giants. This is due to the fact that the distribution of the temperature difference is not centred on zero. Nevertheless, we can safely conclude that the spectral classification obtained by the FOSC method is generally good up to half a spectral class for the majority of \corot{} targets.

\section{Comparison with spectroscopic data}\label{sec:spectro}

To further test the FOSC, we have chosen catalogues of spectral types, in the available literature, containing a large number of targets and obtained with homogenous methods. As a result, we compare the result of the FOSC with two different spectral catalogues. The first catalogue contains the fundamental stellar parameters for 1127 stars in the LRa01, LRc01, and SRc01 \corot{} fields, for which  intermediate-resolution spectroscopy was acquired with the FLAMES/GIRAFFE multi-fibre facility as ESO-VLT \citep[][hereafter Ga10]{Gazzano2010}. The second catalogue gives an estimate of the spectral type obtained for 11 466 stars in the IRa01, LRa01, LRa02, and LRa06 fields, using the AAOmega multi-object spectrograph mounted on the Anglo-Australian Telescope \citep[][hereafter jointly cited as SG12]{AAOmega1,AAOmega2}. 

There is a third recent catalogue that we considered. It contains stellar parameters for 6832 targets in the IRa01, LRa01, LRa02, LRc01, and LRc02 fields, once again using intermediate-resolution spectroscopy from a GIRAFFE survey, but specifically targeting variable stars \citep{Sarro2013}. This catalogue has a typical uncertainty on its parameters that is almost an order of magnitude greater than that produced by \citet{Gazzano2010}. These authors clearly state that their main objective was to obtain an estimate of the effective temperature for variability classification and they did not require that this estimate would be extremely precise. Moreover, they used a regression model for stellar characterisation that has not been as extensively tested as the model used by \citet{Gazzano2010}. We compared stellar parameters obtained for the same stars by both surveys and we found discrepant results between the two methods. For those evident reasons, it was impossible to use this catalogue to assess the performances of the FOSC.

\subsection{Luminosity class}
\begin{figure}
\begin{center}
\includegraphics[width=0.9\columnwidth]{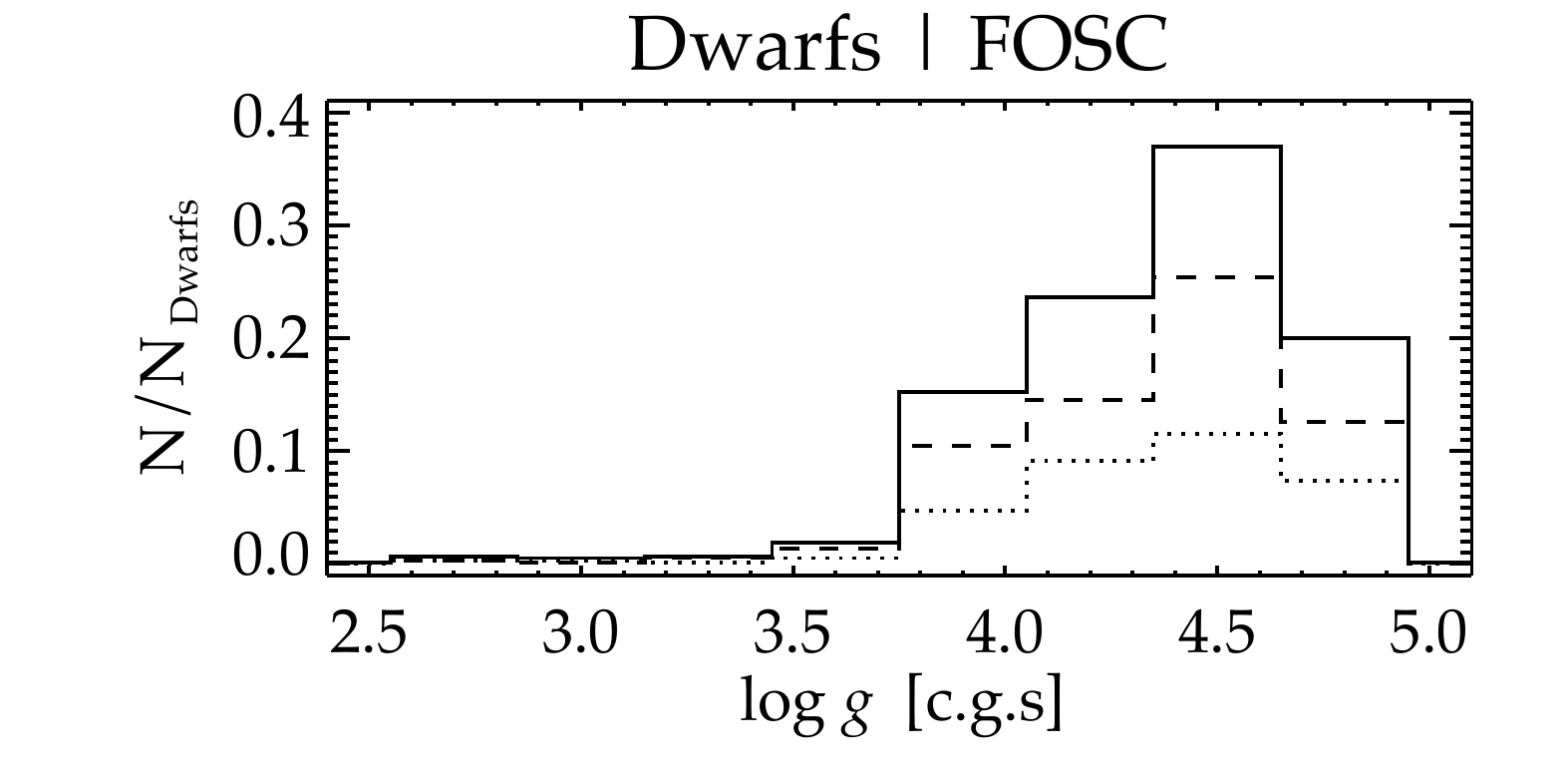}\\
\includegraphics[width=0.9\columnwidth]{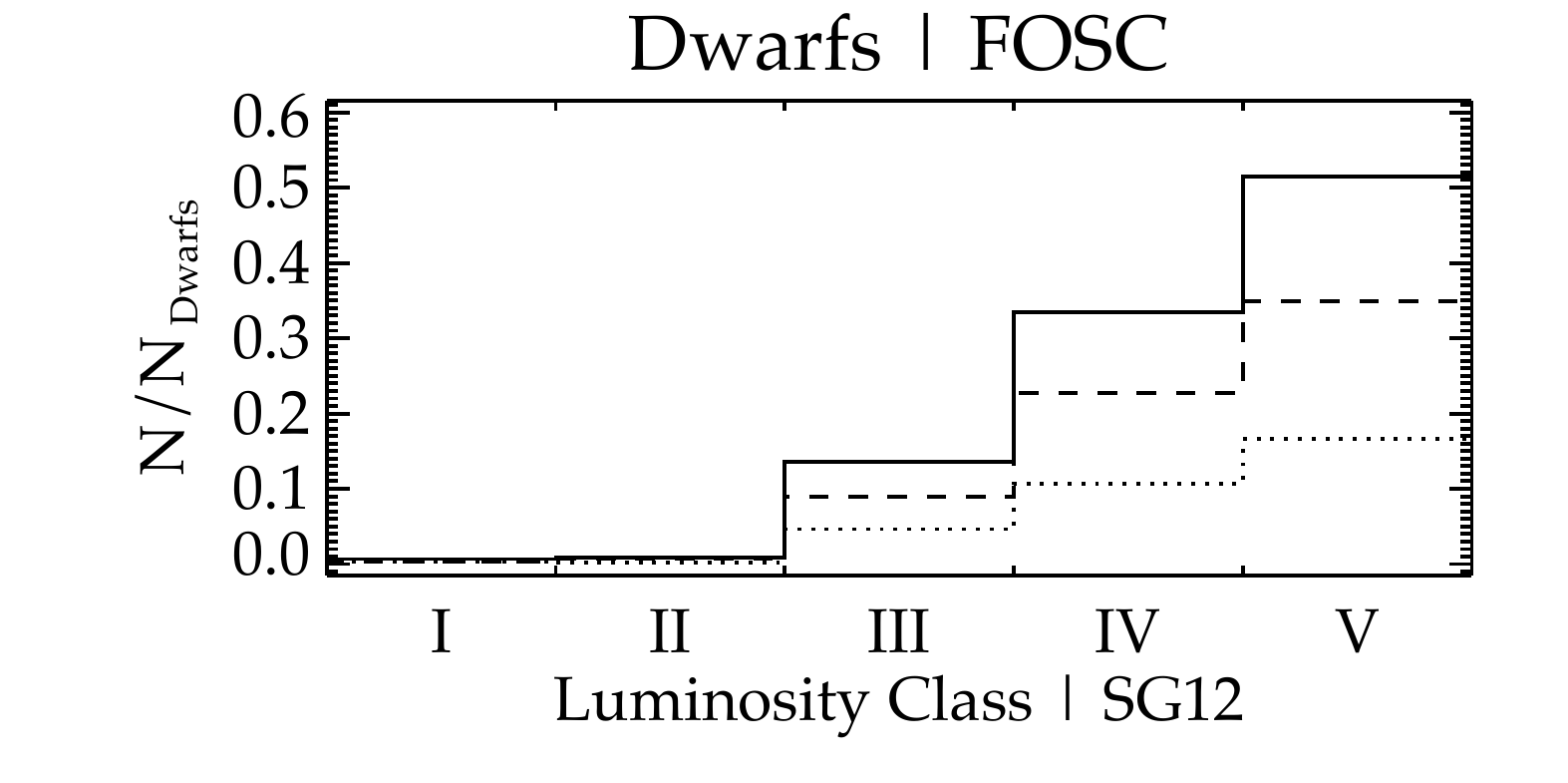}
\caption{Distribution of indicators of the luminosity class found by other authors for stars classified as dwarfs by the FOSC. Top: distribution of $\log g$ found by Ga10. Bottom: distribution of luminosity classes found by SG12. The solid lines show the relative frequency of the $\log g$ value (luminosity class, respectively, for the bottom panel) in the whole sample stars classified as dwarfs by the FOSC. The contribution from stars in the sample that have the luminosity class V with the FOSC is shown by the dashed line, and the contribution of those of luminosity class IV is indicated by the dotted line. The dotted and dashed curves add up to the solid curve.}
\label{fig:histLCdwarf}
\end{center}
\end{figure}

\begin{figure}
\begin{center}
\includegraphics[width=0.9\columnwidth]{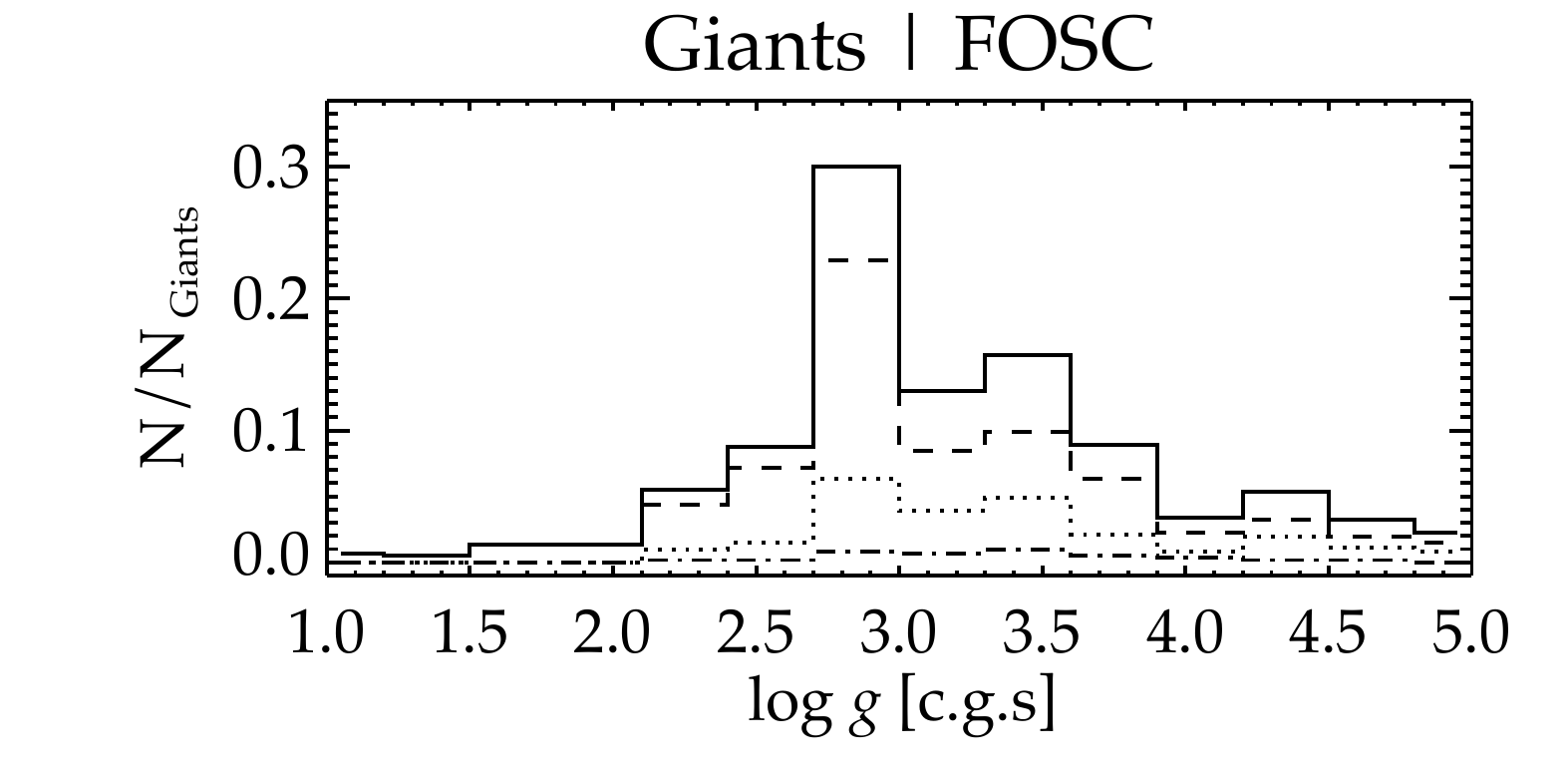}\\
\includegraphics[width=0.9\columnwidth]{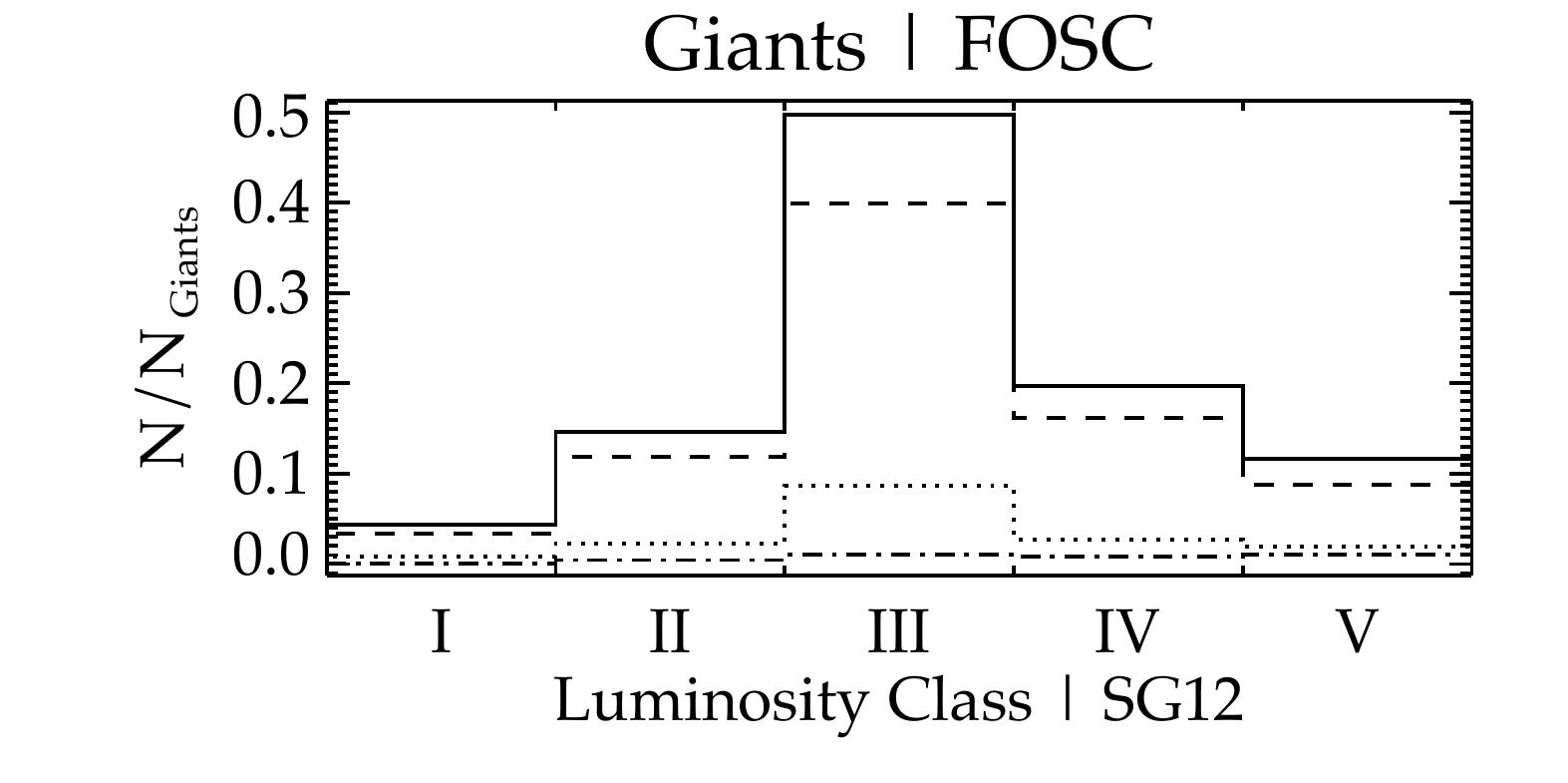}
\caption{Distribution of indicators of the luminosity class found by other authors for stars classified as giants by the FOSC. Top: distribution of $\log g$ found by Ga10. Bottom: distribution of luminosity classes found by SG12. The solid lines show the relative frequency of the $\log g$ value (luminosity class, respectively, for the bottom panel) in the whole sample stars classified as giants by the FOSC. The contribution from stars in the sample that have the luminosity class III with the FOSC is shown by the dashed line; the contribution of those of class II is indicated by the dotted line and the contribution of class I is indicated by the dashed-dotted line. The dashed, dotted, and dashed-dotted curves add up to the solid curve.}
\label{fig:histLCgiant}
\end{center}
\end{figure}

The luminosity class attribution based on a CMD is very powerful when comparing it to the $\log g$ of synthetic stars. On the other hand, the spectral indicators used to attribute the luminosity class in the MK systems are rather sensitive to other variables, such as microturbulence or metallicity. We compare the luminosity class attributed with the FOSC to the one found by spectral analysis of SG12 and the $\log g $ found by Ga10. The results are displayed in Fig.~\ref{fig:histLCdwarf} for dwarfs and in Fig.~\ref{fig:histLCgiant} for giants. To draw those plots, we have selected the sample of dwarfs (respectively giants) as found by the FOSC, and we have counted the proportions of such targets having different values of the luminosity class indicators found be other method.

For dwarfs stars, the agreement between the FOSC and spectral methods is very good. We find that more than 97\% of stars classified as dwarfs are also attributed a $\log g \geq 3.5$ by Ga10, and 85\% of dwarfs also have the luminosity class IV or V according to SG12. 

On the other hand, in the sample of stars that are classified as giants by the FOSC, only 75\% also have $\log g < 3.5$ from the Ga10 analysis and 70\% that agree with the classification of SG12. Assuming that the analysis from Ga10 and SG12 are exact, this would mean that $\sim25-30$\% of the giant stars found by the FOSC are actually dwarfs. This is a greater rate of misclassification than what was estimated from the synthetic sample (see Sec.~\ref{sec:LCgal}), for which we estimated the proportion of misclassified giants to be between 3 and 15\% depending on the field. However, factoring in the typical error bars given by Ga10 and SG12 on their parameters, we find that our estimated rate of misclassified giants is still consistent with the proportion of giants found in agreement between the FOSC and the other methods. 
\subsection{Spectral class}
\begin{figure*}
\begin{center}
\includegraphics[width=0.49\textwidth]{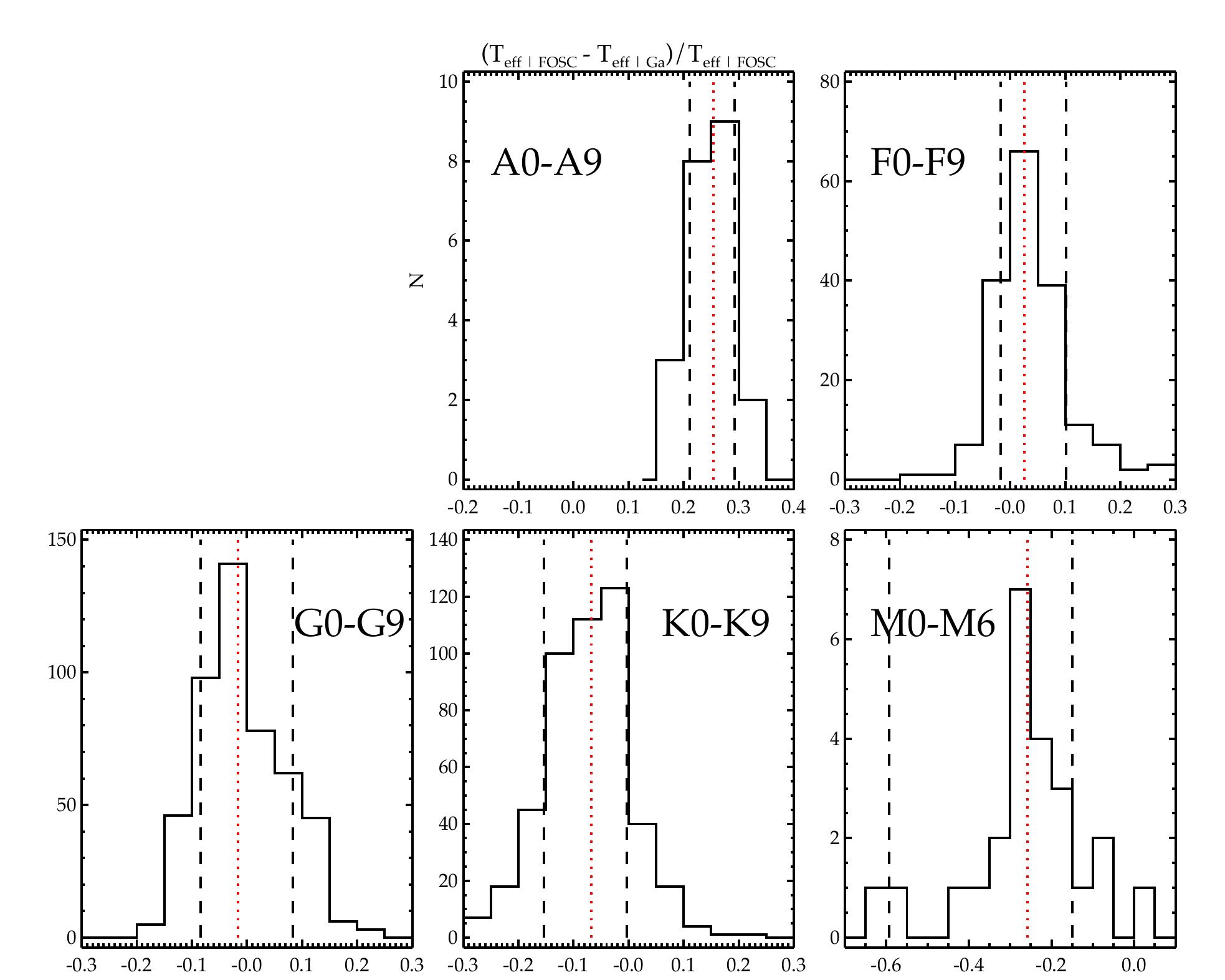}
\includegraphics[width=0.49\textwidth]{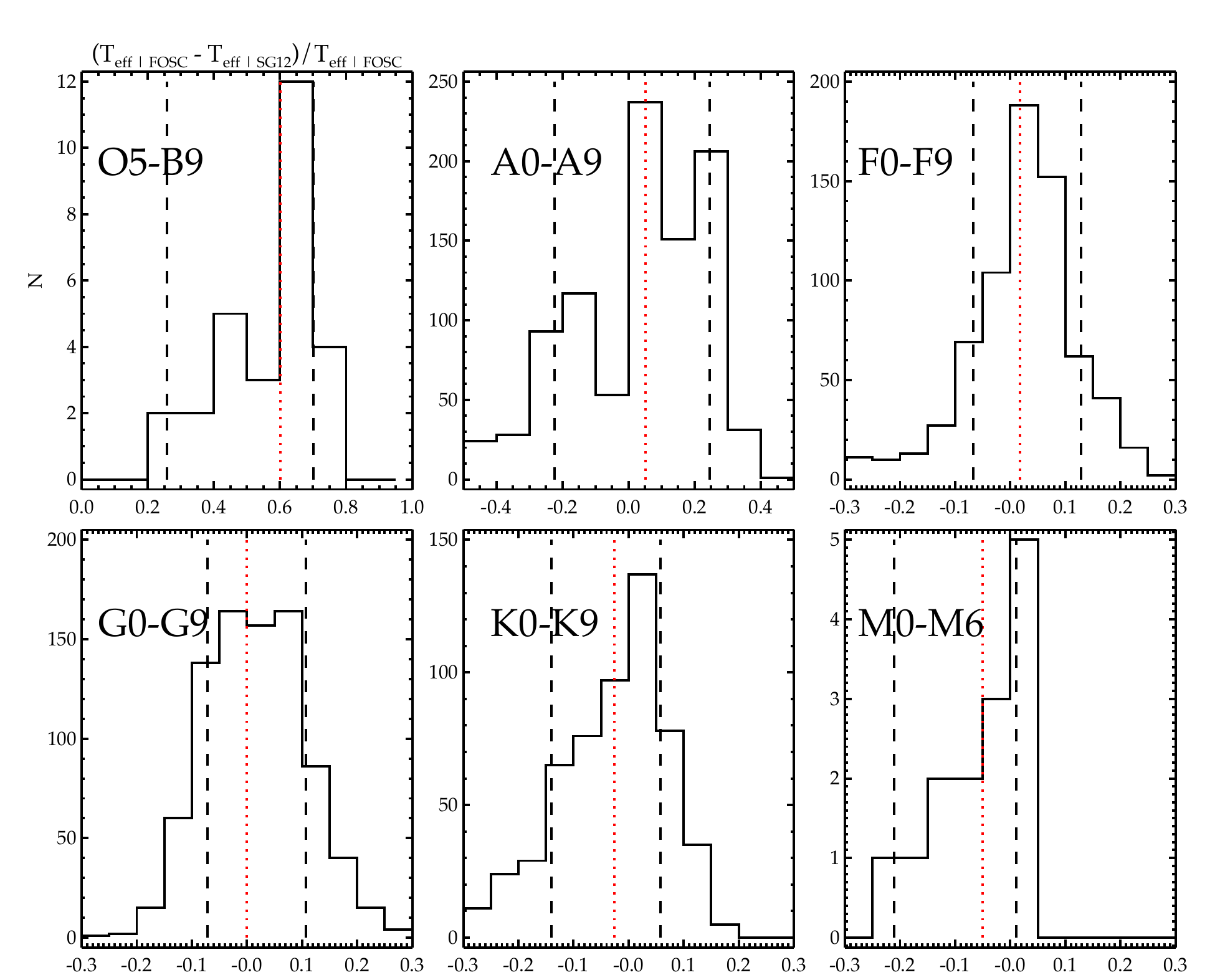}
\caption{Left: distribution of relative temperature difference found by Ga10 and the FOSC. Right : distribution of relative temperature difference between SG12 and the FOSC using the same temperature calibration. Each plot corresponds to a subsample selected by spectral class as determined by the FOSC and labelled on the plot. The vertical red dotted line shows the median value and the black dashed lines give the value of the 15th and 85th percentiles.}
\label{fig:histsc}
\end{center}
\end{figure*}

Next we compared the $\log T_{\rm eff}$ obtained by Ga10 and the spectral class found by SG12 to that found by the FOSC for all targets, disregarding the possible mismatch in luminosity class. We have organised the results by spectral class, as found by the FOSC and, for each class, computed the histogram of the relative difference in effective temperature. To do so, as previously, we attributed the effective temperature given by the Pickles library to the corresponding spectral class. To compare our results with those of SG12, we proceeded in the same way and we discarded from the sample the spectral types of SG12 that do not have an effective temperature calibration in the Pickles library; the results are shown in Fig.~\ref{fig:histsc}. There is no star classified by the FOSC as an O-type star in the sample of Ga10, and there is just one B-type star, so these classes are not included in the figure.

Interestingly, we see that there are consistent systematics when comparing the FOSC to Ga10 and SG12. In both cases, early types found by the FOSC tend to overestimate the effective temperature and later types tend to underestimate it. This trend was clearly found with the synthetic catalogue for hot stars, but was not so clear for cooler stars. For M stars, there is a much better agreement between the FOSC and SG12 than between the FOSC and Ga10. The FOSC in both cases tend to return a cooler spectral type than other methods. This is characteristic of stars that are wrongly classified as giants and is limited to a very small number of targets. Except for those stars, the precision and accuracy of the FOSC are similar when estimated with the synthetic galactic population and with results from spectroscopy. Typically for late-type stars, the median error is less than about $\pm 5\%$ and the interval between the 15th and 85th percentiles has a spread of 10 to 20\% in effective temperature. Those values are doubled for earlier type stars. The comparison with Ga10 indicates that the uncertainty for the effective temperature of M-type stars is of the same order, but it is difficult to see if it is also the case when comparing with SG12 owing to a very small sample size. Considering the limitations of our validation method based on the synthetic catalogue, we recommend exercising caution when using the sample of stars classified as M type with the FOSC.
\begin{figure}[b]
\begin{center}
\includegraphics[width=1.0\columnwidth]{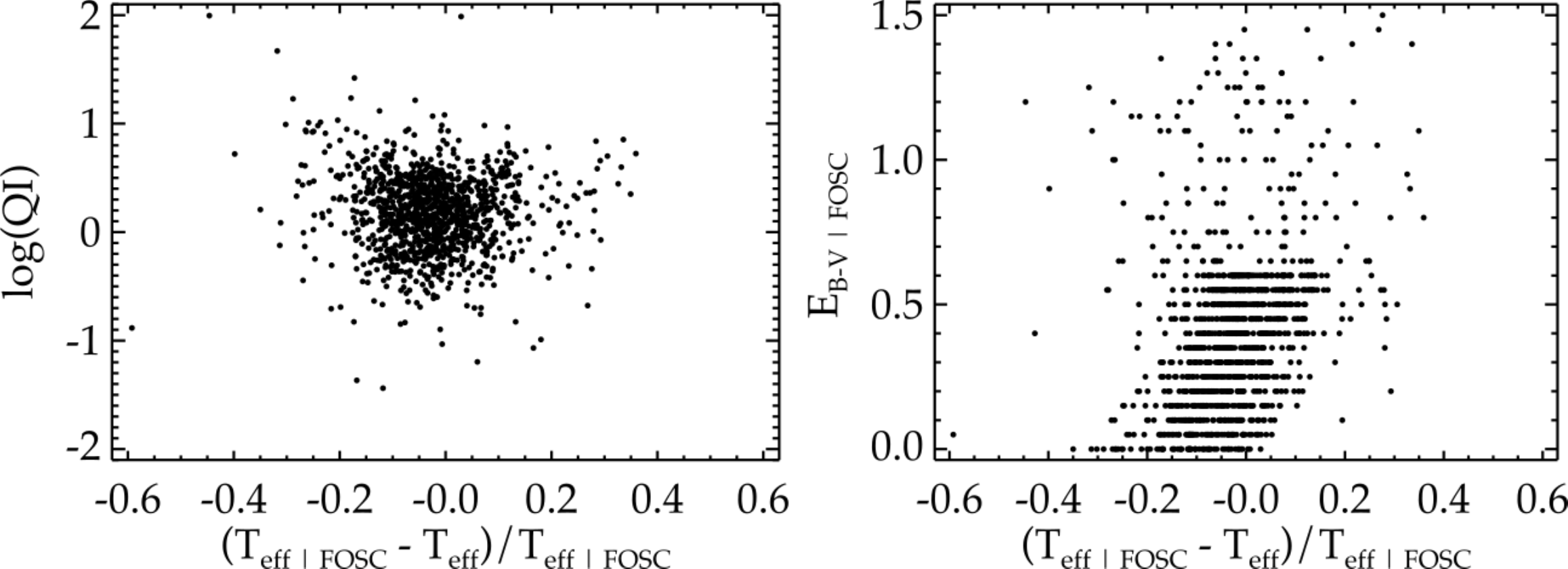}
\caption{Logarithm of the quality index (left) and $E_{B-V}$ (right) given by the FOSC vs. the relative difference in effective temperature compared to Ga10. }
\label{fig:diffteffchi}
\end{center}
\end{figure}

Finally, we note that there is no clear correlation between the quality index of the FOSC and the difference in effective temperature relative to Ga10 (Fig.~\ref{fig:diffteffchi}). There is however a weak correlation between $E_{B-V}$ and the relative error. This was expected from our analysis, since we have shown that the error on the spectral type increases with increasing absorption. Moreover, the systematics over the whole range of temperature are visible in the correlation between $E_{B-V}$ and the temperature error. This makes sense because methods based on stellar spectra are less sensitive to reddening. Thus, stars that are classified as early spectral type (O to F) are systematically found with a lower temperature by Ga10 and SG12. This means that the template of the early type has been reddened to match the magnitudes of a target that appears as a later type to both Ga10 and SG12 . This produces a positive correlation between the $E_{B-V}$ found by the FOSC and the temperature difference. On the other hand, stars that are classified as late type (G to M) are systematically found with a higher temperature by Ga10 and SG12. This means that the template of the late type has been left free to match the magnitudes of what is actually a reddened star of earlier type. This also results in a positive correlation between reddening and temperature difference. Those systematics were already discusses in \citet{AAOmega2}. The comparison with Ga10 and our validation of the method on a synthetic sample provide new evidence that confirms this trend.  It is the result of the joint effects of systematics of the method and of the selection of a particular stellar population due to the magnitude limits of \corot{} targets.

\section{Properties of the fields observed by \corot{}}\label{sec:stelpop}
\begin{figure}
\begin{center}
\includegraphics[width=0.49\columnwidth]{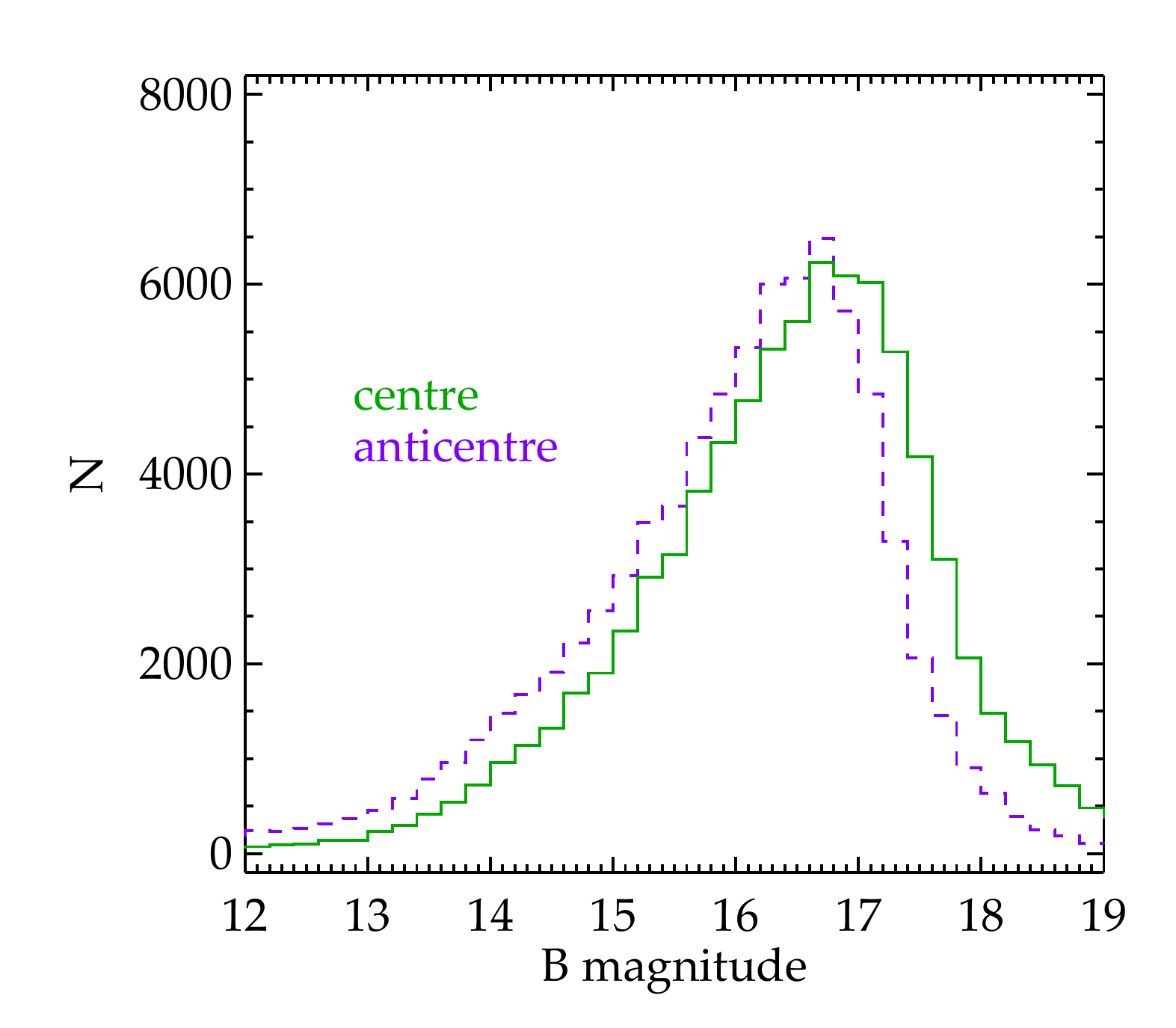}
\includegraphics[width=0.49\columnwidth]{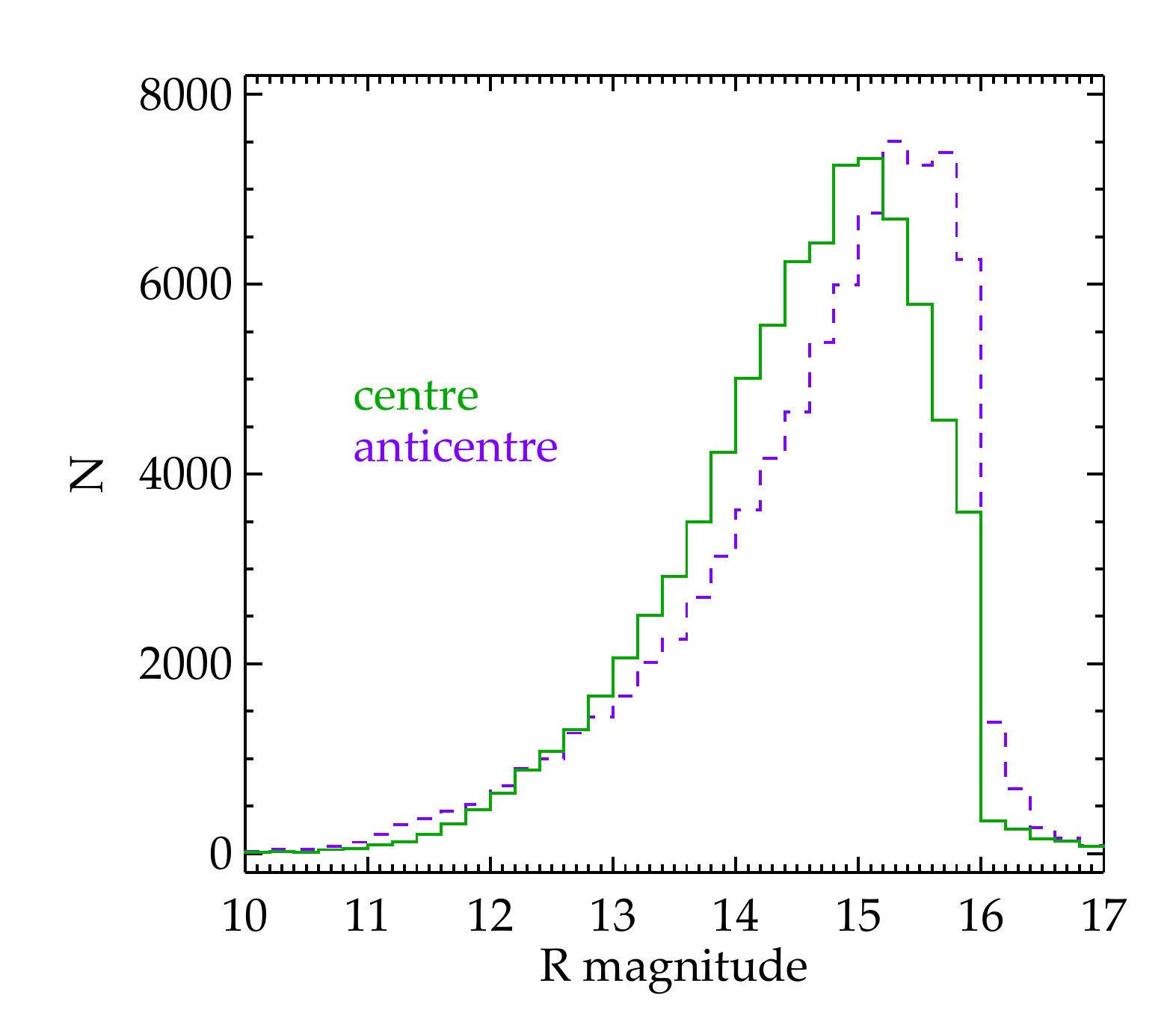}
\caption{Histogram of B (left) and R (right) magnitudes for \corot{} targets observed in the centre direction (green line) and anticentre direction (purple dashed line).}
\label{fig:histmagtarg}
\end{center}
\end{figure}

\corot{} targets are almost homogeneously distributed in the centre and anticentre directions, with 82 386 targets for the former and 81 194 targets for the latter. The FOSC could not be performed for only 4 targets in the centre direction (respectively 27 in the
anticentre) because they did not have magnitudes available in at least 4 different bands. They are generally faint stars at a very small angular distance to a very bright star or, occasionally, failures of the mask attribution algorithm on board the satellite. This also results in a small number of targets that do not comply with the selection criterion $11\leq r \leq 16$, as can be seen in Fig.~\ref{fig:histmagtarg}. This figure also shows that the centre fields have an average extinction that is greater than the anticentre fields, which results on average in fainter blue magnitudes and relatively brighter red magnitudes in the centre.
\subsection{Stellar properties of \corot{} targets}
Consequently, the proportion of dwarfs and giants are different depending on the direction of observation. For the runs observing in the centre direction, there are about 50\% of targets that are classified as dwarfs, whereas this proportion rises to about 75\% in the anticentre direction. This is in good agreement with spectroscopic measurements and can be predicted using galactic models \citep{Gazzano2010}. Indeed, interstellar absorption dims the magnitudes of giant stars to a point where they can be observed within the magnitude range of \corot{}, which increases their number in the target population relative to dwarfs for fields with stronger reddening. 

\begin{figure}
\begin{center}
\includegraphics[width=0.49\columnwidth]{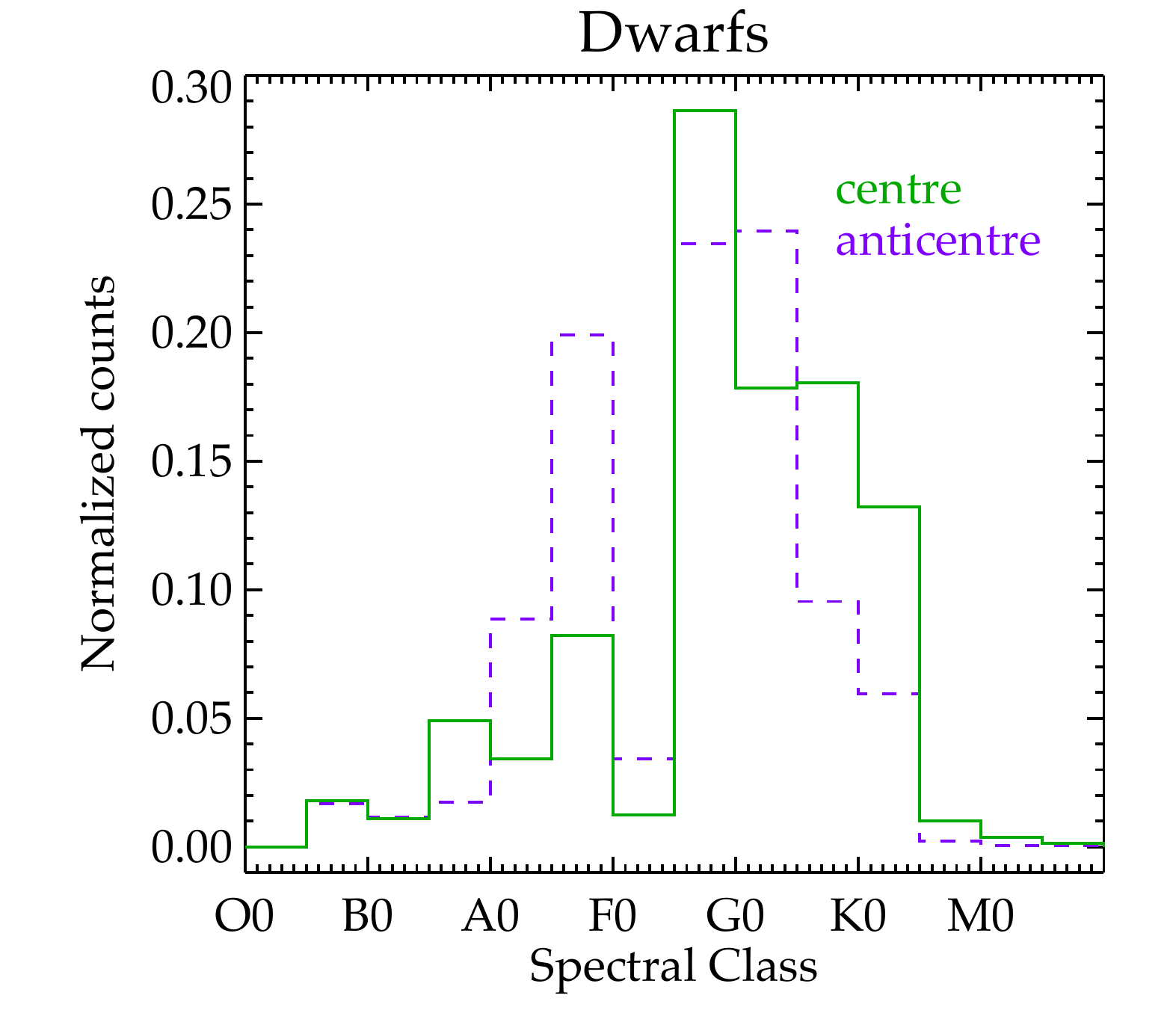}
\includegraphics[width=0.49\columnwidth]{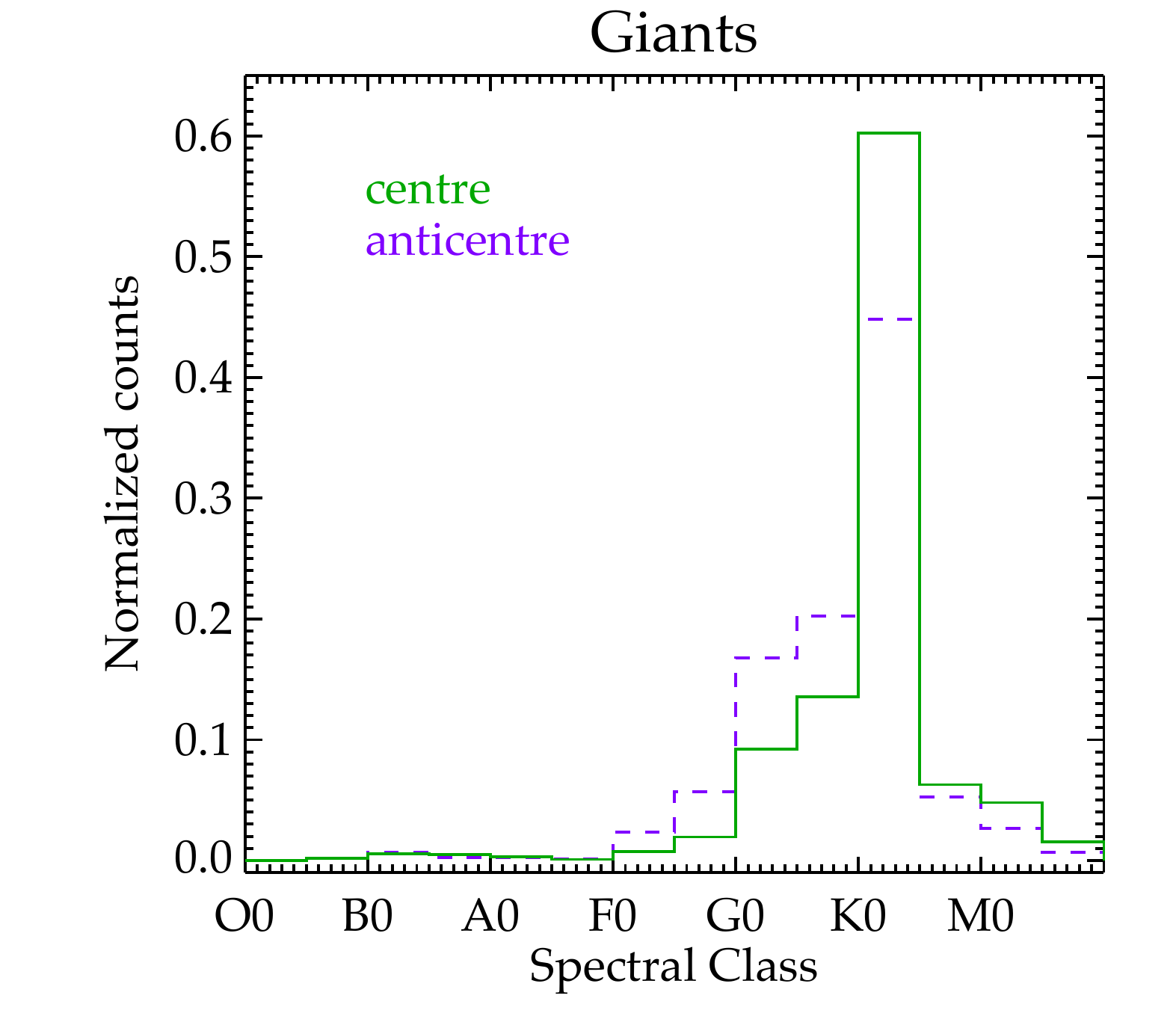}
\caption{Distribution of spectral types in the population of \corot{} dwarfs (left) and giants (right) observed in the centre direction (green line) and anticentre direction (purple dashed line).}
\label{fig:hsc}
\end{center}
\end{figure}

The difference in mean stellar absorption between centre and anticentre is also visible in the distribution of spectral types in the population of \corot{} targets. Figure~\ref{fig:hsc} shows the relative frequency of the different spectral type for dwarfs and giants as a function of the direction of observation.  There are no significant differences for O to mid-B stars, but those stars have a relative error of $70\%$ in effective temperature so we discard them for this discussion. The lack of early F-type dwarfs is mainly spurious and is due to the lack of corresponding templates in the library. This is reinforced by the reddening or spectral type degeneracy, where early F-type stars are systematically classified as A stars when the reddening is weak, but are classified as G stars when the reddening is strong. M stars are rare amongst \corot{} targets, and most of these stars are giants, although about 15\% of them may be misclassified dwarfs. The most abundant type of dwarfs are late-F types. Clearly, there are more observed A-type dwarfs in the anticentre direction, while there are more mid-G to later dwarfs in the centre direction.
This is the result of two concurring effects: one that can be attributed to the limitations of the target magnitude range and the other that is due to the increasing error on the effective temperature with increasing reddening. 

Firstly, early-type targets are indeed more frequent in the anticentre direction. On average, when the interstellar medium is less dense, distant bright blue stars may be dimmed enough, but are not significantly reddened so that their apparent magnitudes enter the range of \corot{} targets. On the other hand, the stronger dust column density in the centre direction further dims the magnitudes of distant bright blue targets, which results in a smaller proportion of those stars that are actually within the targets magnitude range. 

Secondly, weakly reddened blue stars are more often correctly classified than strongly reddened blue stars. This result in the classification of reddened blue stars as less reddened later type stars. Thus in the centre direction, a significant number of early-type stars are lacking in the early-type bins, which decreases their relative frequency, and  they are added to the number of stars classified as late type, which increases their relative frequency.

The higher frequency of bluer targets in the anticentre direction is also seen in the difference in the distribution of early-G dwarfs and late-F dwarfs. But this is a more robust result because those classes have an overall better accuracy and are less affected by reddening. Moreover, this effect also seems to  be seen in \citet{Gazzano2010}, but their sample is biased against F-type stars due to their high angular velocity. Nonetheless, they also find that early-G dwarfs are as abundant as late-F dwarfs in the anticentre direction, while late-F dwarfs are about twice as abundant as early-G dwarfs in the centre direction.

This has an important effect on exoplanet searches. Indeed, the depth of the transit is related to the planet-to-star radius ratio. Late-F dwarfs can have a radius that is 20 to 40 \% bigger than late-G dwarfs. Thus, the proportion of targets offering favourable conditions to detect planetary transits is greater in the anticentre direction. Nevertheless, more planets have been detected and characterised in the centre direction, which is evidence that the occurence rate of planets may indeed be greater around stars in the centre direction.

The distribution of giant targets is clearly dominated by early-K type stars in the centre and anticentre directions. This was clearly found for LRc01 but not for LRa01 in \citet{Gazzano2010}. However, the prevalence of early-K giants is also expected from galactic population synthesis in the centre and anticentre directions. Those stars have radii larger than 10~R$_\odot$, which make them not suitable for a planet search with \corot{}. But the number of such stars observed with \corot{} has revolutionised the field of red giant asteroseismology \citep{Mosser2011,Miglio2013}.

\subsection{Galactic properties of \corot{} targets}
\begin{figure}
\begin{center}
\includegraphics[width=0.49\columnwidth]{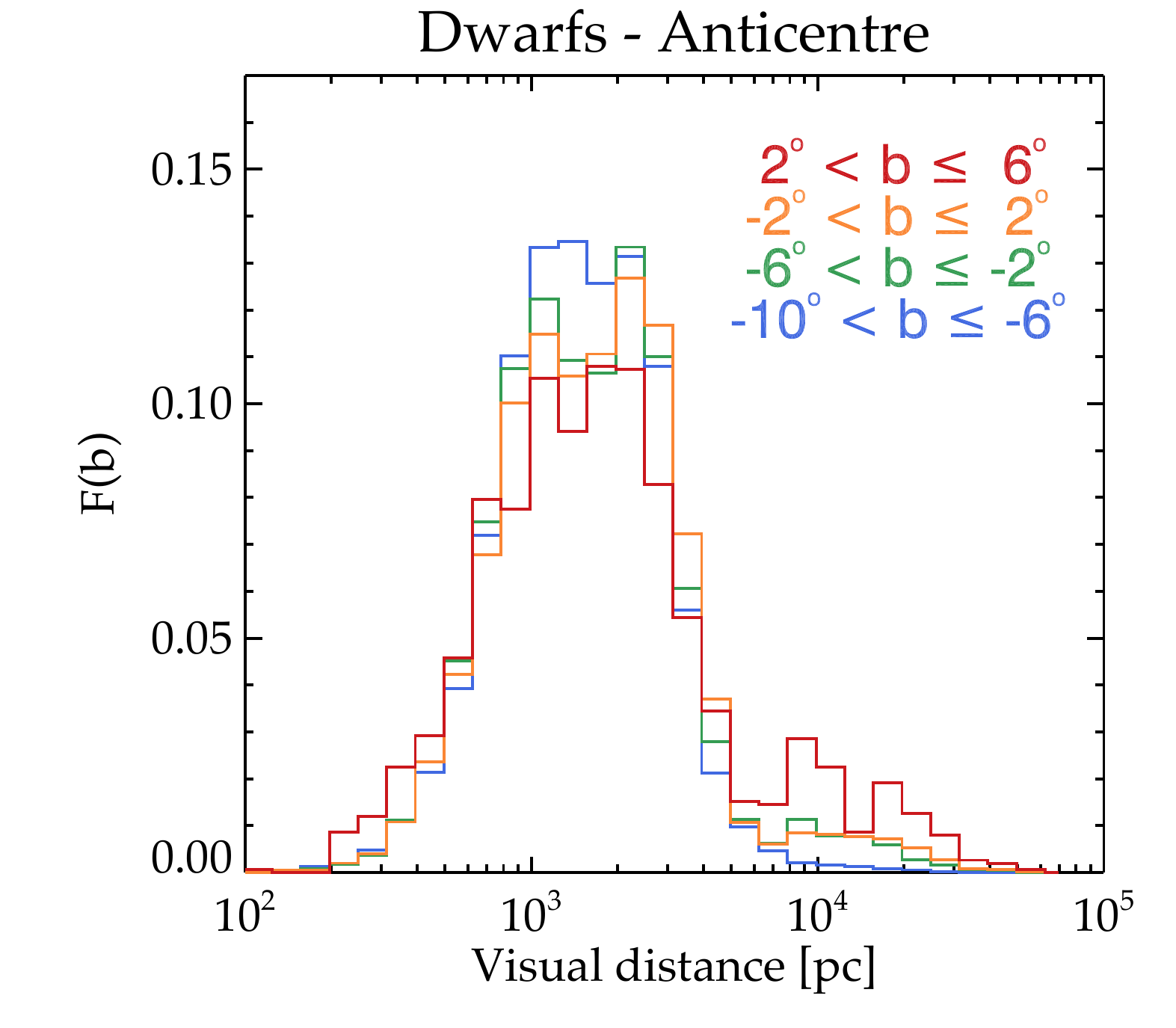}
\includegraphics[width=0.49\columnwidth]{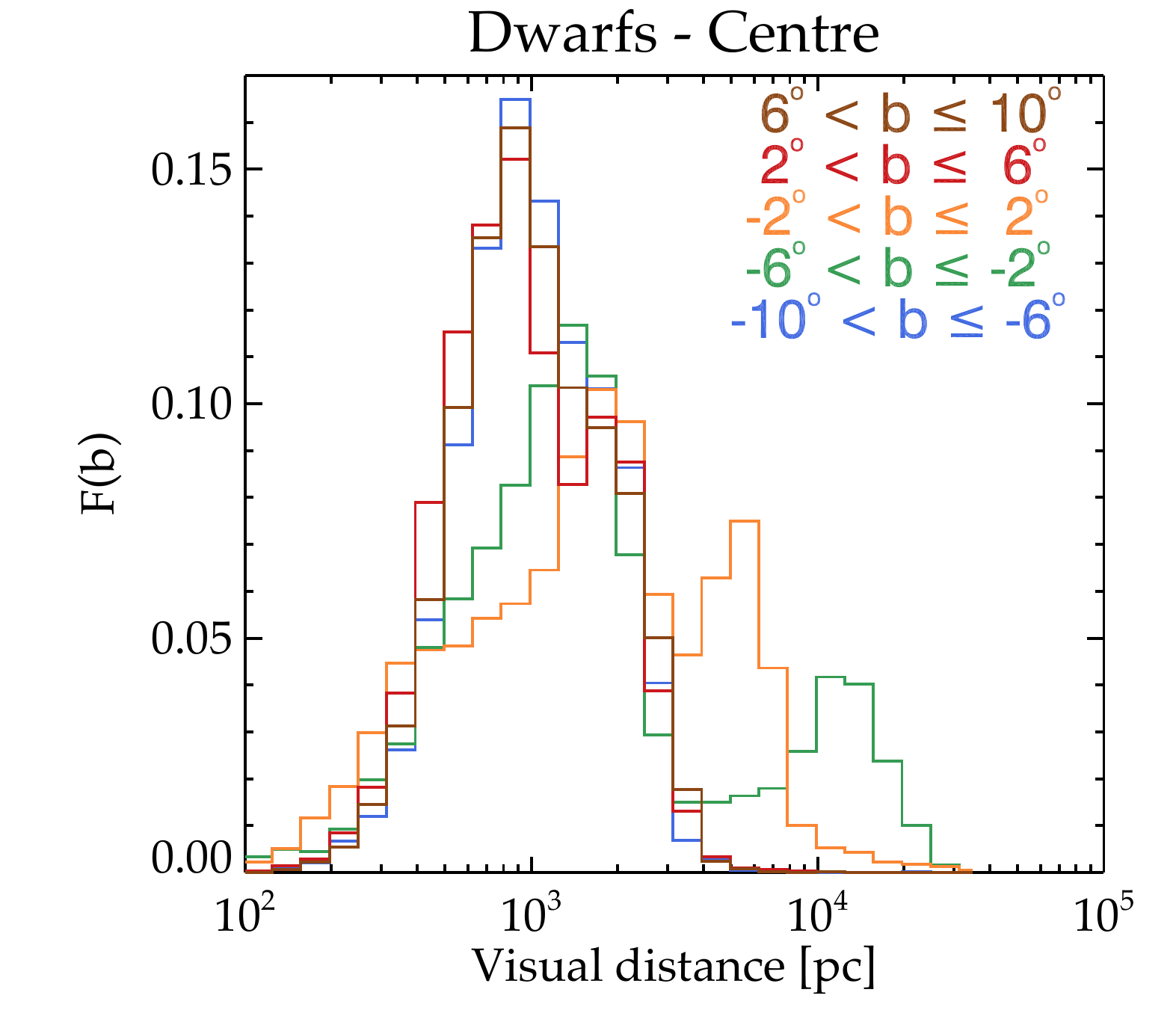}\\
\includegraphics[width=0.49\columnwidth]{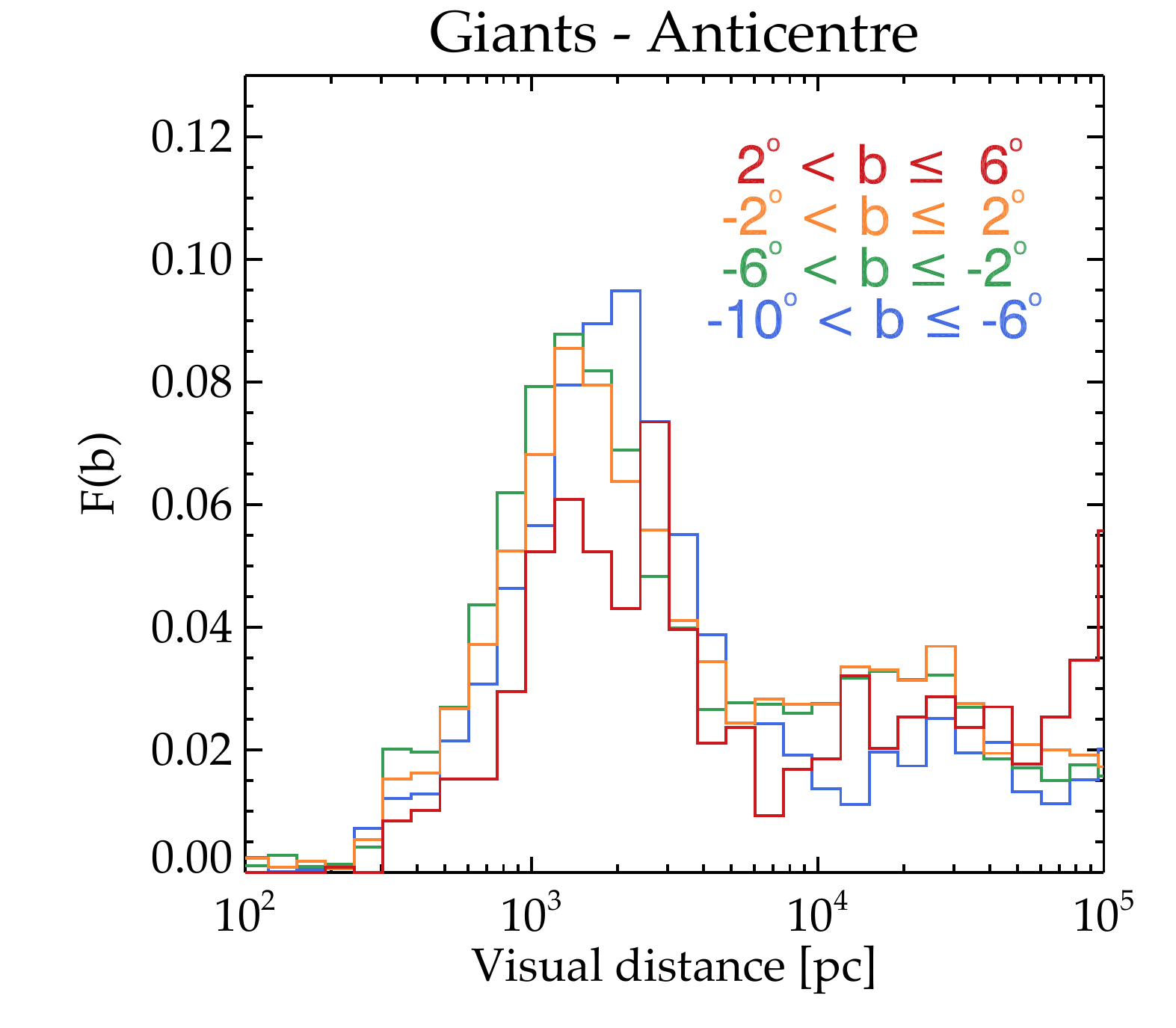}
\includegraphics[width=0.49\columnwidth]{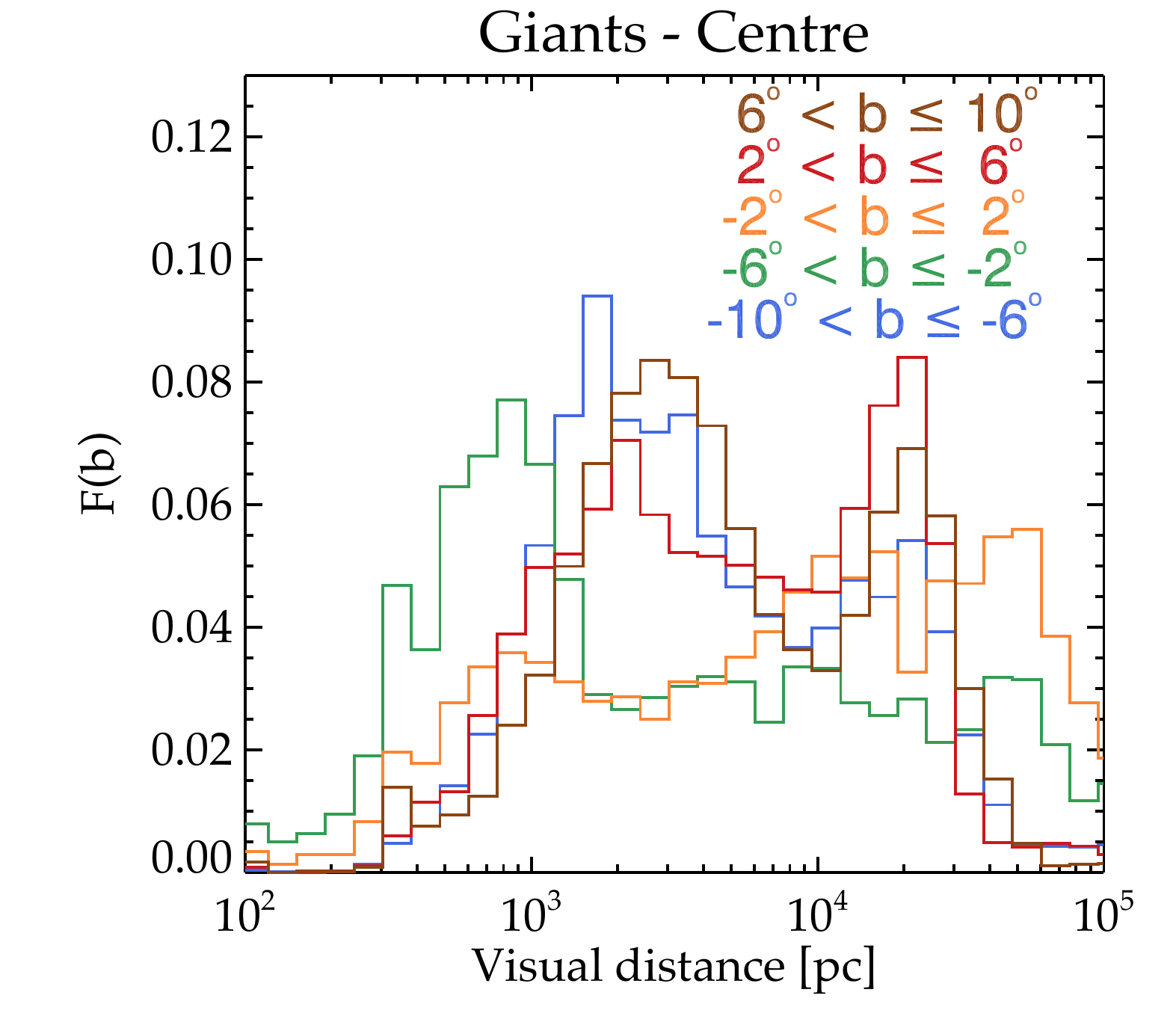}
\caption{Normalised histogram of visual distances for samples of different galactic latitudes $b$ (in colour as indicated on the plots) for dwarfs (top) and giants (bottom) in the anticentre(left) and centre fields (right).}
\label{fig:distlat}
\end{center}
\end{figure}

In addition to an effective temperature calibration, the Pickles library provides the bolometric absolute magnitude and bolometric corrections in several bands as a function of the spectral type. Using this information and the spectral type found by the FOSC, we can compute the visual distance of the targets, i.e the distance obtained using the apparent magnitude uncorrected for interstellar absorption. This is an upper limit on the real distance to the target. Figure~\ref{fig:distlat} shows the distribution of visual distances for dwarfs and giants and for different bins of galactic latitude computed using the K-band magnitude. We see that, in general, the distribution of the distance of dwarfs is found to be shifted to lower values relative to giants. This is expected because the visual distance modulus is set by the magnitude range of \corot{}. Moreover, the similarity in the distribution of distances shows the uniformity of interstellar extinction in the anticentre direction, while the different distributions at high galactic latitude and low galactic latitude exhibits the non-uniform dust distribution in the centre direction. This effect is enhanced as we probe deeper in the galaxy and results in the complex patterns observed in the visual distances of giants in the centre direction.

\begin{figure}
\begin{center}
\includegraphics[width=0.85\columnwidth]{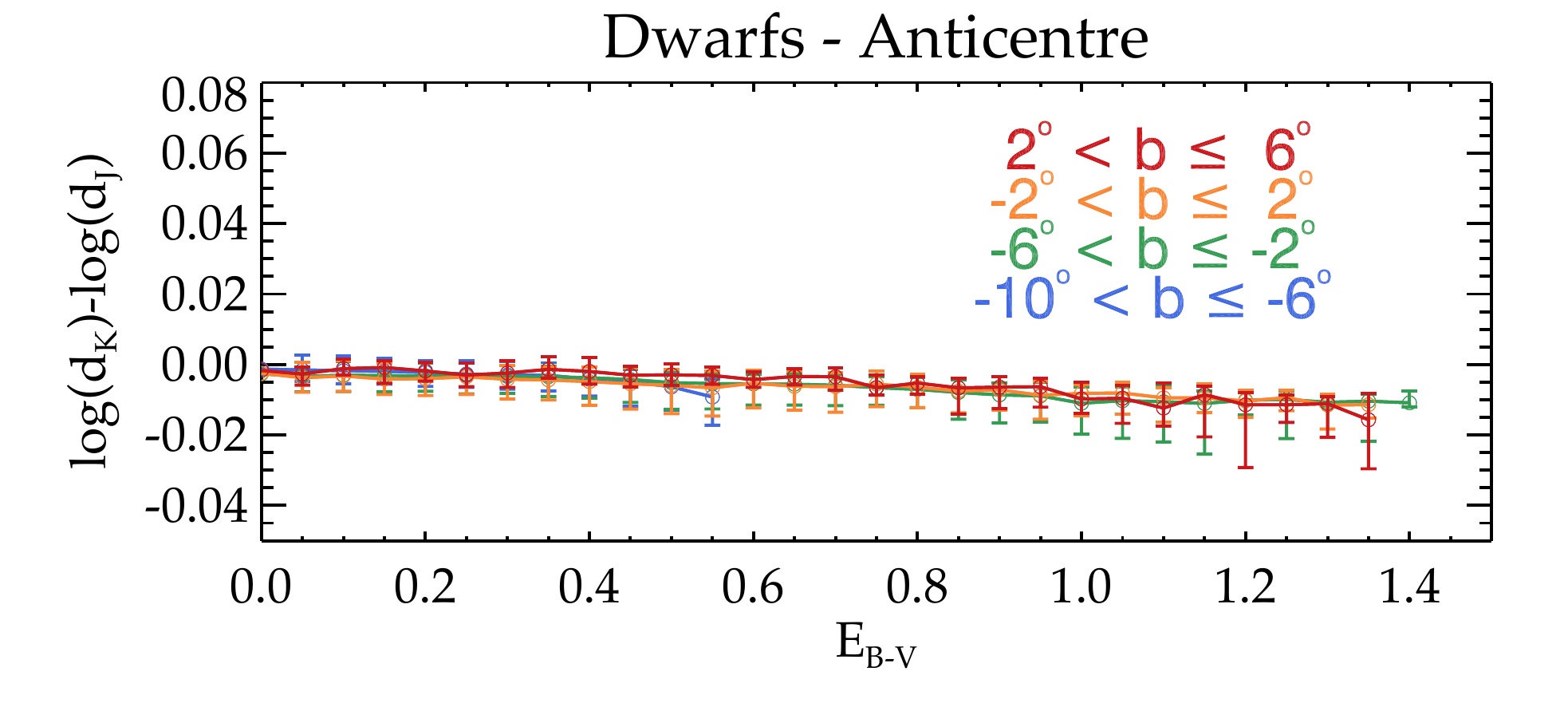}\\
\includegraphics[width=0.85\columnwidth]{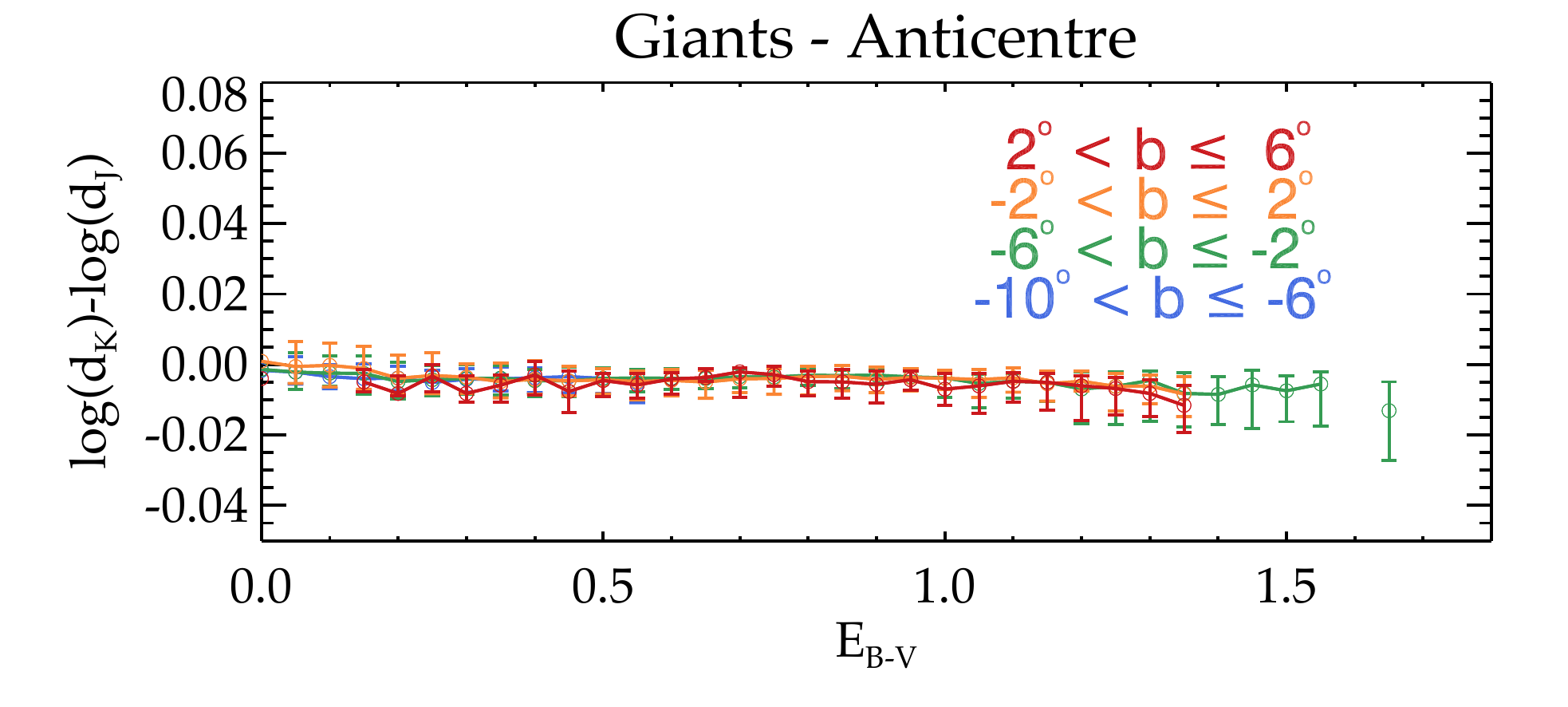}\\
\includegraphics[width=0.85\columnwidth]{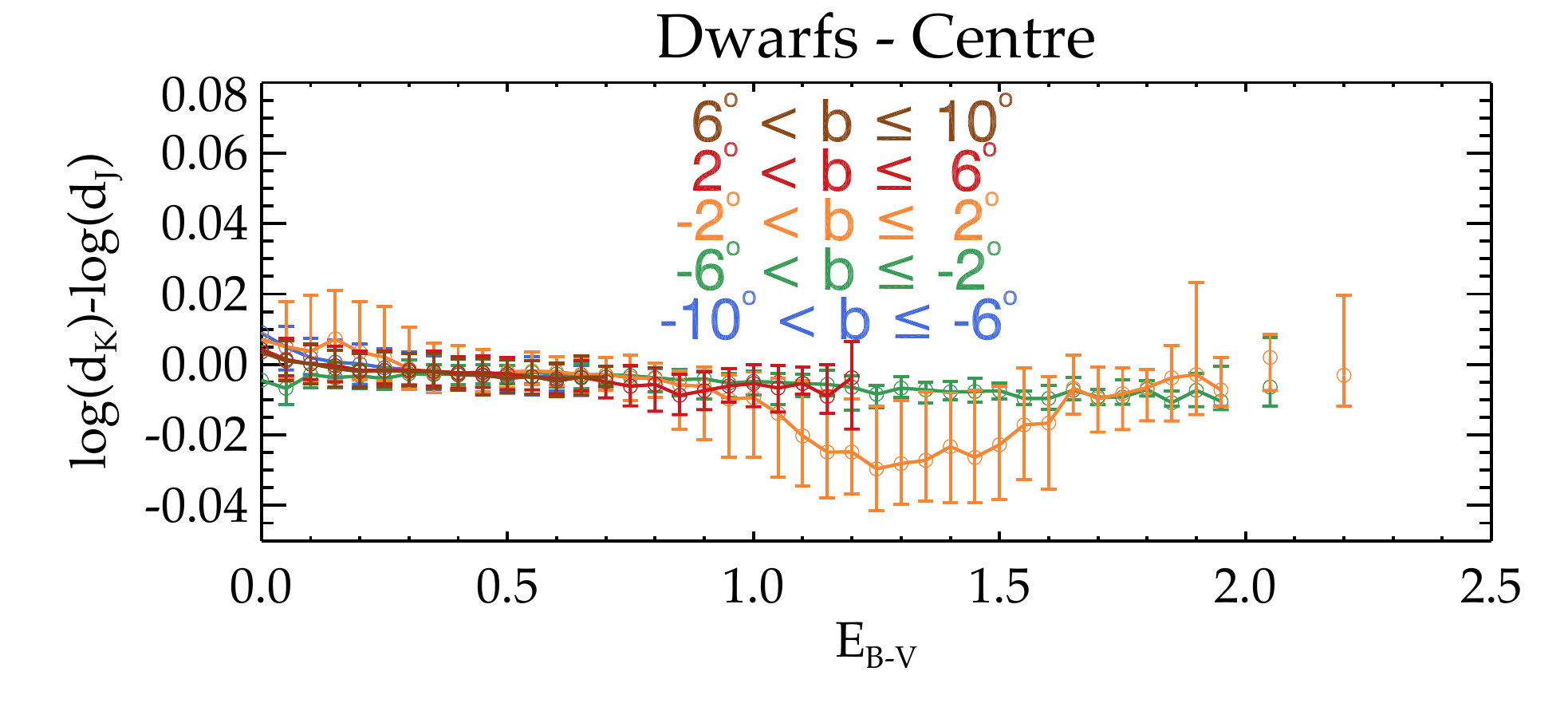}\\
\includegraphics[width=0.85\columnwidth]{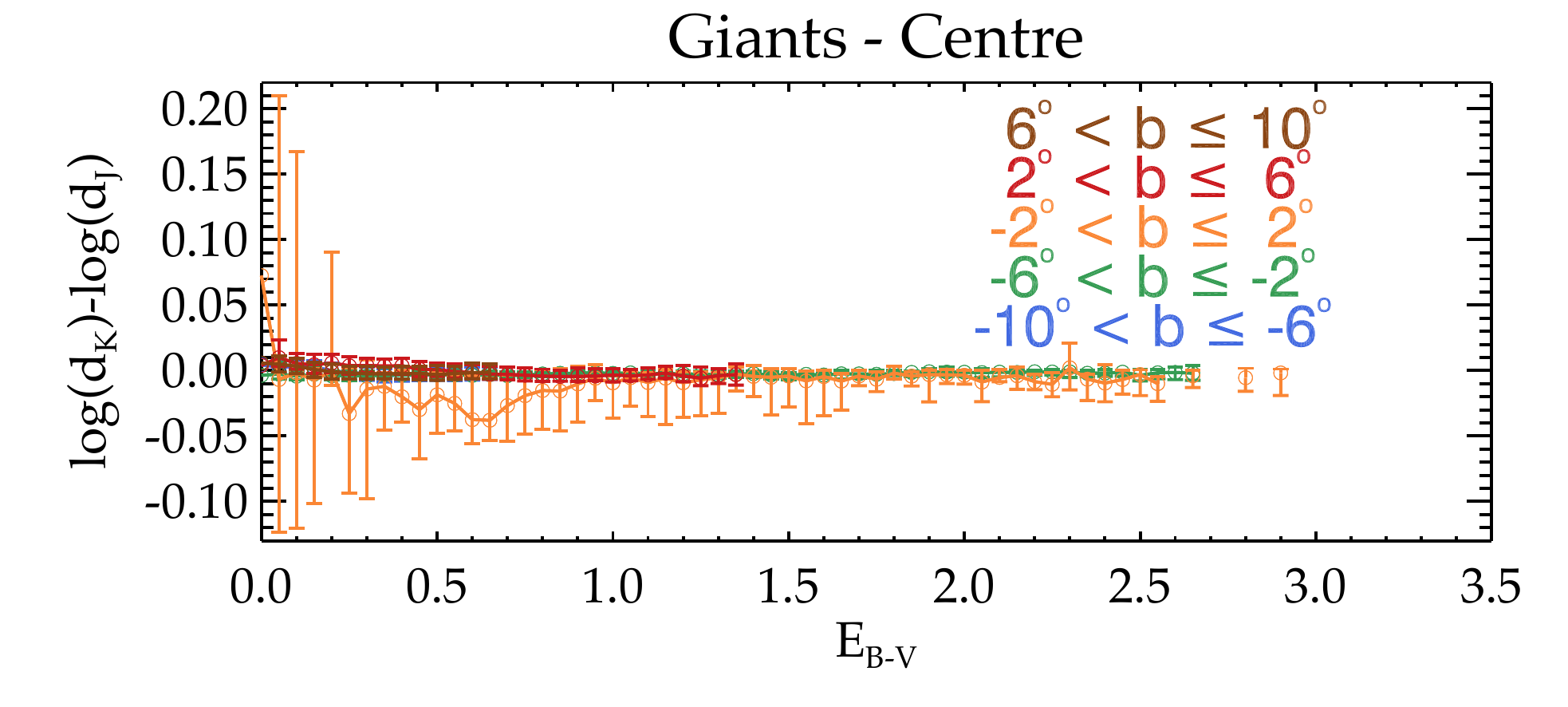}
\caption{Median values of the difference between the distance $d_K$ and  $d_J$ using the $K_s$ band and $J$ band, respectively; these values are computed for bins of $E_{B-V}$, for different galactic latitudes $b$ (in colour as indicated in the plots) in the anticentre fields (two top panels) and centre fields (two bottom panels), and for dwarfs and giants as indicated in the figure. The error bars give the value of the first and third quartile of the distribution for each bin.}
\label{fig:deltad}
\end{center}
\end{figure}

The real distances are obtained by dividing the visual distance by a factor proportional to the interstellar absorption. We used the extinction law of \citet{Fitz1999} with $R_V =3.1$ and the value of $E_{B-V}$ determined by the FOSC. To test whether the distance estimation is sensitive to the value of $R_V$, we computed the difference between the distance computed using two different magnitudes, namely $J$ and $K_s$, as a function of $E_{B-V}$. The results are shown in Fig.~\ref{fig:deltad}.
We see that overall the distances computed using different magnitudes give consistent results for all galactic latitudes and longitudes except in the galactic plane in the centre direction. This suggests that the extinction law that we use may indeed not apply to the densest region of interstellar medium. 

\begin{figure}
\begin{center}
\includegraphics[width=0.48\columnwidth]{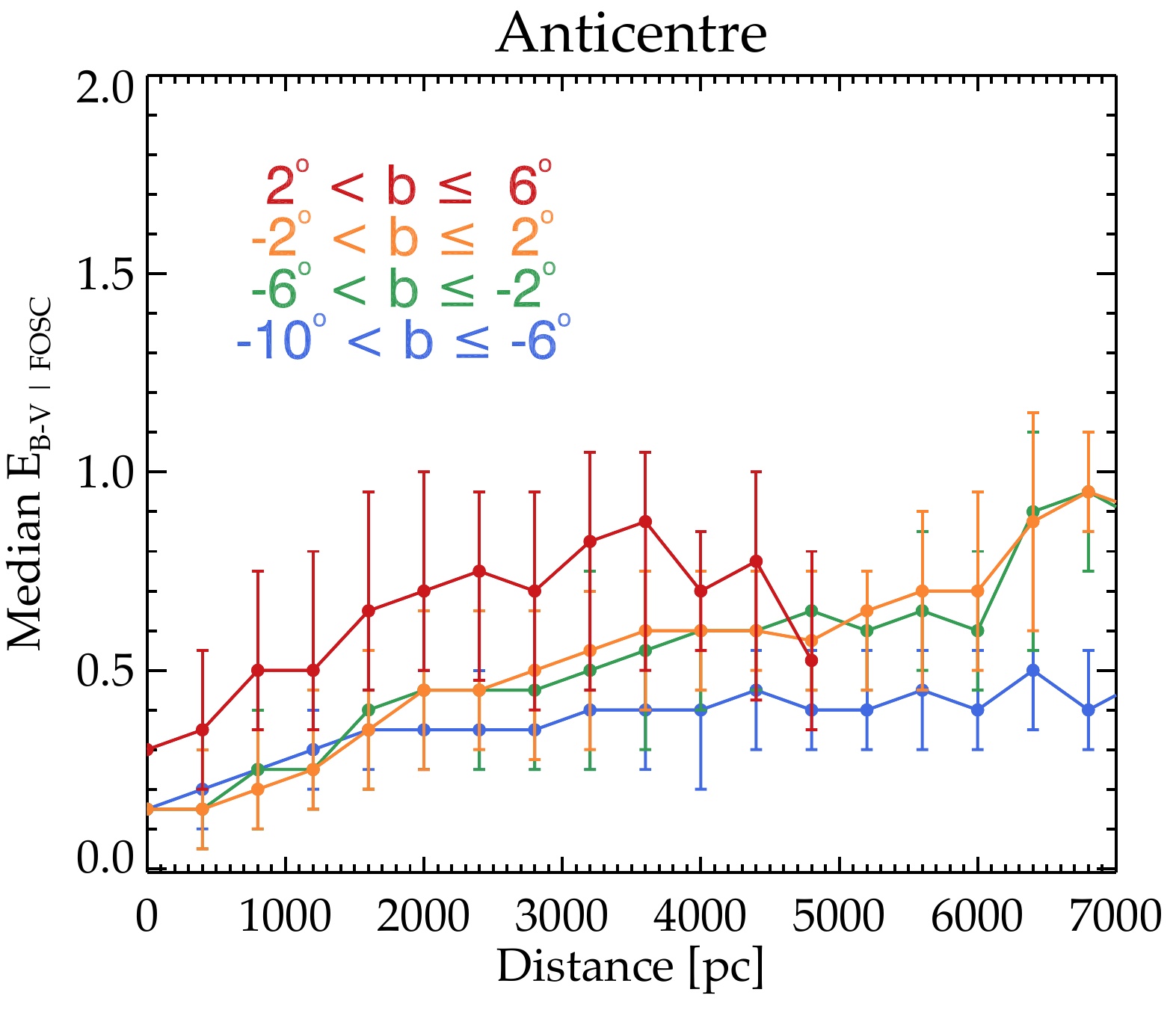}
\includegraphics[width=0.48\columnwidth]{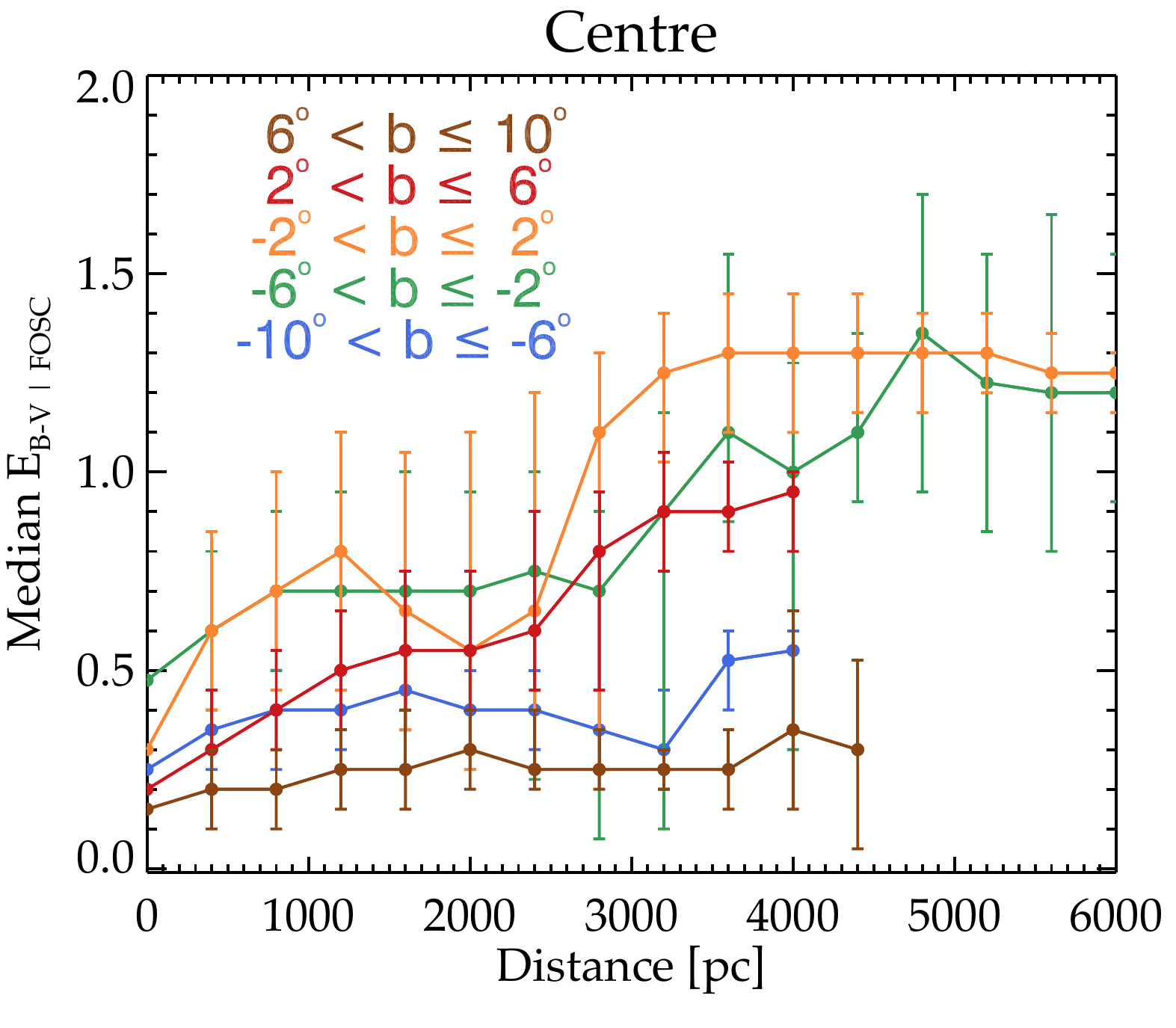}
\caption{Median value of $E_{B-V}$ determined by the FOSC for bins of 500 pc of radial distance and different values of galactic latitude $b$ for the anticentre fields (left) and centre fields (right). The error bars give the value of the first and third quartile of the distribution for each bin.}
\label{fig:reddist}
\end{center}
\end{figure}

Owing to lower values of mean interstellar absorption, fields pointing in the anticentre direction probe more distant populations with a mean distance from the sun of about 8.6 kpc in the anticentre direction (2.3 kpc for dwarfs and 28 kpc for giants) against a mean distance of 7.2 kpc in the centre direction (1.6 kpc and 13 kpc for dwarfs and giants, respectively). There is a general trend to overestimate $E_{B-V}$ with the FOSC, thus one must bear in mind that the distances derived this way tend to be underestimated. On the other hand, when the reddening is strong, the FOSC tend to underestimate it. Thus for giants, distances are generally overestimated. We plot in Fig.~\ref{fig:reddist} the median values of $E_{B-V}$ determined by the FOSC for different bins of radial distance and different values of galactic latitude. We limit our sample to the dwarf stars to avoid the complexity expected from the deeper probing obtained with giants.
If we suppose that interstellar absorption is proportional to the dust column density, we would expect its value to increase linearly with the distance provided that the dust is homogeneously distributed in the sky. We see that it is generally the case, except once again for low latitudes in the direction of the galactic centre. The global behaviour of the $E_{B-V}$ curve as a function of the radial distance results from the combined effect of the spatial distribution of stellar populations and the non-homogeneous interstellar absorption. Overall, we see that in regions further away from the galactic plane, the median $E_{B-V}$ is in general lower than those that are close to the galactic plane. We find similar values ($0.2 \lesssim E_{B-V} \lesssim 0.4$) for high latitudes ($10 \leq b \leq 6$) in the centre and anticentre directions. Moreover, the centre fields show a remarkable symmetry relatively to the galactic plane that is not as clear in the anticentre direction. This is in good agreement with \textit{Planck} data (see Fig.~\ref{fig:red}). However, the \textit{Planck} maps suggest that the mean $E_{B-V}$ in the anticentre direction should be stronger for negative latitudes ($-6 \leq b \leq 0$). But we see that the $E_{B-V}$ found by the FOSC is the strongest for  ($2 \leq b \leq 6$). This is readily explained by the fact that the only runs that are observing at those latitudes are SRa01 and SRa05. They cover a very small band of longitude and are close to the Rosette Nebula within a dense molecular cloud. Thus the strong value of $E_{B-V}$ there is not representative of this latitude range, but of this very special region in space. Indeed, once we consider only this small region on the \textit{Planck} maps, we find it in good agreement with our results.

\section{Conclusions}\label{sec:conclusion}
In this paper, we have described the latest version of the FOSC, which is the procedure used to estimate the spectral type of \corot{} targets. The spectral type and $E_{B-V}$ are products of the \corot{} mission and are delivered together with the \corot{} light curves in their FITS format. They are also available online through the Exodat database. Because of the faintness and the number of the targets, the FOSC is based on broadband photometry. The photometry obtained during  a mission-dedicated observational campaign is used whenever possible, but it was necessary to resort to USNO-B1.0 data for about a third of the targets. Using a synthetic set of stars, we were able to quantify how the performance of the classification are affected by the use of different photometric catalogues. We also provide a statistical estimate of the precision and accuracy of the classification. We find consistent results by confronting the result of the FOSC to other independent estimates of the spectral type. We find that the luminosity class is estimated with great significance, and that the typical uncertainty on the spectral type given by the FOSC is of half a spectral class. Finally, the spectral classification and corresponding $E_{B-V}$ of \corot{} targets are used to discuss the properties of their galactic environnement.

This study cannot provide a precise error bar for individual \corot{} targets but is valid in a statistical way. In particular, there are two \corot{} runs pointed in directions of particularly high space absorption (SRc01 and LRc03) in which the validity of these results may be limited. However, the strong reddening in this case dims the apparent magnitude to a point where distant stars exit the fainter magnitude limits of \corot{} targets. This effectively removes almost all giants of the target sample, and filters out the dwarfs that would be too severely reddened to ensure good performance of the FOSC. It can be expected thus that the performances of the FOSC on the remaining targets that are not strongly reddened would be equivalent to that shown here. Another shortcoming of this study is that it was assumed that in the case of a binary star target, the spectral type determined by the FOSC would correspond to that of the primary star. This introduces a source of error on the spectral type that has not been taken into account here. We suspect however that this effect may be small compared to other sources of uncertainties. Finally, this study has made clear that interstellar absorption is the main source of uncertainty on the spectral type. The performances of the FOSC are significantly improved by limiting the range of possible values for the reddening. This was only performed in the latest version of the FOSC and previous versions are expected to produce significantly more uncertain results. An accurate estimation of the maximum reddening in the \corot{} fields was made possible by the use of reddening maps produced thanks to a new parametrisation of dust emission that covers the whole sky, which is one of the products associated with the 2013 release of data from the \textit{Planck} mission. Eventually,  the third data release of the GAIA mission that is scheduled in summer 2018\footnote{see \url{http://www.cosmos.esa.int/web/gaia/release}} should give the spectral classification of all stars observed in the faint stars channel, as well as their binary nature and proper motions, to an unprecedented level of precision.

\section*{Acknowledgments}
The authors are grateful to the referee for valuable comments that improved the manuscript. CD acknowledges support by CNES grant 426808 and ANR (Agence Nationale de la Recherche, France) programme IDEE (ANR-12-BS05-0008) "Interaction Des \'Etoiles et des Exoplan\`etes". Figure~\ref{fig:corot_eyes} uses images from the Digitized Sky Survey, which was produced at the Space Telescope Science Institute under U.S. Government grant NAG W-2166. The images of these surveys are based on photographic data obtained using the Oschin Schmidt Telescope on Palomar Mountain and the UK Schmidt Telescope. The plates were processed into the present compressed digital form with the permission of these institutions. The Obscat catalogue was derived from photometry acquired with the Isaac Newton Telescope (INT), operated on the island of La Palma by the Isaac Newton group in the Spanish Observatorio del Roque de Los Muchachos of the Instituto de Astrofisica de Canarias. This publication makes use of data products from the Two Micron All Sky Survey, which is a joint project of the University of Massachusetts and the Infrared Processing and Analysis centre/California Institute of Technology, funded by the National Aeronautics and Space Administration and the National Science Foundation. This work is based in part on services provided by the GAVO data centre. This research has made use of the VizieR catalogue access tool, CDS, Strasbourg, France. This research uses observations obtained with Planck, an ESA science mission with instruments and contributions directly funded by ESA Member States, NASA, and Canada.
The authors thank M.-A. Miville-Deschênes for providing them with the \textit{Planck} data. CD is grateful to P.-Y. Chabaud and F. Agneray for their work on the Exodat database and information system. She wishes to thank J.M. Almenara, S.C.C. Baros and R.F. Diaz and the \corot{} Exoplanet Science Team for helpful discussions.
\bibliographystyle{aa}
\bibliography{exodat}
\newpage
\begin{appendix}
\section{Correspondence between spectral type and effective temperature}\label{appA}
Following the information found through the Vizier Service in table \texttt{J/PASP/110/863/synphot}.\\
\begin{minipage}[t]{\columnwidth}
                \begin{tabular}[t]{ll}
                 \hline
                 \hline
                Spectral Type & $\log T_{\rm eff}$ \\
                 \hline
                 \hline
                O5V     &       4.600   \\
                O9V     &       4.550   \\
                B0V     &       4.450   \\
                        B1V     &       4.350   \\
                B2IV    &       4.300   \\
                B3V     &       4.280   \\
                B6V     &       4.150   \\
                B6IV    &       4.100   \\
                B8V     &       4.070   \\
                B9V     &       4.030   \\
                A0IV    &       3.988   \\
                A0V     &       3.980   \\
                A2V     &       3.950   \\
                A3V     &       3.944   \\
                A5IV    &       3.900   \\
                A5V     &       3.929   \\
                A7V     &       3.906   \\
                F1IV    &       3.847   \\
                F0V     &       3.858   \\
                F2V     &       3.831   \\
                F5IV    &       3.817   \\
                F5V     &       3.815   \\
                F6V     &       3.798   \\
                F8IV    &       3.789   \\
                F8V     &       3.781   \\
                G0IV    &       3.773   \\
                G0V     &       3.764   \\
                G2IV    &       3.755   \\
                G2V     &       3.751   \\
                G5IV    &       3.748   \\
                G5V     &       3.747   \\
                G8IV    &       3.725   \\
                G8V     &       3.727   \\
                K0IV    &       3.700   \\
                K0V     &       3.715   \\
                K1IV    &       3.680   \\
                K2V     &       3.689   \\
                K3IV    &       3.660   \\
                K3V     &       3.653   \\
                K4V     &       3.638   \\
                K5V     &       3.622   \\
                K7V     &       3.602   \\
                M0V     &       3.580   \\
                M1V     &       3.566   \\
                M2V     &       3.550   \\
                M3V     &       3.519   \\
                M4V     &       3.493   \\
                M5V     &       3.470   \\
                M6V     &       3.431   \\
                \hline
                \hline
                \end{tabular}
                \begin{tabular}[t]{ll}
                    \hline
                    \hline
                     Spectral Type & $\log T_{\rm eff}$ \\
                     \hline
                     \hline
                    O8III       &       4.500   \\
                    B0I &       4.415   \\
                    B1I &       4.316   \\
                    B2III       &       4.300   \\
                    B3III       &       4.230   \\
                    B2II        &       4.204   \\
                    B3I &       4.193   \\
                    B5III       &       4.170   \\
                    B5I &       4.127   \\
                    B5II        &       4.100   \\
                    B8I &       4.049   \\
                    B9III       &       4.045   \\
                    A0I &       3.988   \\
                    A0III       &       3.981   \\
                    A2I &       3.958   \\
                    A3III       &       3.953   \\
                    A5III       &       3.927   \\
                    A7III       &       3.906   \\
                    F0II        &       3.900   \\
                    F0I &       3.886   \\
                    F0III       &       3.880   \\
                    F2II        &       3.865   \\
                    F2III       &       3.835   \\
                    F5I &       3.822   \\
                    F5III       &       3.815   \\
                    F8I &       3.785   \\
                    G0III       &       3.749   \\
                    G0I &       3.741   \\
                    G2I &       3.724   \\
                    G5II        &       3.720   \\
                    G5III       &       3.713   \\
                    G5I &       3.703   \\
                    G8III       &       3.700   \\
                    K0II        &       3.700   \\
                    K0III       &       3.686   \\
                    K1III       &       3.668   \\
                    G8I &       3.662   \\
                    K2III       &       3.649   \\
                    K3III       &       3.640   \\
                    K2I &       3.629   \\
                    K3II        &       3.629   \\
                    K4III       &       3.626   \\
                    K3I &       3.616   \\
                    K5III       &       3.603   \\
                    K4I &       3.601   \\
                    M0III       &       3.582   \\
                    M1III       &       3.577   \\
                    M2III       &       3.569   \\
                    M3III       &       3.560   \\
                    M4III       &       3.551   \\
                    M2I &       3.538   \\
                    M5III       &       3.534   \\
                    M3II        &       3.533   \\
                    M6III       &       3.512   \\
                    M7III       &       3.495   \\
                    M8III       &       3.461   \\
                    M9III       &       3.426   \\
                   \hline
                   \hline
                \end{tabular}
        \end{minipage}

\section{Maximum values of $E_{B-V}$ for each run}\label{app:ebv}
\begin{figure}[h]
\begin{center}
\includegraphics[width=1.0\columnwidth]{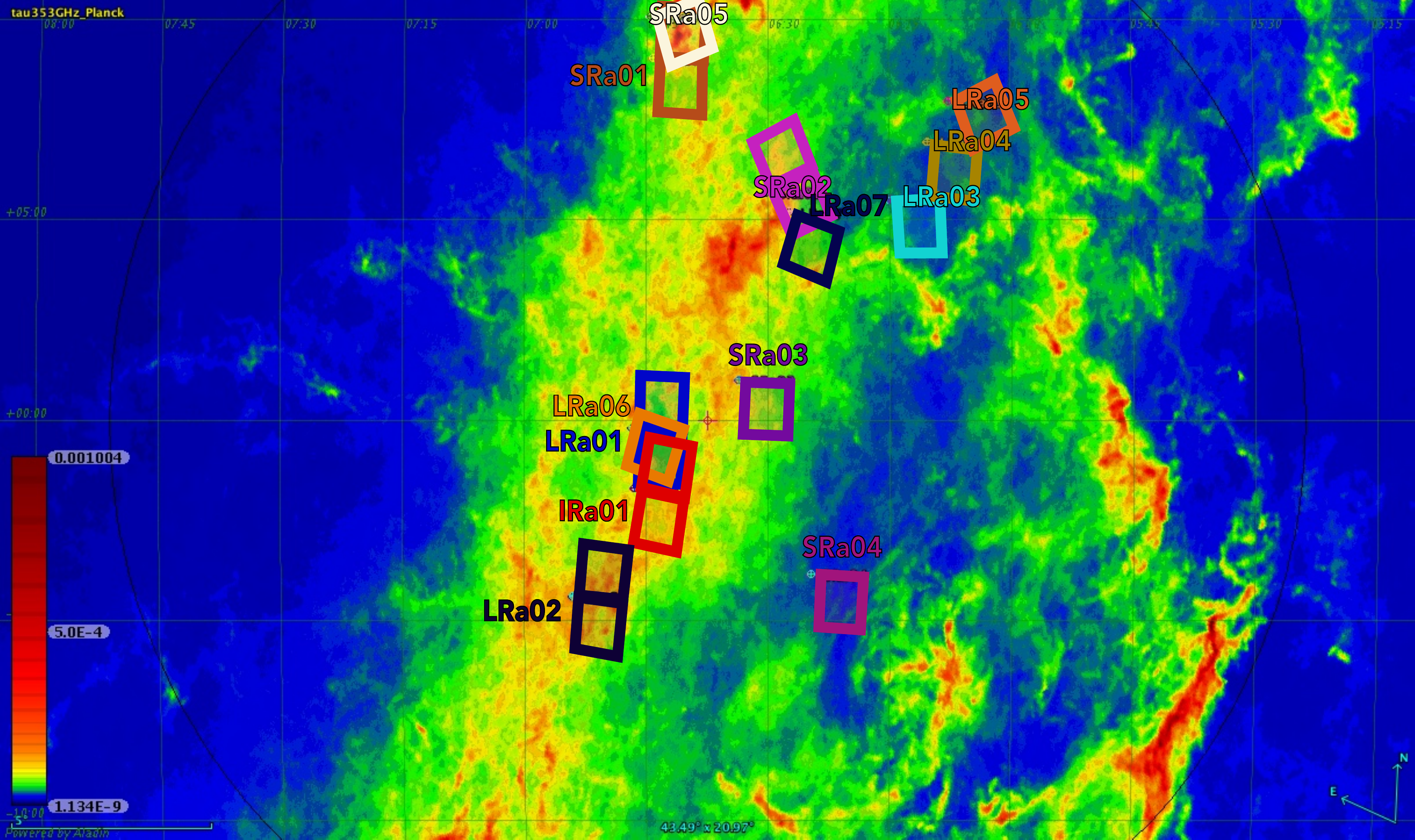}\\
\includegraphics[width=1.0\columnwidth]{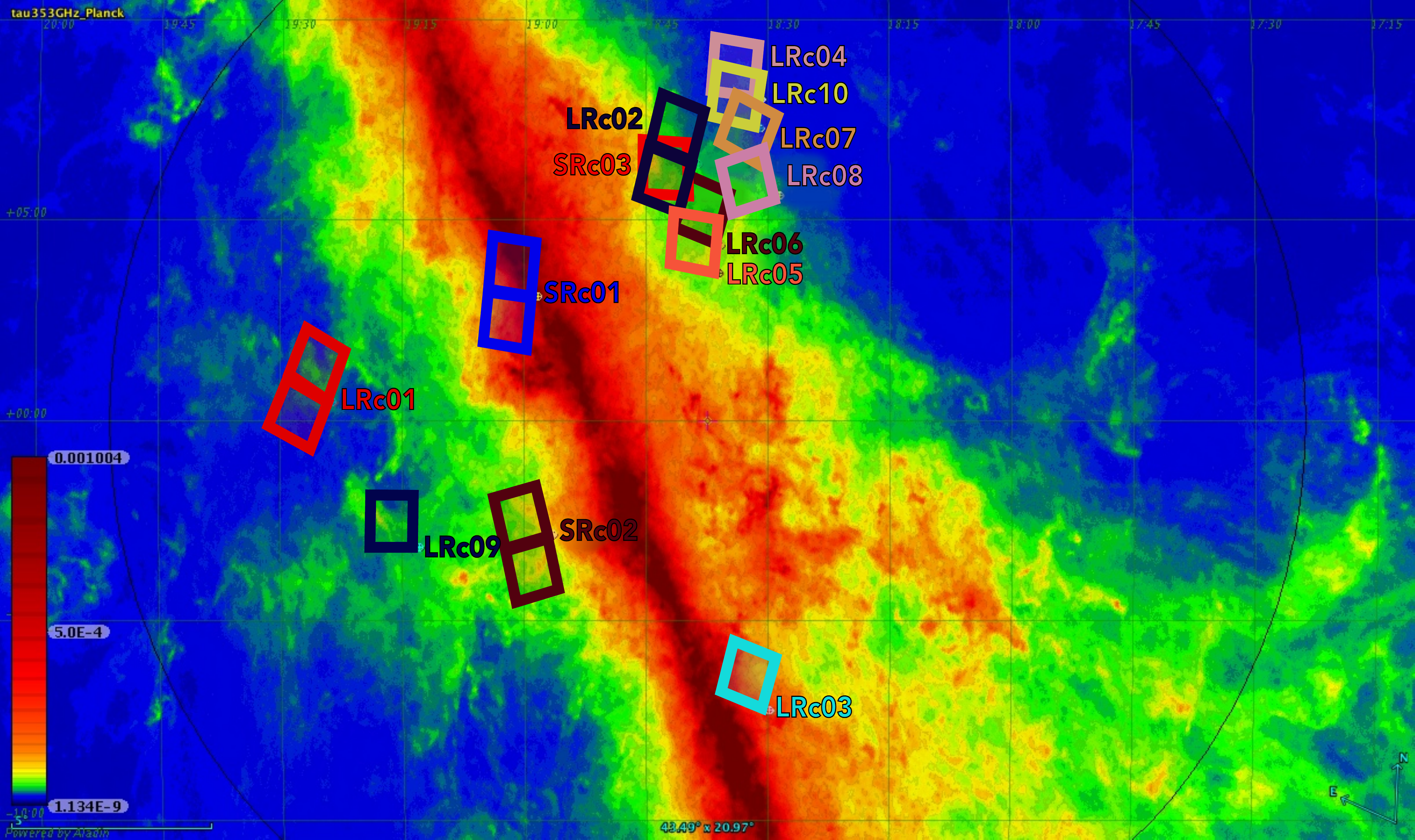}
\caption[Dust optical depth]{Map of dust optical depth at 353 GHz from \textit{Planck} in the CoRoT eyes. Top: anticentre direction. The footprints of the CCDs of the different runs of observation are given in different colours. Bottom: same as above but for the centre direction. The same colour scale is used in both directions.}
\label{fig:red}
\end{center}
\end{figure}

Using dust optical depth at 353 GHz measured by \textit{Planck} and shown in Fig.~\ref{fig:red}, we determined the maximum values of $E_{B-V}$ allowed in the fitting procedure for each run. Table~\ref{tab:red} gives the maximum values of $E_{B-V}$ that were used.

\begin{table}[h]
        \centering
        \caption{Maximum values of $E_{B-V}$ obtained for the \textit{Planck} $\tau_{353}$ maps allowed in the fitting procedure for each run}\label{tab:red}
        \begin{tabular}{lr|lr}
        \hline
        \hline
        Run     &  max($E_{B-V}$) & Run     &  max($E_{B-V}$)\\
        \hline
        \hline
IRa01& 1.3 & SRc01& 4\\
LRa01& 1.4 & LRc01& 0.6\\ 
SRa01& 1.4 & LRc02& 1.4\\  
SRa02& 1.2 & SRc02& 2.7\\
LRa02& 1.7 & LRc03& 4\\ 
LRa03& 0.6 & LRc04& 0.4\\ 
SRa03& 1.6 &  LRc05& 1.2\\ 
LRa04& 0.6 & LRc06& 1.1\\  
LRa05& 0.6 & LRc07& 0.6\\ 
SRa04& 0.6 & SRc03& 1.6\\ 
SRa05& 1.4 & LRc08& 0.7\\ 
LRa06& 1.4 & LRc09& 0.7\\ 
LRa07& 1.1 & LRc10& 0.5\\ 
        \hline
        \hline
        \end{tabular}
\end{table}
\section{Testing the FOSC}\label{App:accuracy}
\begin{figure*}[t]
\begin{center}
\includegraphics[width=0.4\textwidth]{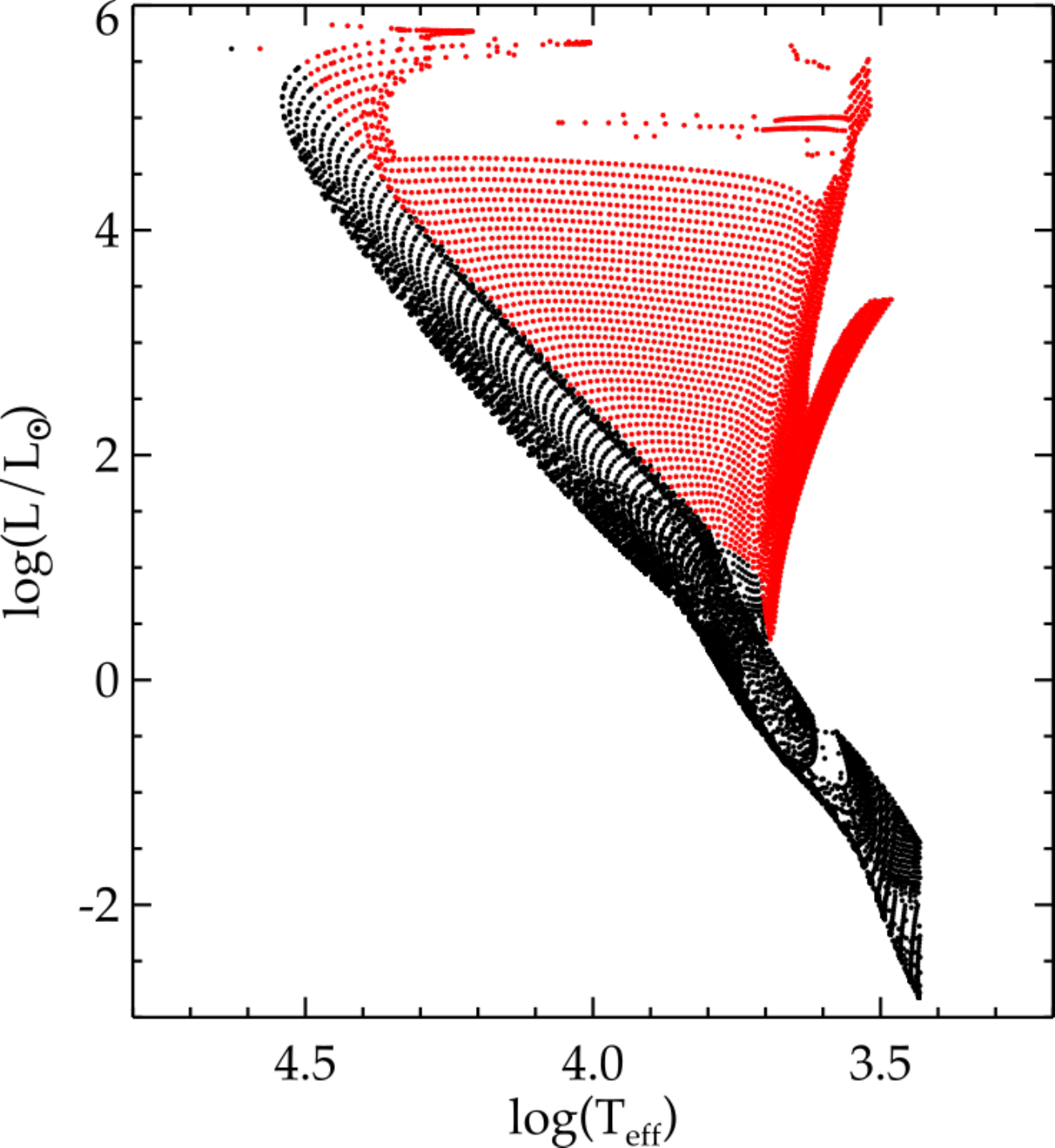}
\includegraphics[width=0.4\textwidth]{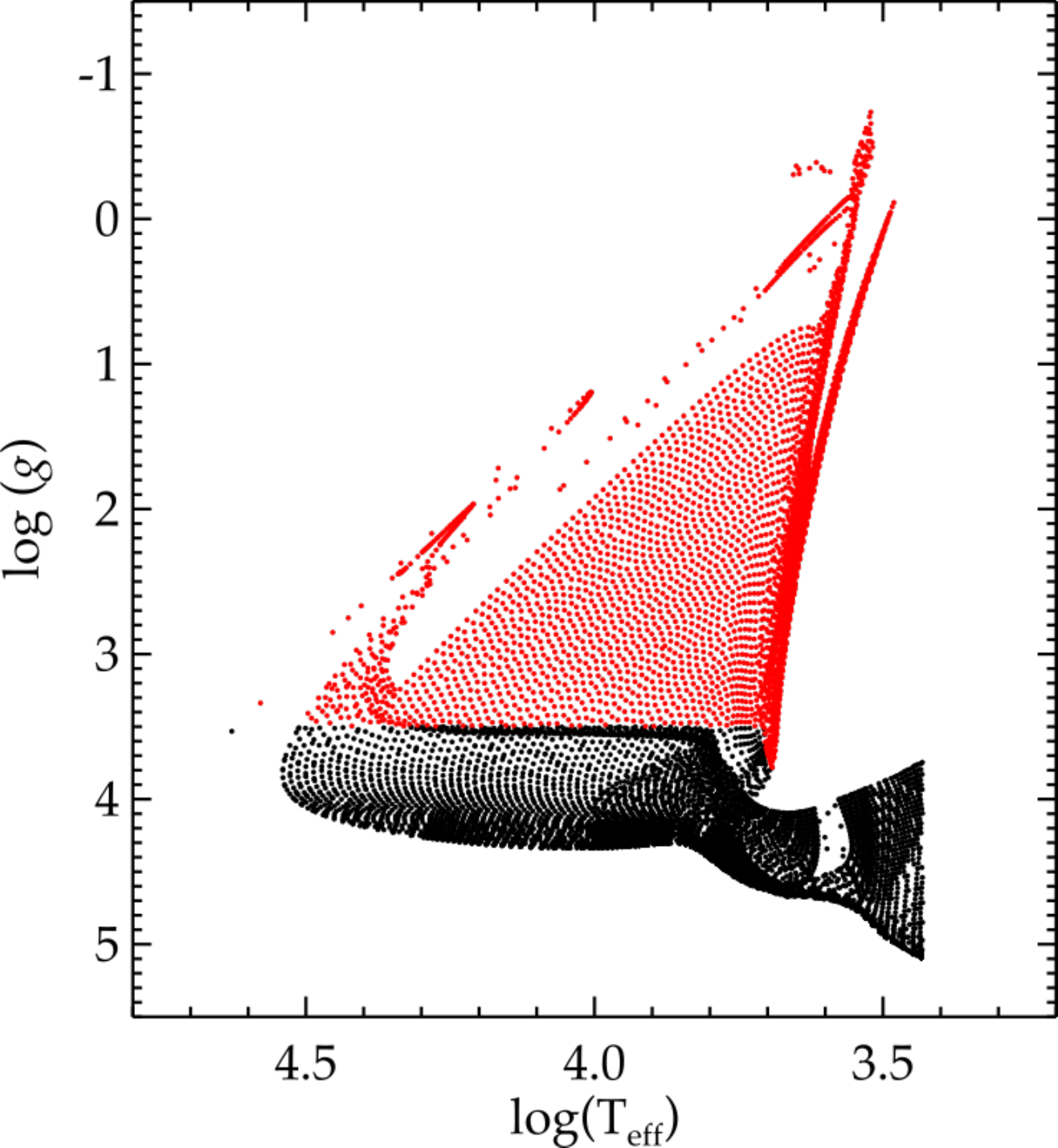}
\caption{Synthetic stellar sample. The black dots correspond to stars with $\log g \geq 3.5$ in c.g.s units that were processed through the dwarf pipeline, and the red dots denote stars that were processed with the giant pipeline. Left: luminosity as a function of effective temperature. Right: logarithm of the surface gravity as a function of effective temperature.}
\label{fig:padHR}
\end{center}
\end{figure*}
As explained in Sec.\ref{sec:accuracy}, the FOSC was validated on a synthetic set of stellar magnitudes. Several sources of systematic effects were studied, as detailed below.
\subsection{Error introduced by the templates}\label{sec:temperror}
To evaluate the sampling effects, we first generated a sample of stars of solar metallicity ($Z=0.019$, $Y=0.282$) with $T_{\rm eff}$ in the range of parameters covered by the Pickles library. Their distribution in the HR diagram is given in Fig.\ref{fig:padHR}.
The Padova isochrones are available for several photometric systems \citep[][and references therein]{Marigo2008}. We used the filters of the Jonhson-Cousin $U, B, V, R, I$ and Jonhson-Glass $J, H, K$ systems because they are very similar to the photometry that is available for \corot{} targets. We set photometric error to 1\% of the magnitude for each band. To ensure that the quality index remains comparable across the parameter space, we have, for each synthetic star, arbitrarily set the band with the lowest magnitude to 10, and shifted the other magnitudes accordingly. In this way, the 1\% error on the photometry has a homogenous absolute value for the whole sample. There are 7095 synthetic stars in the sample with $\log g \geq 3.5$ in c.g.s units and evolutionary stages that are compatible with luminosity classes IV or V of the MK system. These are shown with black symbols in Fig.~\ref{fig:padHR} and were classified using the "dwarf" pipeline of the FOSC. There are 4127 synthetic stars in the sample with $\log g < 3.5$ in c.g.s units and evolutionary stages that are compatible with luminosity classes I, II or III. There are shown with red symbols in Fig.~\ref{fig:padHR} and were classified using the "giant" pipeline of the FOSC. For this test, we did not add reddening to the synthetic magnitudes and we ran the FOSC while fixing $E_{B-V}$ to its known value of 0 for every synthetic target.

The top panel of Fig.~\ref{fig:correltempd} gives the effective temperature determined by the FOSC  (see Appendix \ref{appA}) as a function of the synthetic temperature for dwarfs and subgiants stars ($\log g \geq 3.5$ in c.g.s units) of solar metallicity. 
\begin{figure}
\begin{center}
\includegraphics[width=0.9\columnwidth]{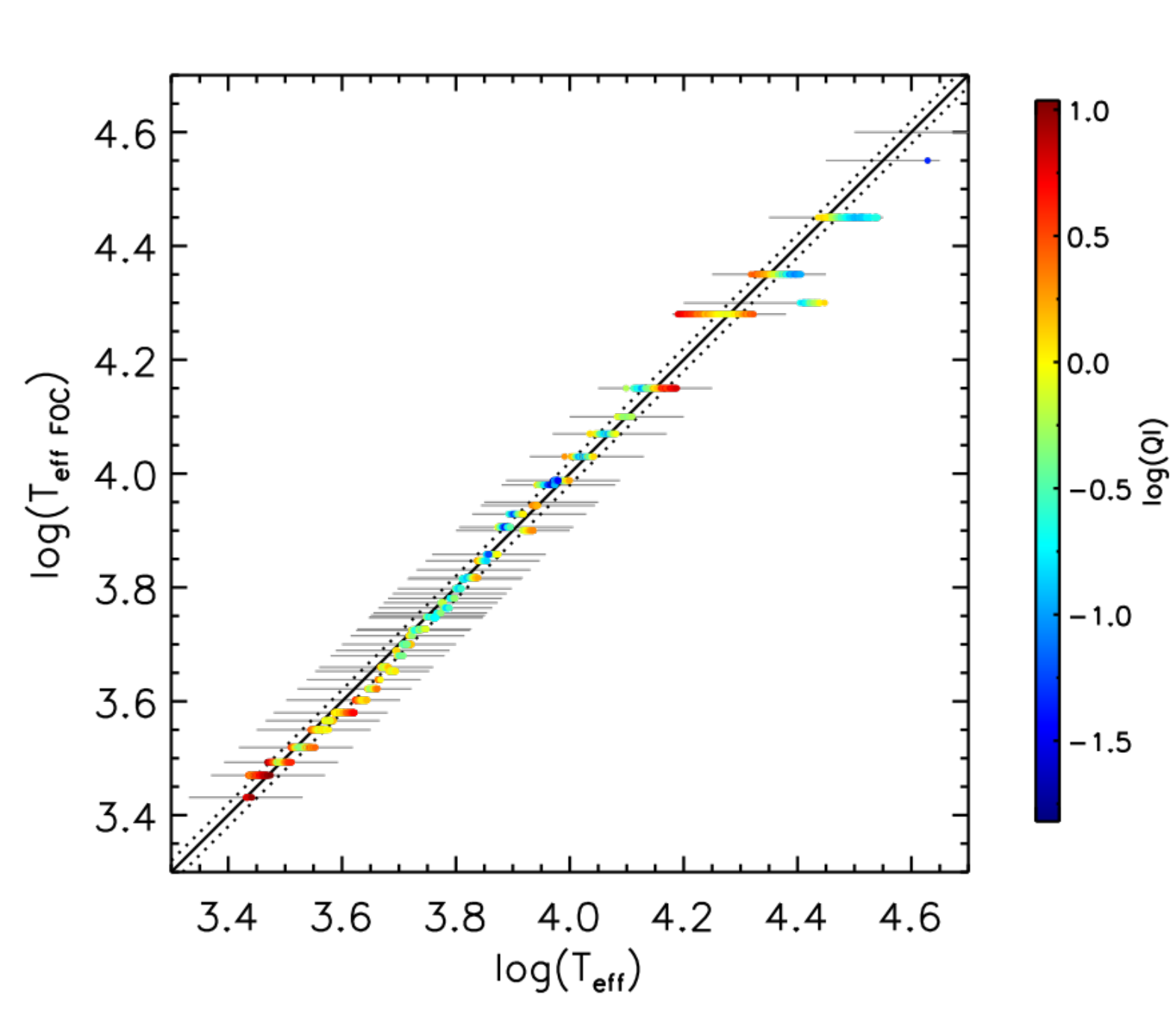}\\
\includegraphics[width=0.9\columnwidth]{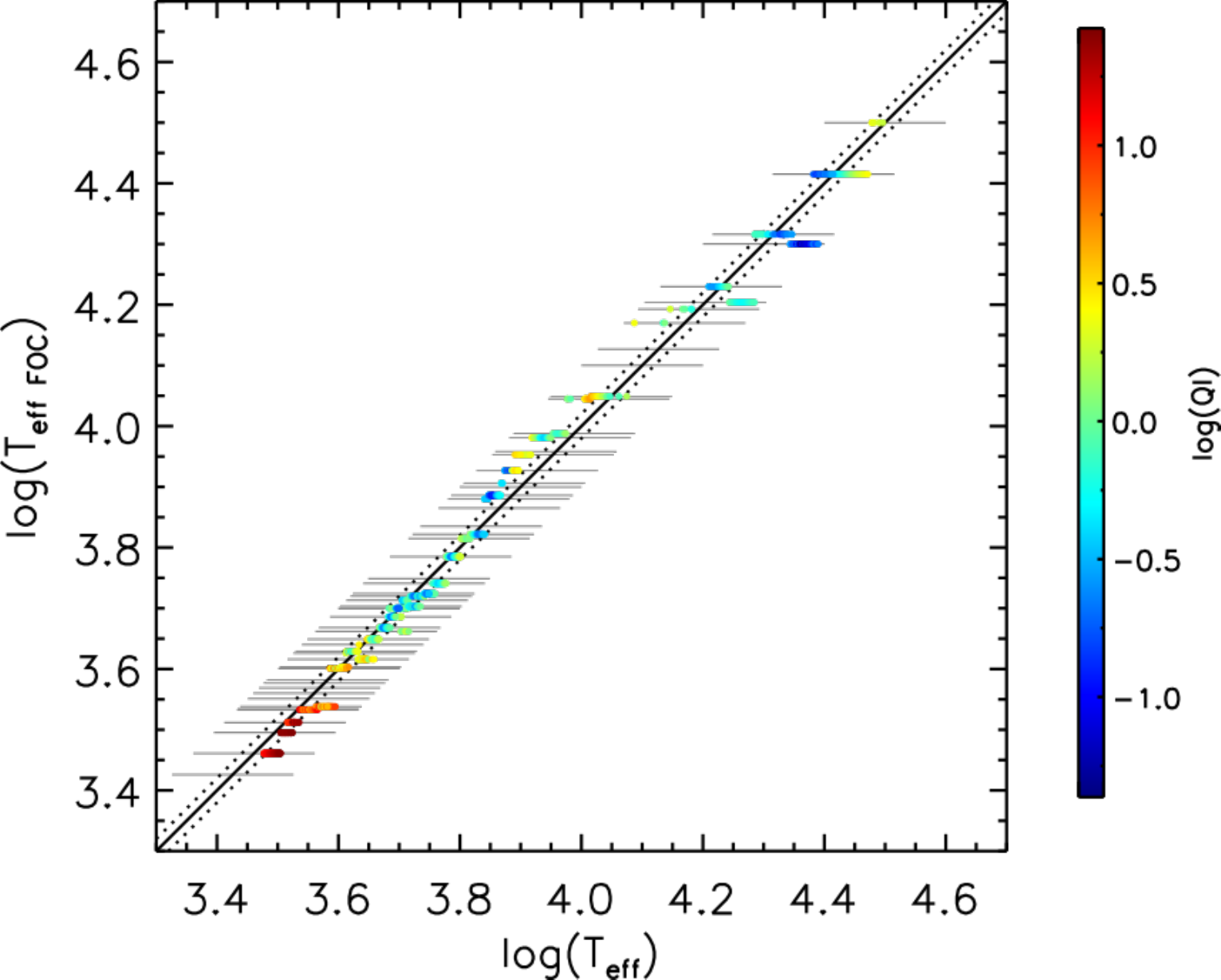}
\caption{Logarithm of the effective temperature corresponding to the spectral class found by the FOSC as a function of the model effective temperature. Each symbol corresponds to a simulated star and their colour depends on the quality index QI (see Eq.\ref{eq:QI}) as indicated by the colour scale on the right-hand side of the plot. The solid black line indicates the exact correlation and the dotted black lines shows the $\pm 5\%$ band. The thin horizontal grey lines indicate the effective temperatures of the templates available in the Pickles library. Top:  for the sample of "dwarfs" stars classified with luminosity classes IV and V. Bottom: for the sample of "giant" stars, luminosity classes I, II, and III.}
\label{fig:correltempd}
\end{center}
\end{figure}
We see that for late-type stars up to early-A stars (with $\log T_{\rm eff} \lesssim $ 4.1), the agreement between the model temperature and the result of the FOSC is good, with a spread of less than $\pm5\%$ in relative effective temperature difference for every spectral class of the templates. The precision on the $T_{\rm eff}$ corresponds to the limit imposed by the sampling of the library. There are small systematic effects for some classes, especially a small negative offset for late-K dwarfs (up to $-5\%$), but this is merely a consequence of the lack of accuracy of the effective temperature calibration of the templates \citep{Pickles1998}. While the effective temperature is recovered with good precision for M stars, the quality index of the fit is particularly bad for the coldest objects. This may be due to a mismatch between the atmospheric models for cool stars and the library of templates. This could also be a consequence of the lack of subgiant templates for the M class in the library (see Fig.~\ref{fig:pickles}). While the quality of the fit is poor, the overall shape of the SED of the models still agrees better with the template of corresponding effective temperature than with others.  This results in the correct estimate of the effective temperature of the star. 

On the other hand, for stars hotter than $\sim12 000$~K, the poor sampling of the temperature scale in the library results in a very wide spread of synthetic effective temperature classified within a single spectral class. For this temperature range, the result of the FOSC is at best uncertain to the level of a few subclasses. Notably, there is a trend to underestimate the temperature of the hottest stars. For example, stars with $4.4 \lesssim \log T_{\rm eff} \lesssim 4.5$ tend to be classified as B2IV subgiants, rather than B0V or B1V. At those high temperatures, the few and shallow absorption lines result in broadband integrated magnitudes that are easily confused with the magnitudes resulting from a cooler, gravity-broadened spectrum.

The same effects are seen for giant stars (with $\log g < 3.5$), shown in the bottom panel of Figure~\ref{fig:correltempd}. In this case the systematic offset seems to be more pronounced for early A-type giants, but the same overall accuracy and precision are reached. 

\subsection{Error introduced by the reddening}\label{app:degstred}
\begin{figure}
\begin{center}
\includegraphics[width=0.9\columnwidth]{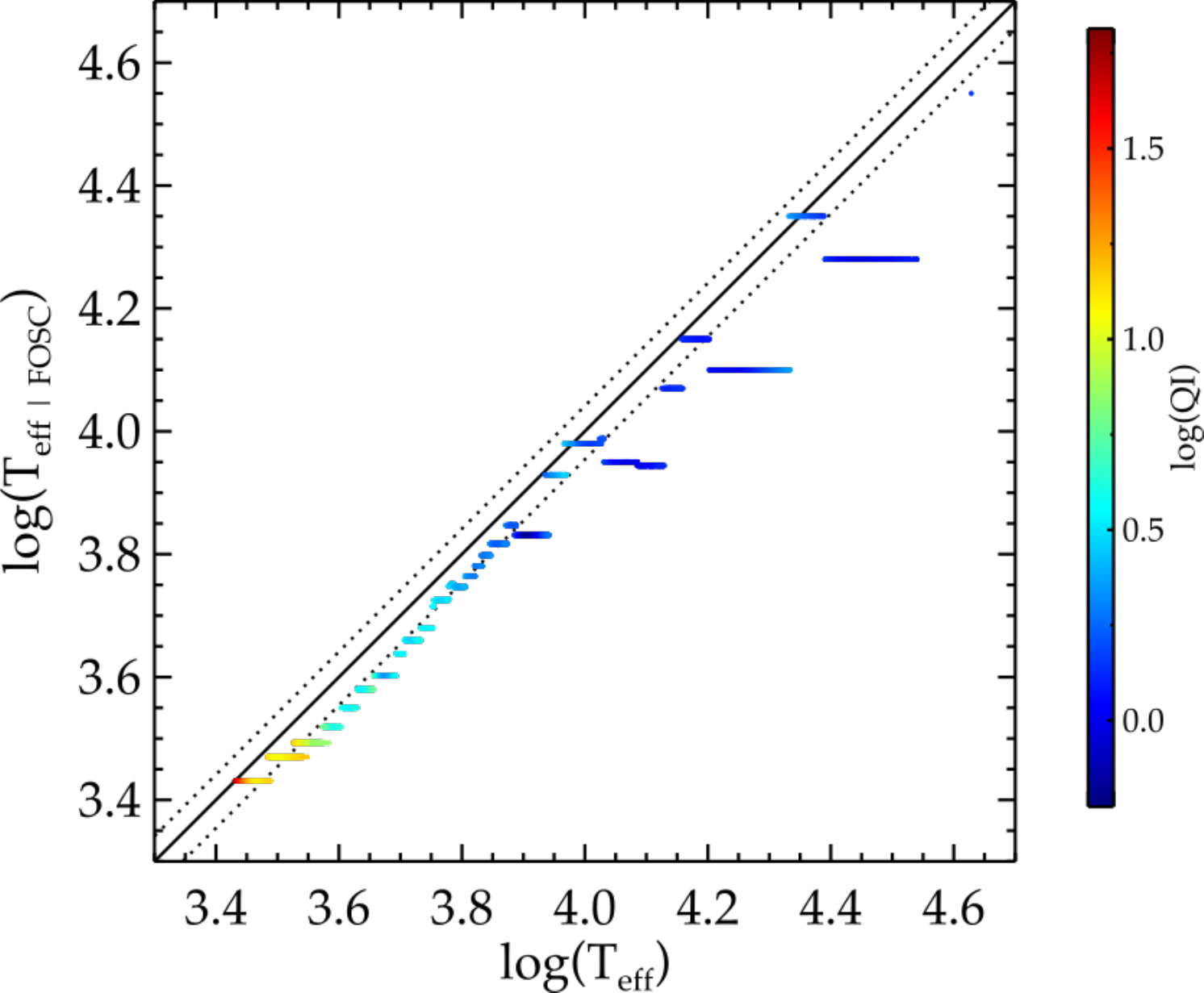}\\
\includegraphics[width=0.9\columnwidth]{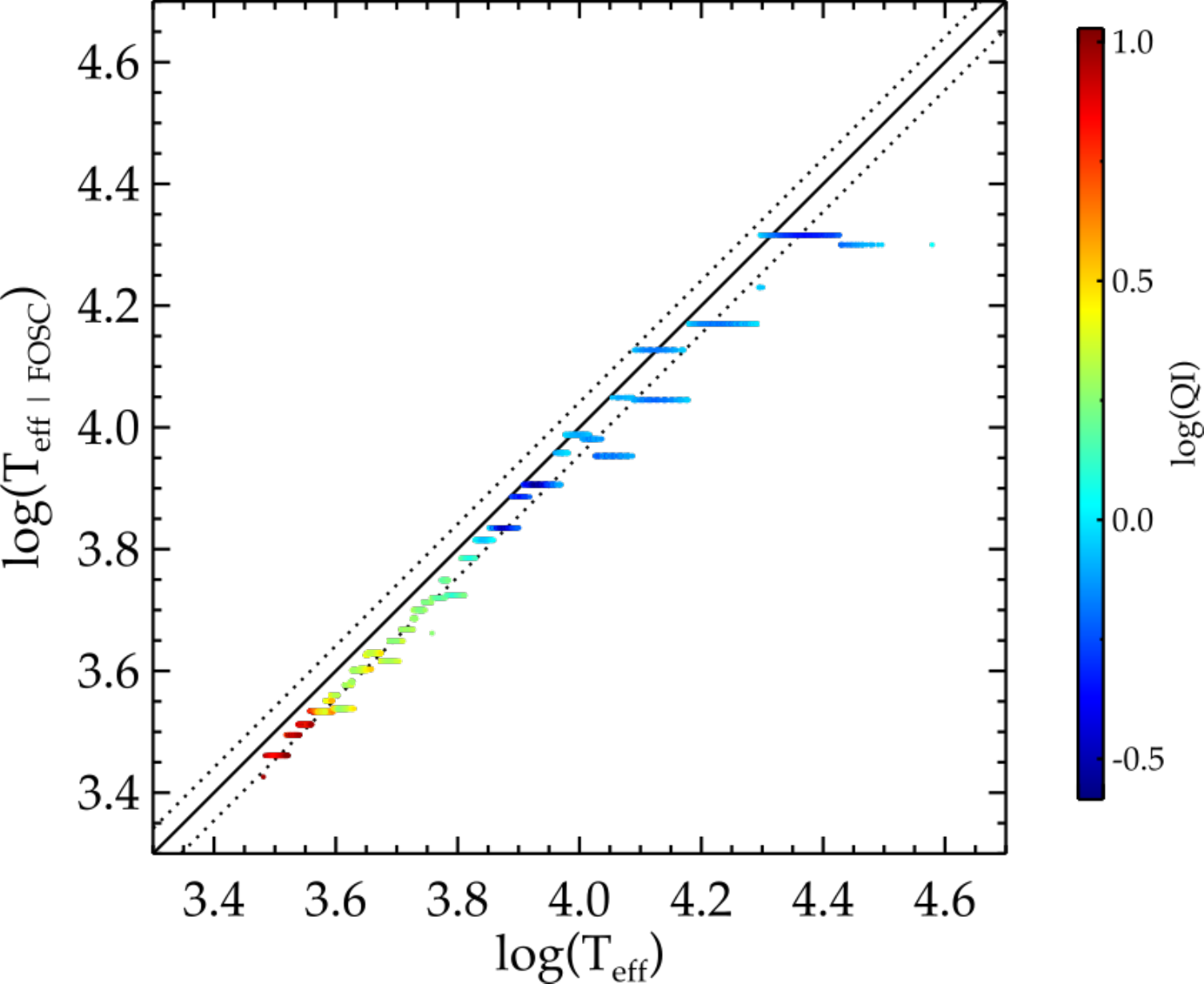}
\caption{Logarithm of the effective temperature corresponding to the spectral class found by the FOSC as a function of the model effective temperature for reddened stars with $E_{B-V}=4$. To test the method, the reddening was not let as a free parameter of the FOSC but was fixed to its actual value. Each symbol corresponds to a simulated star and their colour depends on value of the quality index, as indicated by the colour scale on the right-hand side of the plots. The solid black line is the exact correlation and the dotted black lines shows the $\pm 10\%$ relative effective temperature band. Top: luminosity classes IV and V. Bottom: luminosity classes I, II, and III. }
\label{fig:correltempred4}
\end{center}
\end{figure}

 To test the effect of reddening on our classification scheme, we added space absorption to the magnitudes of the sample of synthetic solar-metallicity stars shown in Fig.~\ref{fig:correltempd}. We used \citeauthor{Fitz1999} extinction law to produce reddened magnitudes for different values of $E_{B-V}$.  
First, we set the value of $E_{B-V}$ to its maximum value considered here, $E_{B-V}=4$, and ran the FOSC while fixing the parameter $E_{B-V}$ to its actual value. The results are shown in Fig.~\ref{fig:correltempred4}.

For early-type dwarfs and giants, a significant reddening causes a systematic underestimation of the effective temperature of up to about $-60\%$. For late-type stars, the results are better with an underestimation of the effective temperature of about $-5\%$ to $-15\%$. As explained in Sec.~\ref{sec:degstred}, this is a limitation imposed by the Pickles library. The difference between the infrared extension of the Pickles library and infrared colours of the target contribute significantly to the result of the fitting procedure. This results in the underestimation of the effective temperature of strongly reddened stars.  The effect is stronger for early-type stars because when the reddening is strong, the most distinctive features of their SED shifts right in the $J$ band, where only eight templates are based on actual observations.

\begin{figure}
\begin{center}
\includegraphics[width=0.9\columnwidth]{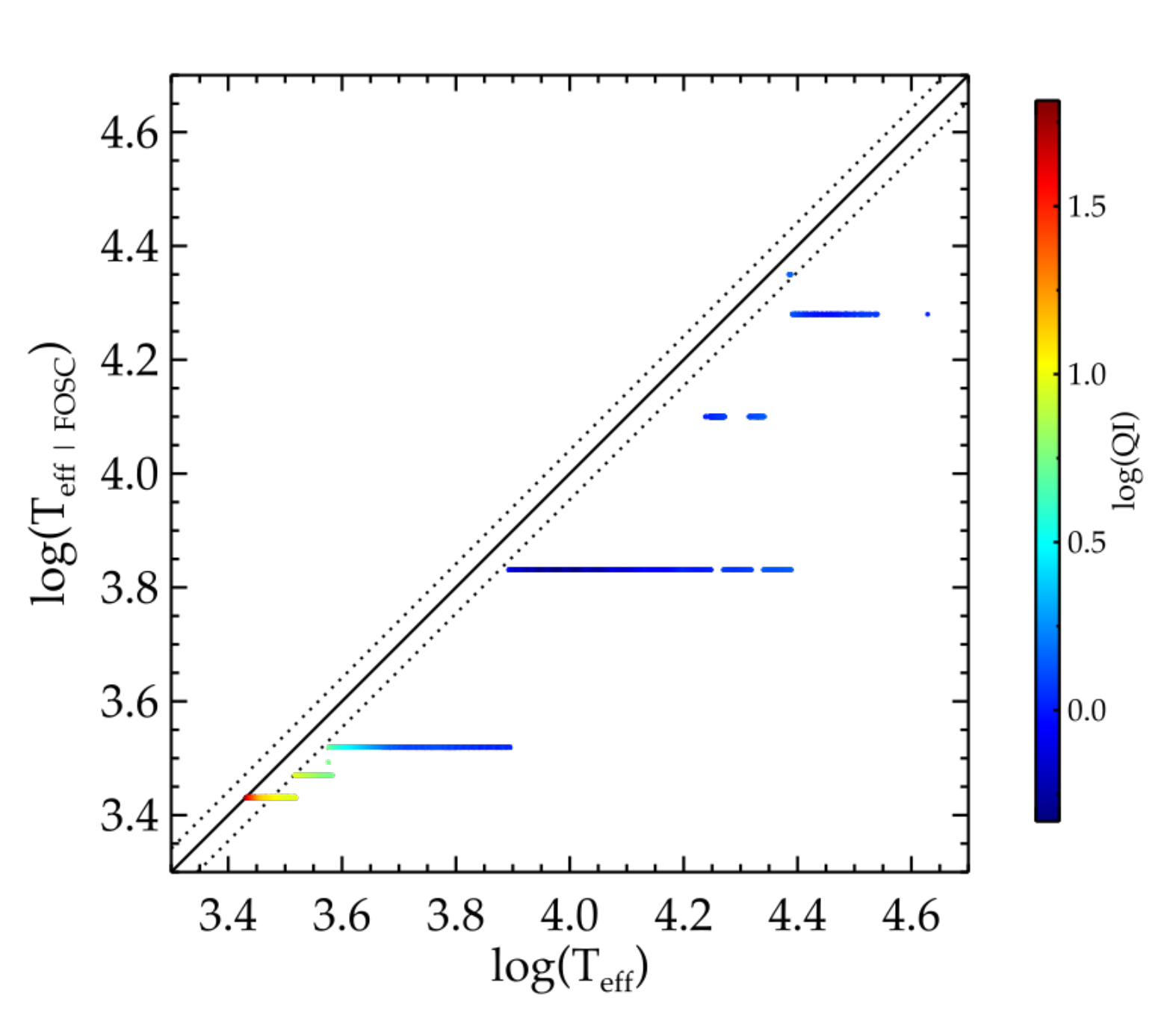}\\
\includegraphics[width=0.9\columnwidth]{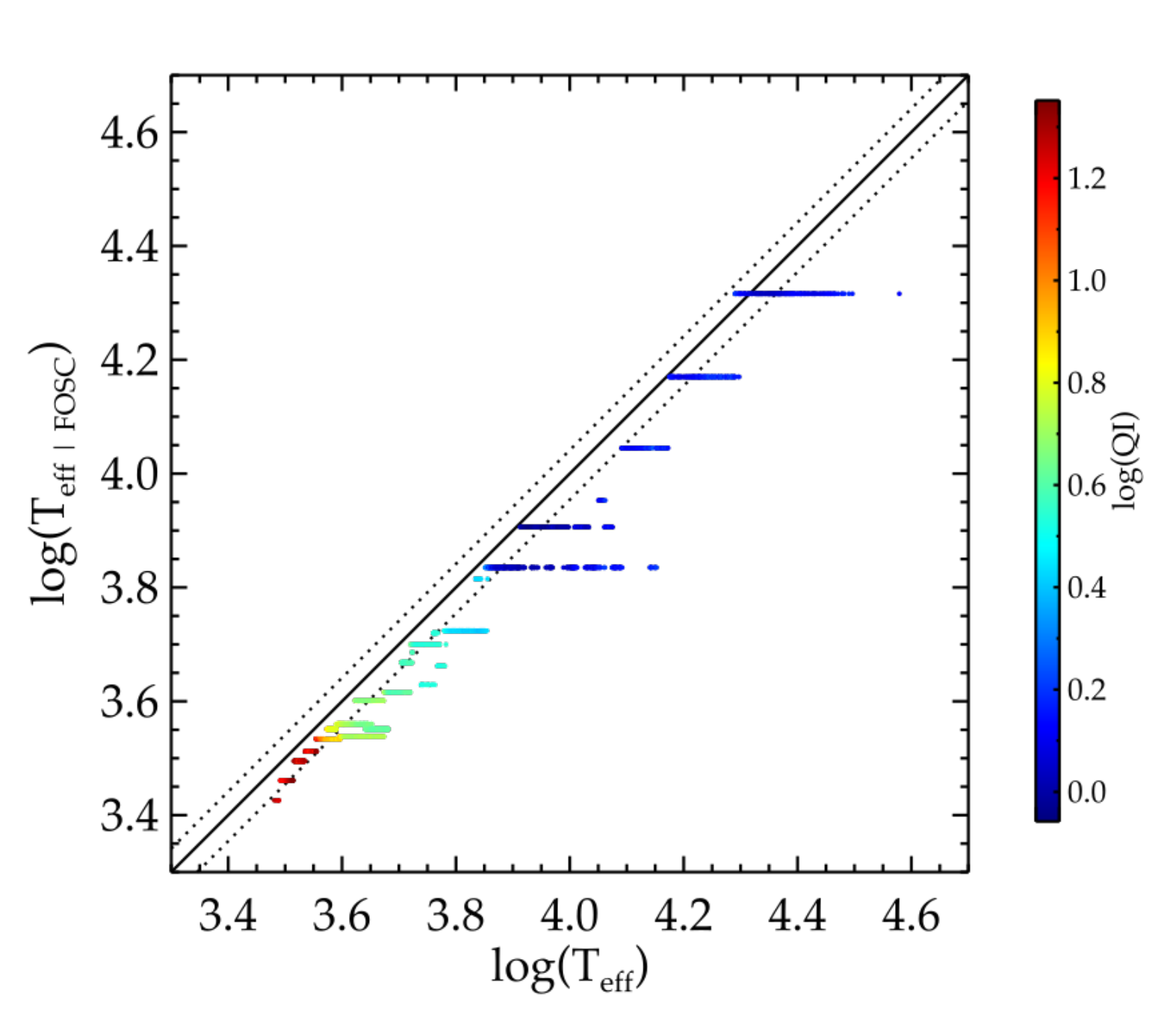}
\caption{Same set of synthetic stars as in Fig.~\ref{fig:correltempred4}, but this time the reddening is a free parameter of the classification.}
\label{fig:correltempred04}
\end{center}
\end{figure}

In the actual classification scheme, the value of the reddening is not known and is a free parameter. The absence of any a priori knowledge on the reddening makes the classification entirely inefficient for the most reddened stars. To illustrate this effect, we took the synthetic catalogue generated with a strong reddening of $E_{B-V} = 4$ and ran the FOSC this time not setting any prior on the value of the reddening. The results are shown in Fig.\ref{fig:correltempred04}.
For giant stars, compared to the case where the correct reddening is not a free parameter (Fig.~\ref{fig:correltempred4}) the performance of the classification is in most cases not significantly different, except for early F-type giants. But for dwarf stars the effect of reddening is very strong; the classification fails for late K- to early B-type stars, and is only marginally good for and M- and O-type stars. In every cases except for very late-M stars, the temperature is systematically underestimated. This cannot be explained by the inherent mismatch between models and templates alone, since in most cases the quality index has  the expected order of magnitude. This limitation is because the SED of a blue reddened star, integrated over broadband filters, is very similar to the SED of a red star with a lower $E_{B-V}$. 
\begin{figure}
\begin{center}
\includegraphics[width=0.9\columnwidth]{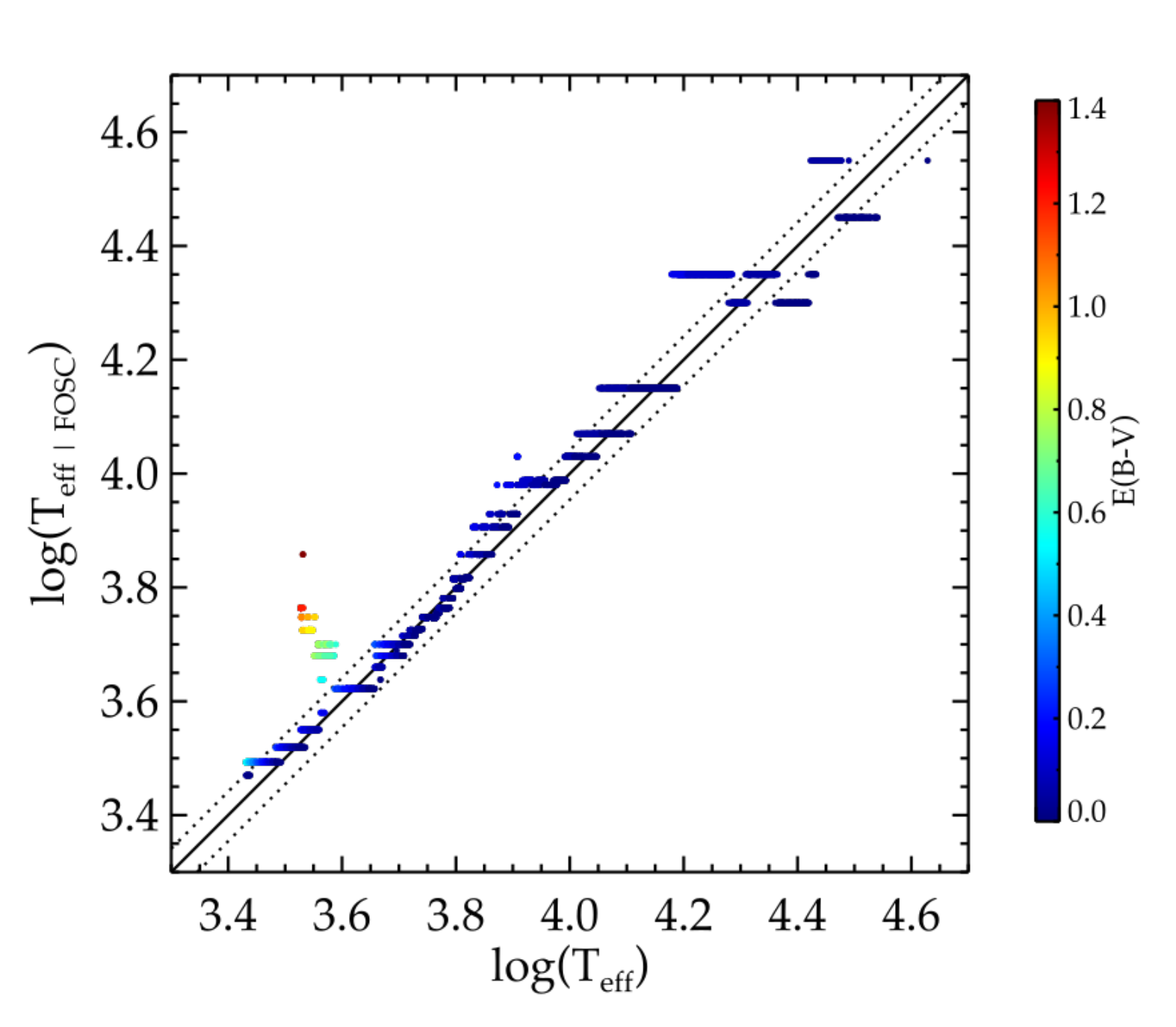}\\
\includegraphics[width=0.9\columnwidth]{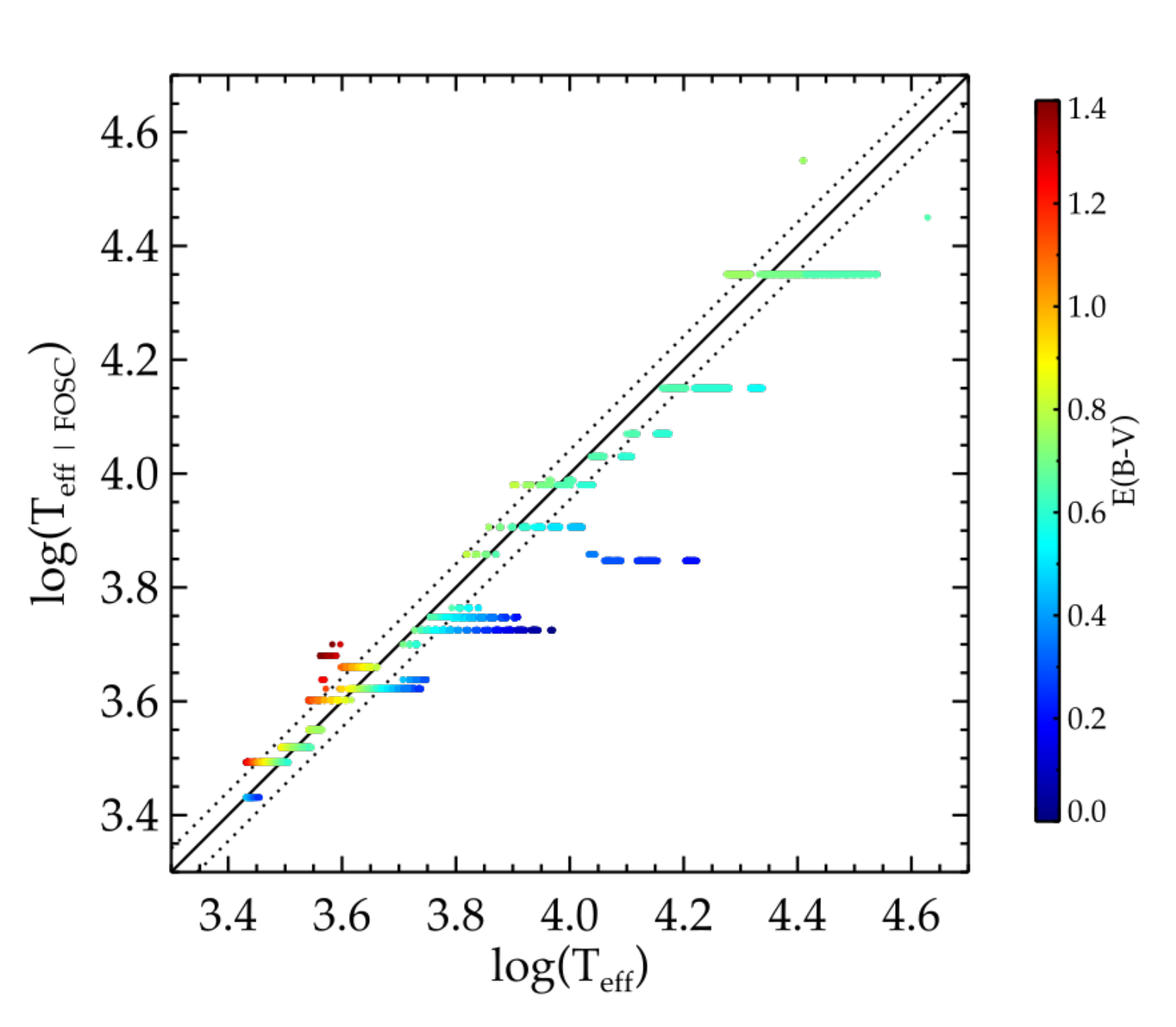}\\
\includegraphics[width=0.9\columnwidth]{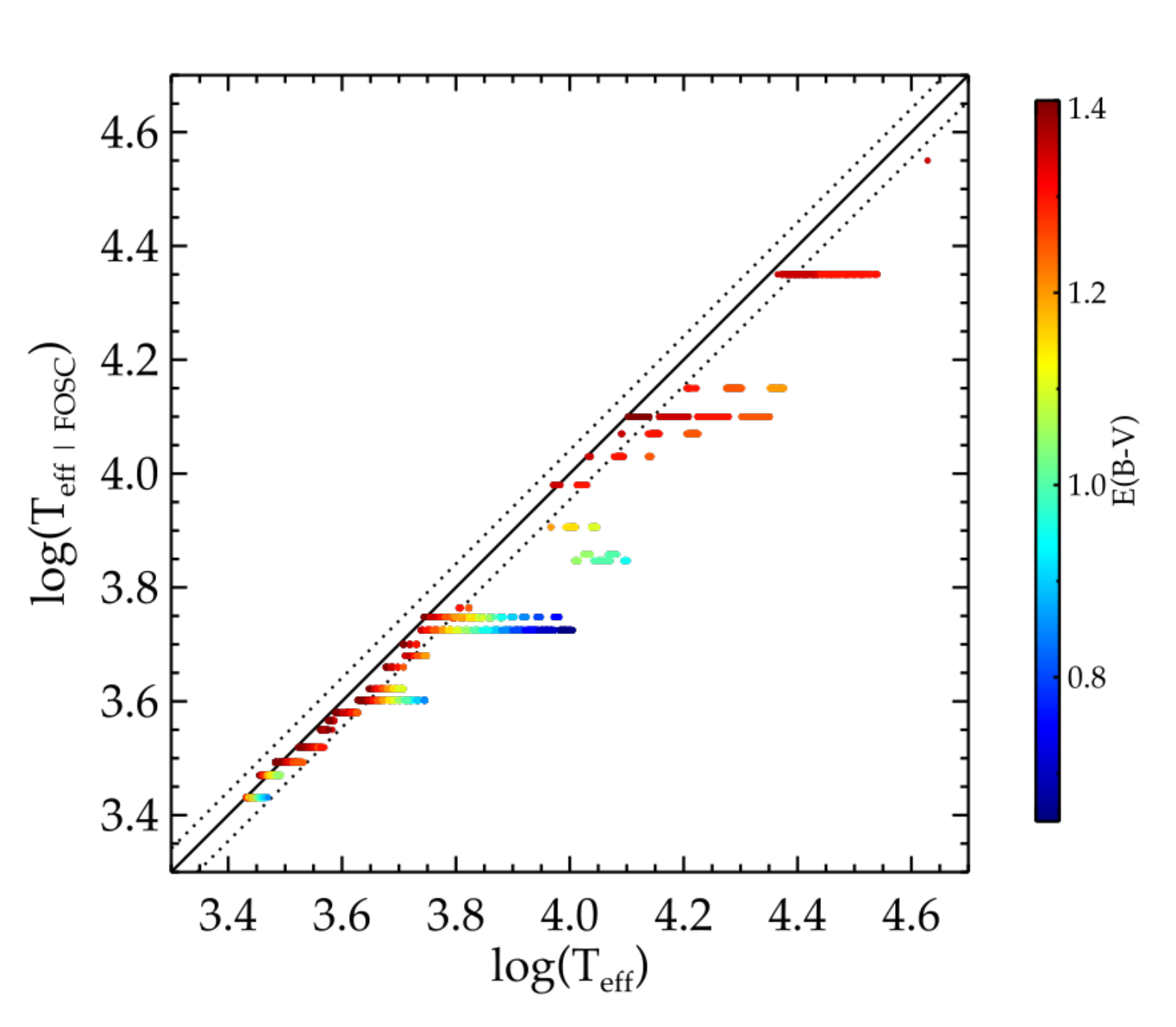}
\caption{Logarithm of the effective temperature found by the FOSC as a function of the known effective temperature of the synthetic star for dwarfs when letting the minimisation be performed on the interval $0 \leq E_{B-V} \leq 1.4$. The colour of the points depends on the value of $E_{B-V}$ determined by the FOSC, as indicated by the colour scale on the right-hand side of the plots. The solid black line is the exact correlation and the dotted black lines show the $\pm 10\%$ relative effective temperature band. Left: simulated magnitudes all have $E_{B-V} =0$. Middle: simulated magnitudes all have $E_{B-V}=0.7$. Right: simulated magnitudes all have $E_{B-V}=1.4$.}
\label{fig:correltempdred014}
\end{center}
\end{figure}
Thus, this effect could only be mitigated by constraining the interval of values of $E_{B-V}$ explored by the classifier. Only the upper boundary of the reddening can be constrained by observations (see Sec.~\ref{sec:reddening}). Nevertheless, the method remains valid because only a small portion of the dwarf targets are affected by such a strong reddening since the observations are limited in magnitude. Moreover, this effect becomes less important as the value of $E_{B-V}$ decreases. In fact, a typical value for the maximum value of $E_{B-V}$ in the fields observed by \corot{} is about $1.4$ (see Sec.~ \ref{app:ebv}). For illustration purposes, we show in Fig.~\ref{fig:correltempdred014} the error in temperature for three set synthetic dwarf stars with different values of $E_{B-V}$ (0, 0.7 or 1.4). For the three sets, the reddening was a free parameter, bounded between $0\leq E_{B-V} \leq 1.4$. Clearly the excursion in temperature is much smaller, but we see that the degeneracy between spectral type and reddening has significant effects. 

\begin{figure}
\begin{center}
\includegraphics[width=0.9\columnwidth]{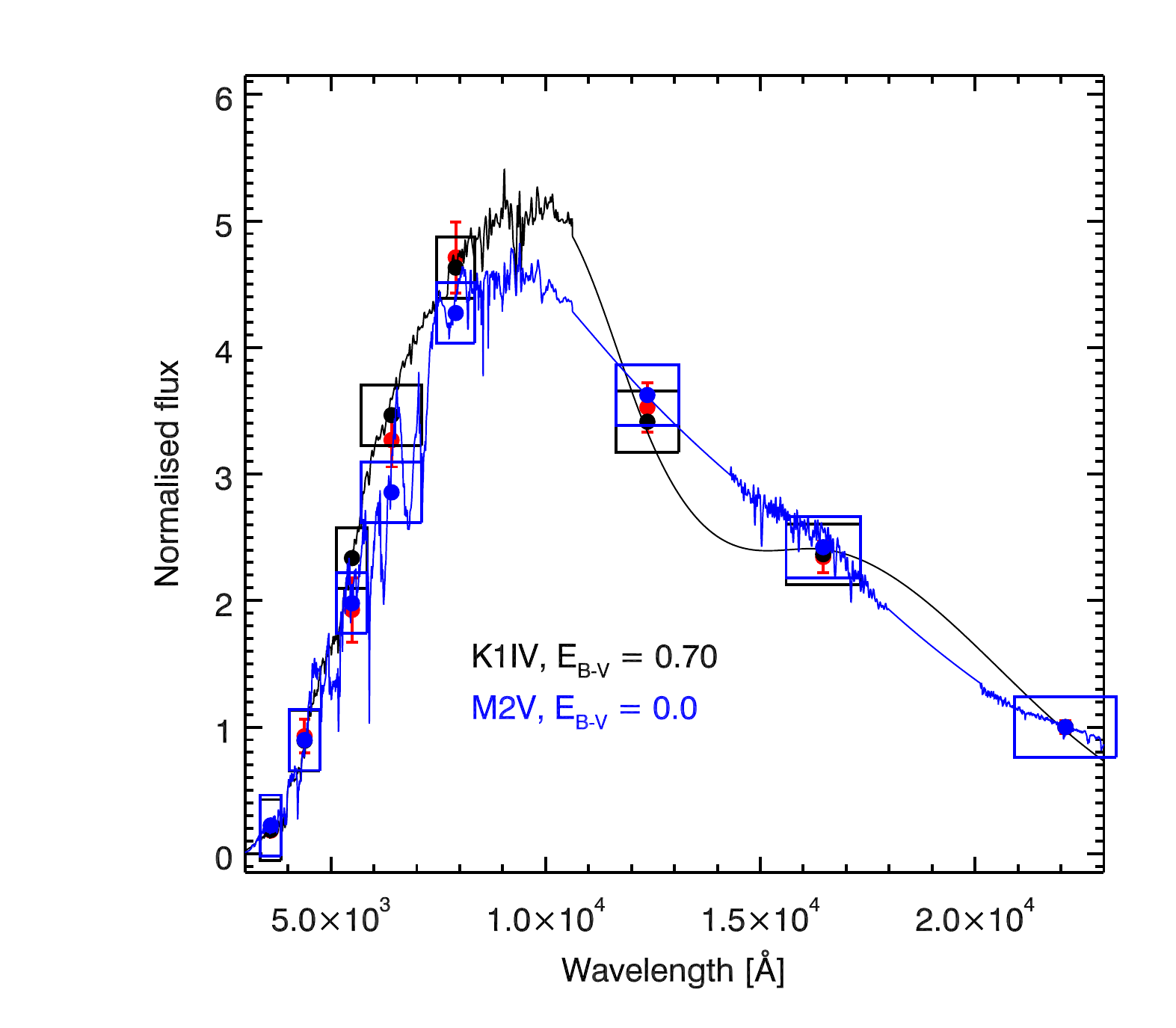}
\caption{Illustration for a case of misclassification. The target is a reddening-free dwarf star with an effective temperature corresponding to a M2V spectral type. Its magnitudes are plotted as red dots with error bars. The best fit given by the FOSC is a K1IV spectral type with $E_{B-V}=0.70$. The corresponding spectrum is plotted in a solid black line. The computed the flux in the bands of the filters is shown as black dots. The black box width gives the bandwidth of the corresponding filter. The same quantities are shown in blue for the correct spectrum.}
\label{fig:MKmis}
\end{center}
\end{figure}

For early-M types ($3.55 \lesssim \log T_{\rm eff} \lesssim3.6$), it appears that the classification returns the spectral type of a hotter but reddened star whenever possible, i.e. when their reddening does not already reach the maximum value explored by the FOSC (Fig.~\ref{fig:correltempdred014}, left and middle). For example, Fig.~\ref{fig:MKmis} presents the case of a non-reddened M2 dwarf that is wrongly classified as a reddened K1 subgiant.
The flux of the synthetic stars is overestimated in the red and near-infrared part of the spectrum compared to the flux of the template of the correct spectral type. Thus, the FOSC returns a reddened subgiant K type. The reddened SED of the subgiant fits better in the red because interstellar absorption dims the bluer part of the spectrum of the K type while retaining enough flux in the $R$ and $I$ bands. The reddened SED also fits well in the infrared magnitudes, where the gravity broadened lines of the K-type subgiant mimics the deep absorption bands of the cooler M type.

For some G-type stars (with $3.7 \lesssim \log T_{\rm eff} \lesssim3.76$), early-F and A stars ($3.8 \lesssim \log T_{\rm eff} \lesssim4.0$), and late-B type stars ($4.0 \lesssim \log T_{\rm eff} \lesssim4.2$), there is clearly a trend to underestimate the effective temperature and attribute a weaker reddening whenever possible (see Fig.~\ref{fig:correltempdred014}, middle and right). Again this is evidence either that the red and near-infrared synthetic magnitudes are overestimated or that the infrared magnitudes are underestimated, compared to the template of the corresponding spectral type. Then, the higher flux of the synthetic star in the red matches better the flux of a cooler, intrinsically brighter in the red template with a low value of $E_{B-V}$.

\begin{figure}
\begin{center}
\includegraphics[width=0.9\columnwidth]{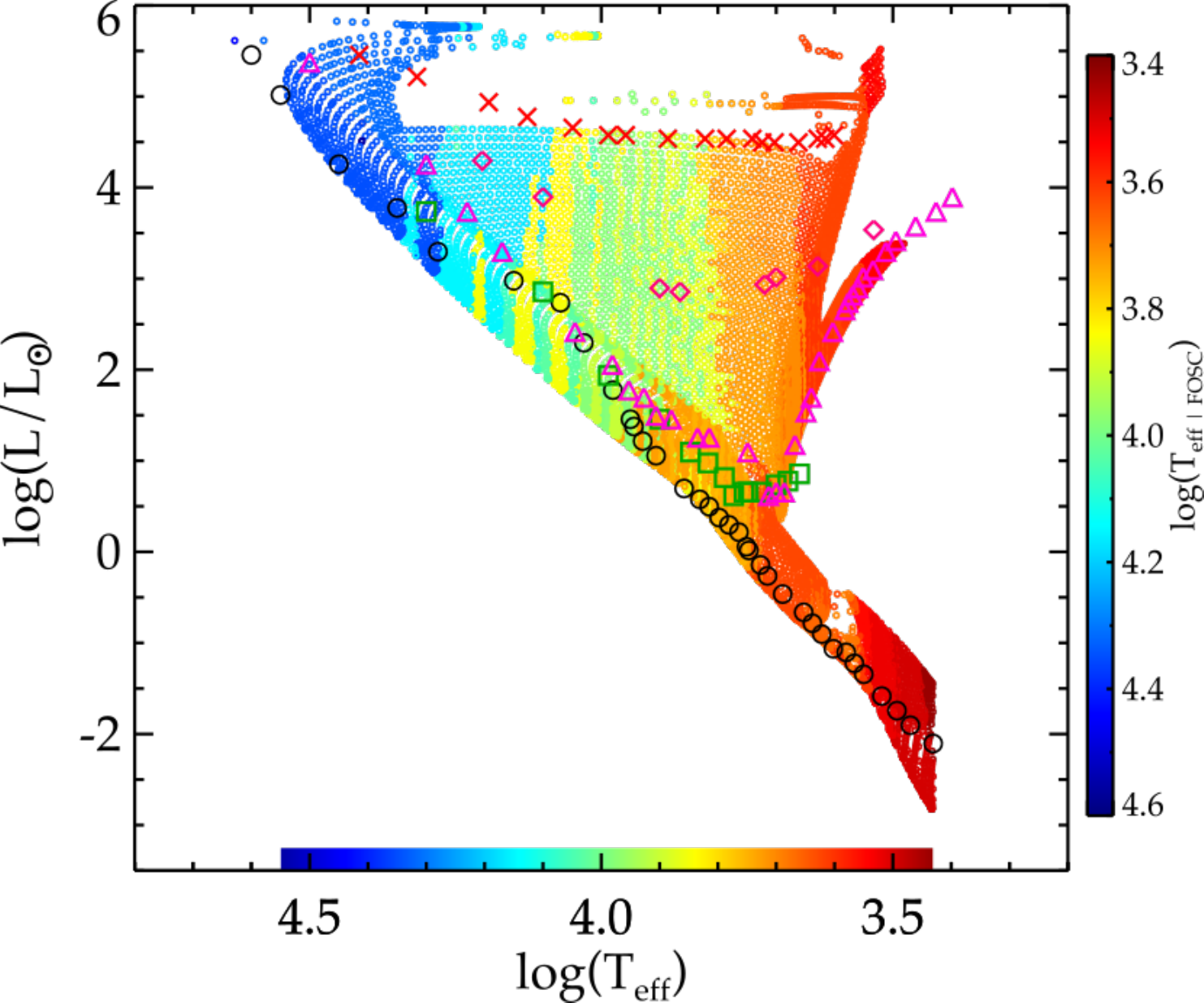}
\caption{Luminosity as a function of effective temperature of the synthetic stellar sample reddened with $E_{B-V}=0.7$. The colour of the points gives the effective temperature found by the FOSC, as indicated on the right-hand side of the plot. The same colour scale is given at the bottom of the plot for the effective temperature range of the synthetic sample to allow visual comparison. The position of the templates of the Pickles library is also indicated, as in Fig.~\ref{fig:pickles}.}
\label{fig:HRdred07}
\end{center}
\end{figure}

For moderate values of the reddening, those systematic effects stem ultimately from the mismatch between the templates of the Pickles library and the models used by the Padova isochrones to simulate magnitudes. As can be seen in Fig.~\ref{fig:HRdred07}, the synthetic stars that have the greater error on the spectral type are distributed along stripes in the HR diagram. Those stripes do not appear to have a direct relationship with the templates' sampling of the parameter space. The same effects are seen for the other luminosity classes, with the notable exception of early-F and A giants, which tend to be better classified than their dwarfs conterparts. Consequently, the misclassification may not be as important for real observed magnitudes, which should be a better match to the Pickles library, especially in the blue where the templates rely on a compilation of observed spectra.  This is investigated in detail by comparing the spectral type assigned by the FOSC to other independent classifications (see Sec.~\ref{sec:spectro}) 

\subsubsection{Missing magnitudes}\label{app:missmags}
\begin{figure}
\begin{center}
\includegraphics[width=0.9\columnwidth]{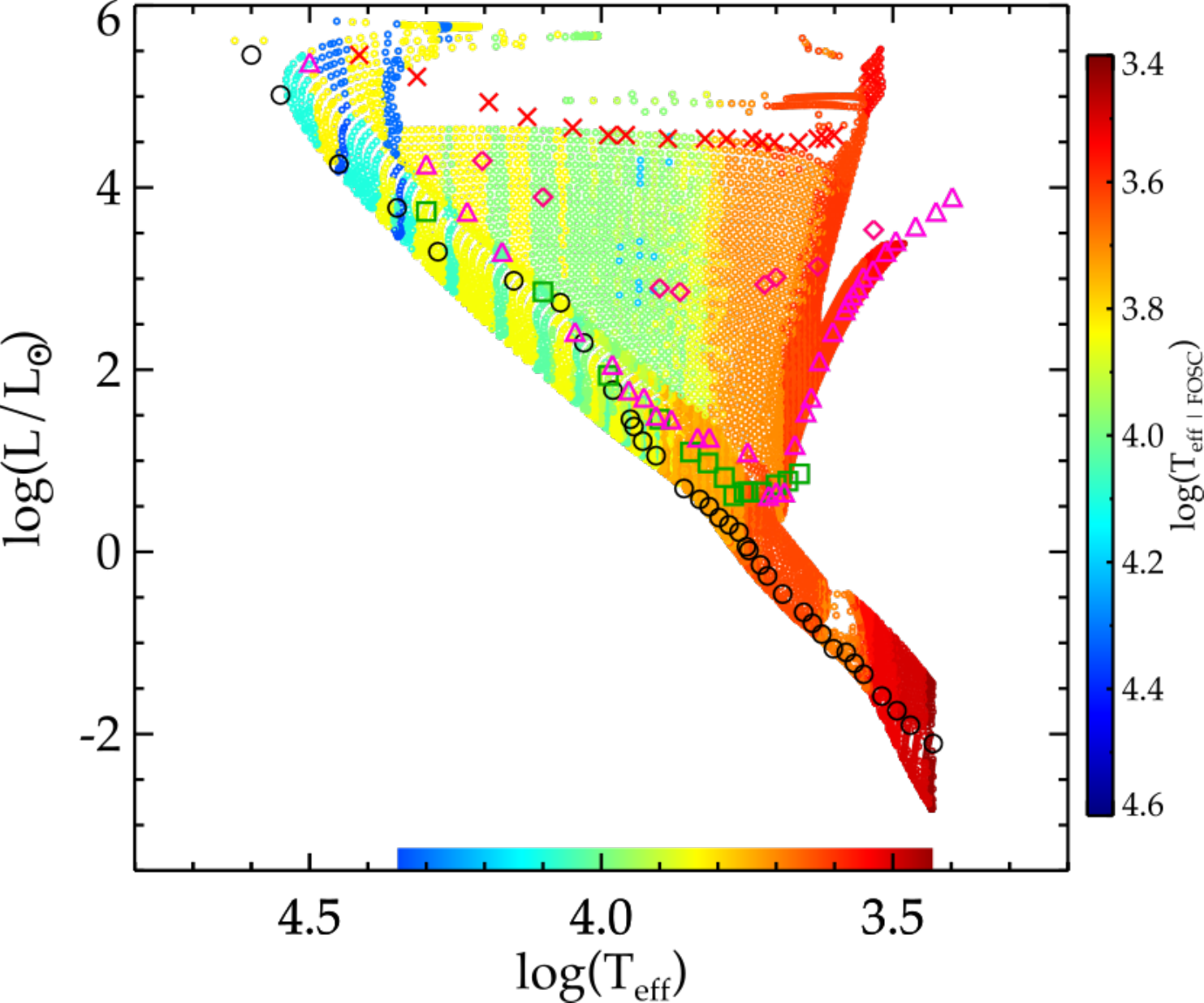}
\caption{Same as Fig.\ref{fig:HRdred07} but only using the $B, R, I, J, H,$ and $K$ bands.}
\label{fig:HRdred07250}
\end{center}
\end{figure}

If the classification is performed with a reduced number of filters, the inherent degeneracies of the method are expected to increase the error on the spectral type. To quantify this effect, we used the same set of synthetic stars described in Sec.~\ref{app:degstred}, but without using the $U$ band, and without both the $U$ and $V$ bands, to perform the classifications. 

\begin{figure}
\begin{center}
\includegraphics[width=0.9\columnwidth]{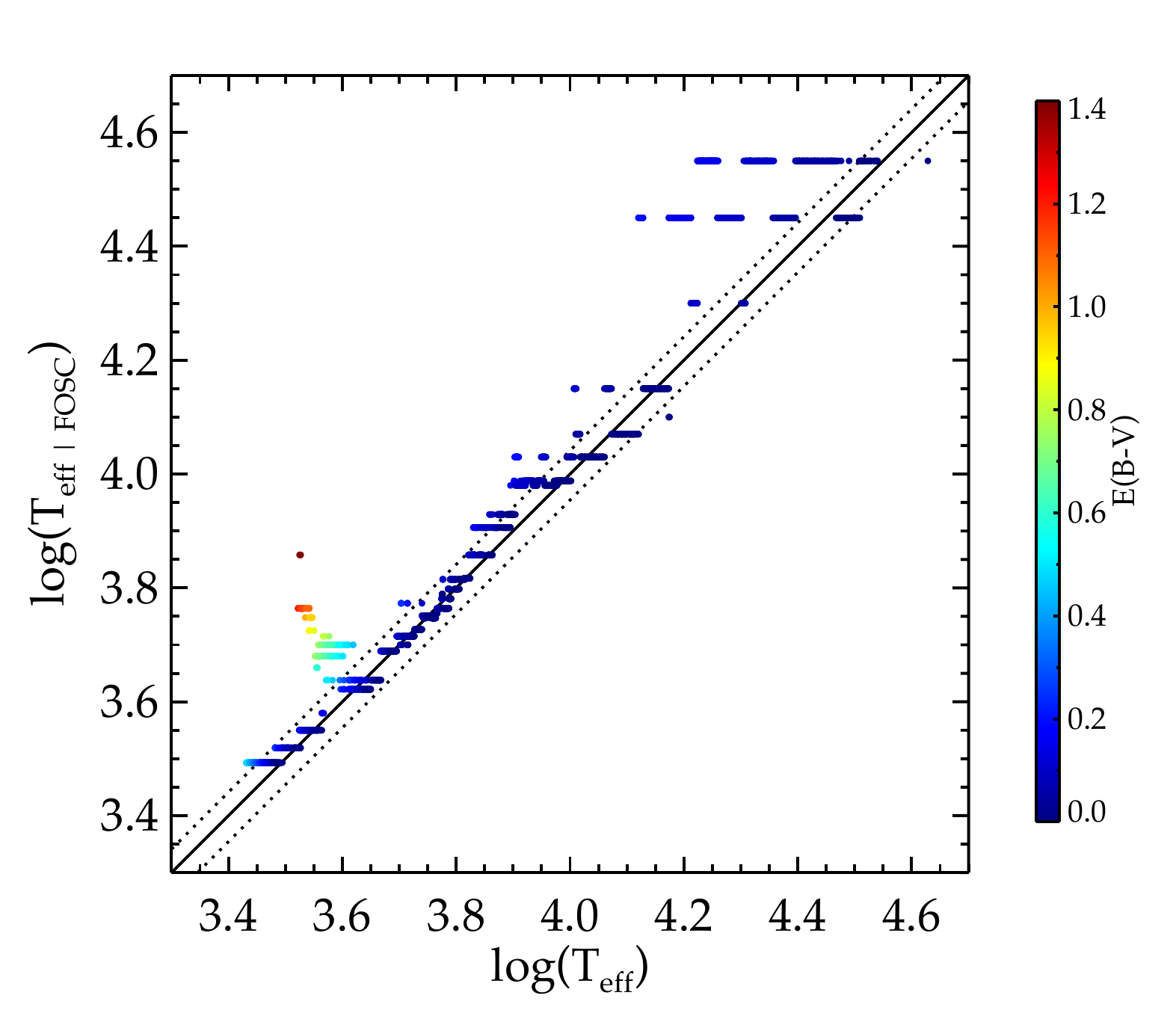}\\
\includegraphics[width=0.9\columnwidth]{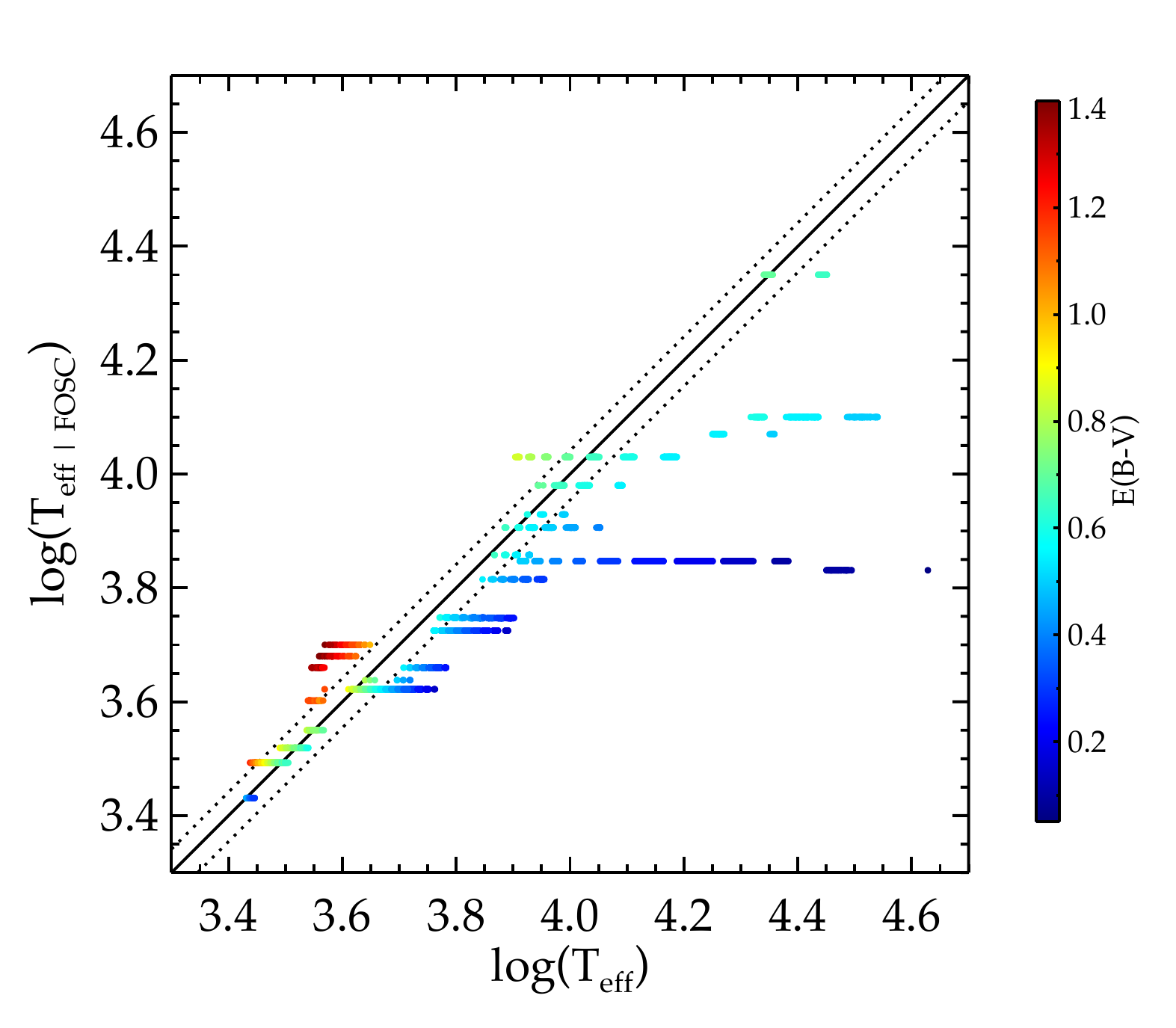}\\
\includegraphics[width=0.9\columnwidth]{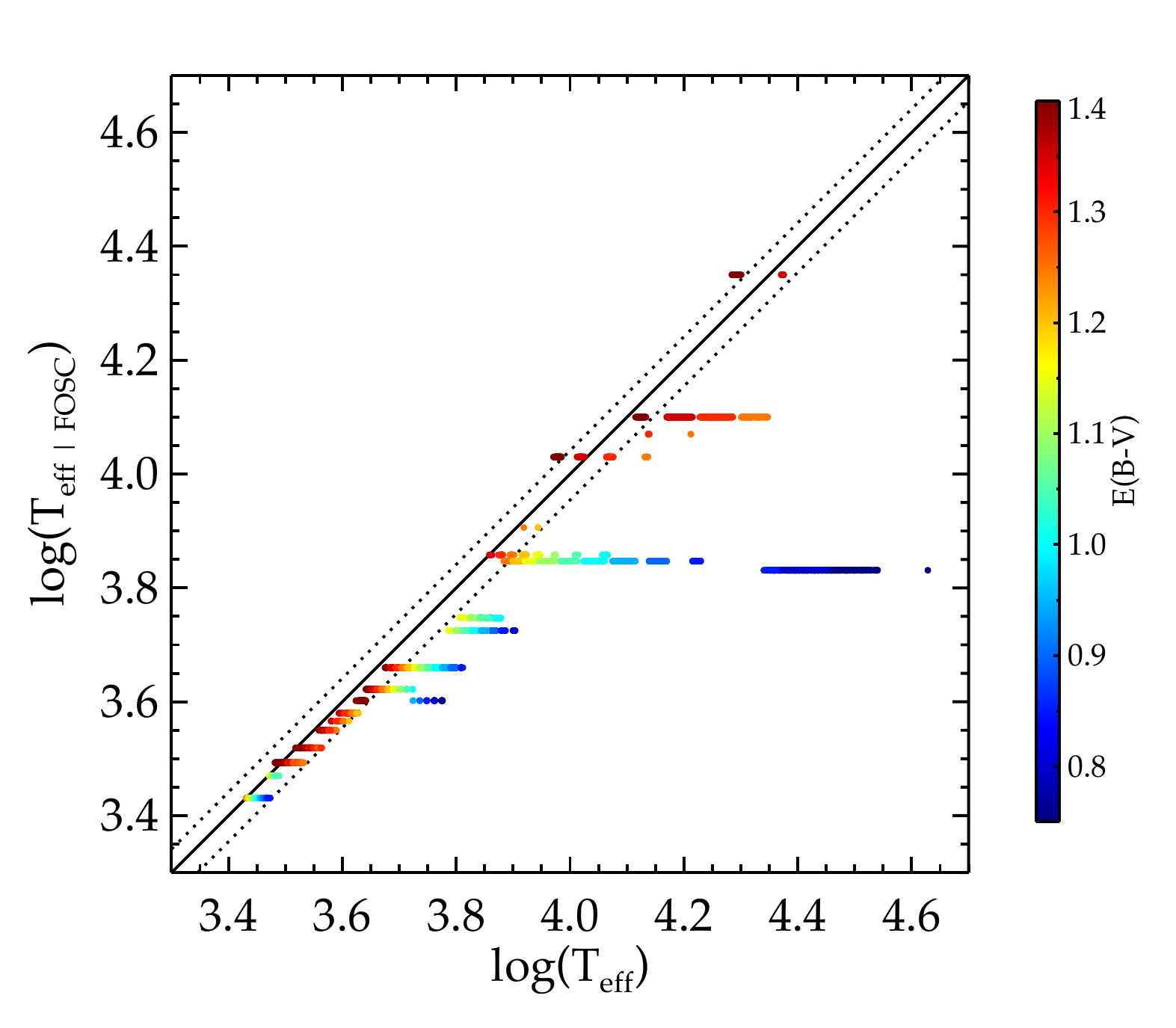}
\caption{Same as Fig.\ref{fig:correltempdred014} but only using the $B, R, I, J, H,$ and $K$ bands.}
\label{fig:correltempdred0146mags}
\end{center}
\end{figure}

When the reddening is close to 0, and when the FOSC is carried out assuming the correct value of the reddening, the effect of missing magnitudes is remarkably negligible for both dwarfs and giants. In fact, even when using just the $R, J, H,$ and $K$ magnitude there is no change in the spectral classification. But again, the main source of error of the method is the degeneracy with reddening. We reproduced the test presented in Fig.~\ref{fig:correltempdred014}, but this time without using the $U$ band and without both the $U$ and $V$ bands. The results are relatively insensitive to the omission of the $V$ magnitude, as long as there is information in the blue provided by the $B$ magnitude. However, as could be expected, the lack of the $U$ magnitude proves to be detrimental to the classification of early-type stars. We show in Fig.~\ref{fig:correltempdred0146mags} the results of the classification using only the $B, R, I, J, H,$ and $K$ bands for different values of the reddening, while letting the minimisation explore the $0\leq E_{B-V}\leq 1.4$ interval.

Not surprisingly, the most striking effect is seen for stars with $\log T_{\rm eff} \geq 4.0$, which have systematically underestimated temperature when the reddening is not negligible. They generally tend to be classified as mid-F dwarfs or subgiants. On the other hand, their temperature is systematically overestimated when $E_{B-V}\sim 0$. This means that O and B stars are impossible to differentiate when the $U$ and $V$ magnitude are missing, even for reddening-free stars (Fig.~\ref{fig:correltempdred0146mags}, left).

For A-type stars, missing $U$ and $V$ magnitudes do not degrade the spectral type determination when $E_{B-V}\sim 0$, and it shifts the interval of misclassification towards a higher temperature when $E_{B-V}>0$. For very reddened A stars, the missing magnitudes actually improve the mean error on the effective temperature. Indeed, for $E_{B-V}=1.4$ and with all magnitudes available, the whole A range tends to be classified as G8 subgiants with very low value of the reddening. But when the $U$ and $V$ magnitudes are missing, the classification returns a F1 subgiant with a reddening closer to its actual value.

For F- and G-type stars, once again, missing magnitudes do not have a significant effect when $E_{B-V}\sim 0$. For non-null values of $E_{B-V}$, the classification is significantly affected with a more systematic attribution of a cooler, less reddened spectral type with an error on the temperature ranging from -10\% up to -30\% when the reddening is strong.

Missing $U$ and $V$ magnitudes do not affect the classification of early-K type stars, whatever the value of the reddening. For late-K dwarfs the classification tends to return a slightly more reddened subgiant KI or K0-type whenever possible (Fig.~\ref{fig:correltempdred0146mags}, left and middle). This trend extends to early M types ($3.55 \lesssim \log T_{\rm eff} \lesssim3.6$). This was already the case where all magnitudes were used, but it is reinforced by the lack of the $V$ magnitude. For stars with $\log T_{\rm eff} \lesssim3.55$, the missing magnitudes do not affect the determination of the spectral type whatever the reddening.   

Finally, the same study has been carried out for giant stars. The effect of missing magnitudes show the same trends and amplitudes except for late-A giants, which can have a relative error in temperature up to $100\%$, while the F giants tend to be better classified than the dwarfs of the same types with a relative error reduced to about $-25\%$. 

\end{appendix}

\label{lastpage}
\end{document}